\documentclass[prd,aps,showpacs,floats,letterpaper,floatfix,groupedaddress,eqsecnum]{revtex4}

\usepackage{amssymb,amsmath}
\usepackage[dvips]{graphicx}

\usepackage{dcolumn,epsfig}

\newcommand{\hL}{\hat{\mathbf{L}}}
\newcommand{\hS}{\hat{\mathbf{S}}}

\newcommand{\vz}{\mathbf{z}}

\newcommand{\vS}{\mathbf{S}}
\newcommand{\vL}{\mathbf{L}}
\newcommand{\vn}{\mathbf{n}}

\newcommand{\beq}{\begin{equation}}
\newcommand{\eeq}{\end{equation}}
\newcommand{\bes}{\begin{subequations}}
\newcommand{\ees}{\end{subequations}}
\newcommand{\bea}{\begin{eqnarray}}
\newcommand{\eea}{\end{eqnarray}}
\newcommand{\ba}{\begin{array}}
\newcommand{\ea}{\end{array}}
\newcommand{\beqn}{\begin{eqnarray*}}
\newcommand{\eeqn}{\end{eqnarray*}}
        
\newcommand{\p}{\partial}
\newcommand{\f}[2]{\frac{#1}{#2}}
\newcommand{\lisa}{{\em LISA}}

\def\bt{\bar \theta}
\def\bph{\bar \phi}

\def\nn{\nonumber}

\def\ii{{\rm i}}

\newlength{\sizeonefig}
\newlength{\sizetwofig}
\begin{document}

\title{Estimating spinning binary parameters\\
and testing alternative theories of gravity with LISA}

\author{Emanuele Berti}

\altaffiliation[Present address: ]{McDonnell Center for the Space Sciences, Department of
Physics, Washington University, St. Louis, Missouri 63130}
\email[Email: ]{berti@wugrav.wustl.edu}

\author{Alessandra Buonanno}

\altaffiliation[Also at: ]{F\'{e}d\'{e}ration de Recherche Astroparticule et
Cosmologie,
Universit\'{e} Paris 7, 2 place Jussieu, 75251 Paris, France}
\email[Email: ]{buonanno@iap.fr}

\author{Clifford M. Will}

\altaffiliation[Permanent address: ]{McDonnell Center for the Space Sciences, Department of
Physics, Washington University, St. Louis, Missouri 63130}
\email[Email: ]{cmw@wuphys.wustl.edu}

\affiliation{Groupe de Gravitation et Cosmologie (GReCO), 
Institut d'Astrophysique de Paris (CNRS), \\
98$^{\rm bis}$ Boulevard Arago, 75014 Paris, France}

\begin{abstract}
We investigate the effect of spin-orbit and spin-spin couplings on the
estimation of parameters for inspiralling compact binaries of massive
black holes, and for neutron stars inspiralling into intermediate-mass
black holes, using hypothetical data from the proposed Laser
Interferometer Space Antenna ({\em LISA}).  We work both in Einstein's
theory and in alternative theories of gravity of the scalar-tensor and
massive-graviton types.  We restrict the analysis to non-precessing
spinning binaries, i.e. to cases where the spins are aligned normal to
the orbital plane.  We find that the accuracy with which intrinsic
binary parameters such as chirp mass and reduced mass can be estimated
within general relativity is degraded by between one and two orders of
magnitude.  We find that the bound on the coupling parameter
$\omega_{\rm BD}$ of scalar-tensor gravity is significantly reduced by
the presence of spin couplings, while the reduction in the
graviton-mass bound is milder.  Using fast Monte-Carlo simulations of
$10^4$ binaries, we show that inclusion of spin terms in massive
black-hole binaries has little effect on the angular resolution or on
distance determination accuracy.  For stellar mass inspirals into
intermediate-mass black holes, the angular resolution and the distance
are determined only poorly, in all cases considered.
We also show that, if {\em LISA}'s low-frequency noise sensitivity can
be extrapolated from $10^{-4}$~Hz to as low as $10^{-5}$~Hz, the
accuracy of determining both extrinsic parameters (distance, sky
location) and intrinsic parameters (chirp mass, reduced mass) of
massive binaries may be greatly improved.
\end{abstract}

\vspace{0.5cm}
\date{\today}


\maketitle

\section{Introduction}
\label{sec1}

The Laser Interferometer Space Antenna ({\em LISA}) is being designed
to detect gravitational-wave (GW) signals in the frequency band between
$10^{-4}$ Hz and $10^{-1}$ Hz~\cite{danzmann}. Ground-based
interferometers such as LIGO, GEO, VIRGO and TAMA are sensitive in a
higher frequency band, between $10$ Hz and $10^3$ Hz. Operating at
these low frequencies, {\em LISA} can detect, among other sources,
inspirals and mergers of massive black holes (MBH) with masses in the
range $10^4 \mbox{-} 10^7 M_\odot$.  Another promising source is the
inspiral and capture of stellar-mass compact objects -- such as
neutron stars (NS) or black holes (BH) -- by intermediate-mass black
holes with masses in the range $10^2 \mbox{-} 10^5 M_\odot$.

Gravitational radiation reaction drives the inspiral of these
binaries.  The amplitude and phase of the gravitational-wave signal
carry information about binary parameters, such as masses and spins,
and about the location and distance of the binary.  They may also be
different in different theories of gravity.  Therefore {\em LISA} can
provide important astrophysical information, yield interesting tests
of fundamental physics, and place bounds on alternative theories of
gravity.  In this paper, we consider, along with standard general
relativity, theories of the scalar-tensor type (the simplest exemplar
being that of Brans and Dicke) and theories with an effective mass in
the propagation of gravitational waves (which we call massive graviton
theories, for short).  In scalar-tensor theories the phasing evolution
is modified predominantly by the presence of dipole gravitational
radiation reaction in the orbital evolution (in general relativity
the lowest radiative multipole moment is the quadrupole).  In massive
graviton theories the gravitational wave propagation speed depends on
wavelength: this generates a distortion in the time of arrival (and in
the wave phasing) with respect to general relativity,
similar to the dispersion of radio waves by interstellar plasma.

Previous
papers~\cite{willST,willgraviton,scharrewill,willyunes,damourfarese}
derived bounds on the graviton mass, on the Brans-Dicke parameter
$\omega_{\rm BD}$ and on parameters describing more general
scalar-tensor theories under the assumption that the compact objects
do not carry spin. In this paper we investigate the effect of
spin-orbit and spin-spin couplings both on the estimation of
astrophysical parameters within general relativity, and on bounds that
can be placed on alternative theories.  We restrict our analysis to
non-precessing spinning binaries, i.e.  binaries whose spins are
perpendicular to the orbital plane.  The effect of non-aligned spins
and the resulting precessions will be considered in future work.

Within Einstein's general relativity, various authors have
investigated the accuracy with which {\em LISA} can determine binary
parameters including spin effects. Cutler~\cite{CC} determined {\em
LISA}'s angular resolution and evaluated the errors on the binary
masses and distance considering spins aligned or anti-aligned with the
(orbital) angular momentum.  Hughes~\cite{SH} investigated the
accuracy with which the redshift can be estimated (if the cosmological
parameters are derived independently), and considered the black-hole
ring-down phase in addition to the inspiralling signal. 
Seto~\cite{seto} included the effect of finite armlength (going beyond
the long wavelength approximation) and found that the distance and
angular resolution accuracy improve. This happens because the response
of the instrument when the armlength is finite depends strongly on the
location of the source, which is tightly correlated with the distance
and the direction of the orbital angular momentum. 
Vecchio~\cite{vecchio} provided the first estimate of parameters for
precessing binaries when only one of the two supermassive black holes
carries spin.  He showed that modulational effects decorrelate the
binary parameters to some extent, resulting in a better estimation of
the parameters compared to the case when spins are aligned or
antialigned with angular momentum. More recently, Hughes and
Menou~\cite{HM} studied a class of binaries, which they denoted {\it
golden} binaries, for which the inspiral and ring-down phases could be
observed with good enough precision to carry out valuable tests of
strong-field gravity. 

These earlier works (except for~\cite{HM}) adopted analytical
approximations to {\em LISA}'s instrumental noise~\cite{CC}, augmented
by an estimate of white-dwarf confusion noise~\cite{BH} in the
low-frequency band.  In this paper we model the {\em LISA} noise curve
by a similar -- albeit slightly updated -- analytical
approximation~\cite{BC}.  This noise curve has the advantage of being
given in analytical form, and reproduces very well the salient
features of numerical noise curves available online from the {\em
LISA} Sensitivity Curve Generator (SCG)~\cite{SCG}, a tool sponsored
by the {\em LISA} International Science Team.

Our central conclusions are as follows.  Inclusion of non-precessing
spin-orbit and spin-spin terms in the gravitational-wave phasing
generally reduces the accuracy with which the parameters of the binary
can be estimated.  This is not surprising, since the parameters are
highly correlated, and adding parameters effectively dilutes the
available information.  Such an effect has already been described
within Einstein's general relativity in the context of ground-based
detectors of the LIGO/VIRGO type~\cite{poissonwill,KKS}. For example,
for massive black-hole binaries at 3 Gpc, we find that including
spin-orbit terms degrades the accuracy in measuring chirp mass by
factors of order 10, and in measuring the reduced mass parameter by
factors of order 20 -- 100; while including spin-spin terms further
degrades these accuracies by factors of order 3 and 5, respectively.
For neutron stars inpiralling into intermediate-mass black holes
(IMBH) with masses between 1000 and $10^4$ solar masses, the
corresponding reductions are factors of order 20 and 5 -- 30 in chirp
mass and reduced mass parameter, respectively, when spin-orbit is
included, and additional factors of order 4 and 7, respectively, when
spin-spin terms are included.

When we consider placing bounds on alternative theories of gravity,
for technical reasons, we treat only spin-orbit terms.  The source of
choice to place bounds on the coupling parameter $\omega_{\rm BD}$ of
scalar-tensor gravity is the inspiral of a neutron star into an
intermediate-mass black hole. We first reproduce results of earlier
work~\cite{willyunes}, apart from small differences arising from
corrected normalization of the \lisa ~noise curve.  Including
spin-orbit effects reduces the bound on $\omega_{\rm BD}$
significantly, by factors of order 10 -- 20.  For example for a $1.4
\, M_\odot$ neutron star inspiralling into a $400 \, M_\odot$ black
hole, the bound on $\omega_{\rm BD}$ goes from $8\times 10^5$ to
$40,000$ when spin-orbit terms are included.  The latter bound should
be compared with the bound of $40,000$ from {\em Cassini} measurements
of the Shapiro time delay~\cite{bertotti}.

The effect of including spin on bounding the graviton mass is more
modest. In this case, the source of choice is the inspiral of binaries
of massive black holes.  For masses ranging from $10^5 M_\odot$ to
$10^7 M_\odot$, the reduction in the bound induced by the inclusion of
spin-orbit terms is only a factor of 4 to 5.

We consider the effect of spin terms on the angular and
distance resolution of \lisa. We find that spin couplings have a mild
effect on the angular resolution, on the distance and, as a consequence,
on the redshift determination for massive black-hole binaries. By
contrast, for stellar mass objects inspiralling into intermediate-mass black
holes, neither distance nor location on the sky is very well determined.

{\em LISA} can observe massive black-hole binaries with large SNR
out to large values of the cosmological redshift. If the corresponding
mass and distance determinations are accurate enough, {\em LISA} will
be an invaluable tool to study structure formation in the early
Universe. Using Monte Carlo simulations we find that {\em LISA} can
provide accurate distance determinations out to redshift $z\sim 2$ for
source masses $\sim 10^7 M_\odot$, and out to $z\sim 4$ for source
masses $\sim 10^6 M_\odot$. Mass determinations strongly depend on an
accurate treatment of spin effects.

Finally, we study the effect of {\lisa}'s low-frequency sensitivity on
the accuracy of estimating parameters for massive black-hole binaries
(similar investigations can be found in~\cite{HH} and~\cite{Baker}).
Below $10^{-4}$ Hz, the noise characteristics of \lisa~are uncertain.
We show, however, that if {\lisa}'s noise spectral density can be
uniformly extrapolated from $10^{-4}$ to $10^{-5}$ Hz, then the
accuracy of estimating both extrinsic parameters such as distance and
sky position and intrinsic parameters such as chirp mass and reduced
mass, as well as the graviton mass, can be significantly higher,
especially for higher-mass systems.

The paper is organized as follows. In Sec.~\ref{subsec1} we discuss
the procedure for estimating binary parameters and the parameters of
alternative theories when we average over all sky directions and
binary orientations; this essentially ignores modulational effects due
to the motion of the spacecraft.  In Sec.~\ref{subsec2} we relax this
assumption, and present the relevant equations for estimation for a
given source direction and orientation.  In Sec.~\ref{noisecurves}, we
discuss the \lisa ~noise curve to be used.  Section~\ref{results}
presents our results.  In Sec.~\ref{results1} we show the results for
estimates assuming averaging over directions.  In Sec.~\ref{results2}
we carry out a Monte Carlo analysis of $10^4$ binaries distributed
over the angles describing the relative orientation of the binary with
respect to {\em LISA} and discuss the accuracy with which binary
parameters can be estimated.  Though somewhat more accurate, this
procedure is still affected by various approximations.
In Sec.~\ref{results2b} we use Monte Carlo simulations to
investigate the dependence of parameter estimation on the redshift of
the source.
In Sec.~\ref{results3} we study the effect of the \lisa ~low-frequency
noise.  Section~\ref{conclusions} summarizes our main conclusions.  In
Appendix \ref{app2} we summarize for completeness the main equations
used in this paper to describe {\em LISA}'s configuration, orientation
and response, as derived in Ref.~\cite{CC}.  In Appendix \ref{app1} we
discuss some subtleties in estimating binary parameters within the
Fisher matrix formalism in alternative theories of gravity when we
include spin effects.

Throughout this paper we use units in which $G=c=1$.

\section{Estimation of parameters in non-precessing spinning compact binaries}
\label{sec2}

We assume that two independent Michelson outputs $h_\alpha(t)$ with
$\alpha = {\rm I,II}$ can be constructed from the readouts of the
three {\em LISA} arms if the noise is totally symmetric (see, for
example,~\cite{CC}).  We take two approaches to estimating parameters.
In the first approach (Sec.~\ref{subsec1}), we assume that the
orientation and location of the source and the orientation of {\em
LISA} are unimportant in estimating intrinsic parameters such as
masses or theory-dependent parameters.  These orientation dependences
are contained in a number of angle-dependent functions, called pattern
functions.  Accordingly, we derive results using only one Michelson
output, and we average over the pattern functions.

In the second approach (Sec.~\ref{subsec2}), we are also interested in
the accuracy of determination of direction and distance to the source,
and thus we do not wish to average {\em a priori} over pattern
functions.  Instead, we carry out Monte Carlo simulations of
measurements using a population of sources across the sky, and we
study the distribution of accuracies of parameter estimation.  In this
case, we use both one and two Michelson detectors.  We use the by now
standard machinery of parameter estimation in matched filtering for
gravitational wave detection that has been developed by a number of
authors~\cite{finn,FinnChernoff,CutlerFlanagan,poissonwill}.  

\subsection{Parameter estimation using pattern-function averaging}
\label{subsec1}

The Fourier transform of the waveform for one Michelson {\em LISA}
detector, in the stationary phase approximation (SPA), and after
averaging over the pattern functions, is given by
\begin{subequations}
\bea
\tilde h_\alpha(f) &=&\frac{\sqrt{3}}{2}\,{\cal A}\,f^{-7/6}\,e^{\ii \psi(f)}\,, 
\quad \quad \alpha = {\rm I,II}\,, \label{hnah}
\\
{\cal A} &=& \frac{1}{\sqrt{30}\pi^{2/3}} \frac{{\cal M}^{5/6}}{D_{\rm L}}\,,
\eea
\label{hrestrict}
\end{subequations}
where $f$ is the frequency of the gravitational waves, ${\cal M} =
\eta^{3/5} M$ is the ``chirp'' mass, with $M=m_1+m_2$ and $\eta =
m_1m_2/M^2$, and $D_{\rm L}$ is the luminosity distance to the source.

We have adopted the standard ``restricted post-Newtonian
approximation'' for the waveform, in which the amplitude is expressed
to the leading order in a post-Newtonian expansion (an expansion for
slow-motion, weak-field systems in powers of $v \sim (M/r)^{1/2} \sim
(\pi{\cal M}f)^{1/3}$), while the phasing $\psi(f)$, to which laser
interferometers are most sensitive, is expressed to the highest
post-Newtonian (PN) order reasonable for the problem at hand.  For
binaries with spins aligned (or anti-aligned) and normal to the
orbital plane, this is a valid approximation because the amplitude
varies slowly (on a radiation reaction timescale) compared to the
orbital period.  But when the spins are not aligned, modulations of
the amplitude on a precession timescale must be included.  Such
modulations are beyond the scope of this paper.

The phasing function $\psi(f)$ is known for point masses up to 3.5 PN
order~\cite{blanchet1,blanchet2}.  But spin terms are known only up to
2PN order, so to be reasonably consistent, we will include in the
phasing point-mass terms only up to this same 2PN order.  The needed
expression for the phasing is
%
%
\bea
\psi(f)&=&2\pi f t_c-\phi_c+ \frac{3}{128}\,(\pi {\cal M}f)^{-5/3}\,
\left\{1- \frac{5 {\cal S}^2}{84 \omega_{\rm BD}}\,\eta^{2/5}\,(\pi {\cal M} f)^{-2/3} 
-\frac{128}{3}\frac{\pi^2 D\,{\cal M}}{\lambda_g^2\,(1+z)}\,(\pi {\cal M} f)^{2/3}\right.\nn\\
&+&\left.\left(\frac{3715}{756}+\frac{55}{9}\eta\right)\,\eta^{-2/5}\,(\pi {\cal M} f)^{2/3}
-16\pi\,\eta^{-3/5}\,(\pi {\cal M} f) + 4 \beta\,\eta^{-3/5}\,(\pi {\cal M} f) \right.\nn\\
&+&\left. \left(\frac{15293365}{508032}+\frac{27145}{504}\eta+\frac{3085}{72}\eta^2\right)\,
\eta^{-4/5}\,(\pi {\cal M} f)^{4/3} 
-10 \sigma\,\eta^{-4/5}\,(\pi {\cal M} f)^{4/3} \right\}\,.
\label{phase}
\eea
%
The structure of the phasing function is as follows: the first two
terms are related to the time $t_c$ and phase $\phi_c$ of coalescence;
they are parameters that essentially establish where the waveform
begins or ends.  The prefactor of the expression in braces, together
with the first term (``1'') inside the braces, is the standard phasing
from the lowest-order quadrupole approximation of general relativity.
Inside the braces is a post-Newtonian expansion in powers of $v \sim
(\pi{\cal M}f)^{1/3}$.  
The second term is the contribution of dipole gravitational radiation
in Brans-Dicke theory. Let us define the scalar charge of the $i-$th
body by $\alpha_i=\bar \alpha \hat \alpha_i=\bar \alpha (1-2s_i)$,
where $\bar \alpha^2=1/(2\omega_{BD}+3) \sim (2\omega_{BD})^{-1}$ in
the limit $\omega_{BD}\gg 1$, and $s_i$ is called the {\it sensitivity}
of the $i-$th body (a measure of the self-gravitational binding energy
per unit mass). Then the coefficient in the dipole term is ${\cal
S}=(\hat \alpha_1-\hat \alpha_2)/2$. The fact that it is dipole
radiation means that it is proportional to $v^{-2}$ compared to the
quadrupole term, but the small size of $\cal S$ and the large current
solar-system bound on $\omega_{\rm BD}$ make this a small correction,
nevertheless.  The third term in the braces is the effect of a massive
graviton, which alters the arrival time of waves of a given frequency,
depending on the size of the graviton Compton wavelength $\lambda_g$
and on a distance quantity $D$, defined below.  The remaining terms in
the braces are the standard general relativistic, post-Newtonian
terms, including spin effects.

The quantities $\beta$ and $\sigma$ represent spin-orbit and spin-spin
contributions to the phasing, given by
\begin{subequations}
\bea 
\beta &=& \frac{1}{12} \sum_{i=1}^2 \chi_i \left [113 \frac{m_i^2}{M^2} + 75
\eta \right ] \hL \cdot \hS_i \,,
\\
\sigma &=& \frac{\eta}{48} \chi_1 \chi_2 \left (-247 \hS_1 \cdot \hS_2 + 721 \hL
\cdot \hS_1 \hL \cdot \hS_2 \right )\,,
\eea
\end{subequations}
where $\hS_i$ and $\hL$ are unit vectors in the direction of the spins
and of the orbital angular momentum, respectively, and $\vS_i=\chi_i
m_i^2 \hS_i$.  For black holes, the dimensionless spin parameters
$\chi_i$ must be smaller than unity, while for neutron stars, they are
generally much smaller than unity.  
It follows that $|\beta|\lesssim 9.4$ and $|\sigma|\lesssim 2.5$.

We assume that any modifications to the post-Newtonian general
relativistic terms listed in the phasing formula that might be
generated in Brans-Dicke theory or in massive graviton theories will
be of order $1/\omega_{\rm BD} \ll 1$ or $1/(f\lambda_g) \ll 1$
relative to those terms, and hence we will ignore such corrections.

In this paper we denote by ${\cal M}$ and $M$ the {\it observed} chirp
and total masses. They are related to masses measured in the source
rest frame by
\beq
\label{source}
{\cal M} = (1+z) {\cal M}_{\rm source}\,, \qquad 
M = (1+z)M_{\rm source}\,,
\eeq
where $z$ is the cosmological redshift. 

Henceforth, to simplify the notation we define
\begin{subequations}
\bea
\label{bd}
&& \varpi \equiv \f{1}{\omega_{\rm BD}}\,,
\\
\label{betag}
&& \beta_{g} \equiv \frac{\pi^2 D\,{\cal M}}{\lambda_g^2\,(1+z)}\,.
\eea
\end{subequations}
To estimate the binary and gravitational theory parameters, we use the
standard technique of parameter estimation in matched filtering.  By
maximizing the correlation between a template waveform that depends on
a set of parameters $\theta^a$ (for example, the chirp mass $\cal M$)
and a measured signal, matched filtering provides a natural way to
estimate the parameters of the signal and their errors.  With a given
noise spectral density for the detector, $S_n(f)$, one defines the
inner product between two signals $h_1(t)$ and $h_2(t)$ by
\begin{equation}
(h_1|h_2) \equiv 2 \int_0^{\infty} \frac{ {\tilde{h}_1}^*\tilde{h}_2 +
{\tilde{h}_2}^*\tilde{h}_1 }{S_n(f)}df \,,
\label{innerproduct}
\end{equation}
where $\tilde{h}_1(f)$ and $\tilde{h}_2(f)$ are the Fourier transforms
of the respective gravitational waveforms $h(t)$.  The signal-to-noise
ratio (SNR) for a given $h$ is given by
\begin{equation}
\rho[h] \equiv  (h|h)^{1/2} \,.
\label{rho}
\end{equation}
If the waveforms may be characterized by a set of parameters $\theta^a$,
then one defines the ``Fisher matrix'' $\Gamma_{ab}$ with components
given by
\begin{equation}
\Gamma_{ab} \equiv \left( \frac{\partial h}{\partial\theta^a} \mid
\frac{\partial
h}{\partial\theta^b} \right) \,.
\label{fisher}
\end{equation}
In the
limit of large SNR, if the noise is stationary and Gaussian, the
probability that the GW signal $s(t)$ is characterized by a given set of
values of the source
parameters $\theta^a$ is
\beq
p(\mbox{\boldmath$\theta$}|s)=p^{(0)}(\mbox{\boldmath$\theta$}) 
\exp\left[-\frac{1}{2}\Gamma_{ab}\Delta \theta^a \Delta \theta^b\right]\,.
\eeq
where $\Delta \theta^a = \theta^a - {\hat \theta}^a$, and
$p^{(0)}(\mbox{\boldmath$\theta$})$ represents the distribution of prior
information.
An estimate of the rms error, $\Delta\theta^a$, in measuring the
parameter $\theta^a$ can then be calculated, in the limit of large
SNR, by taking the square root of the diagonal elements of the inverse
of the Fisher matrix,
\begin{equation}
\Delta\theta^a = \sqrt{\Sigma^{aa}} \,, \qquad  \Sigma = \Gamma^{-1} \,.
\label{errors}
\end{equation}
The correlation coefficients between two parameters $\theta^a$ and
$\theta^b$ are given by
\begin{equation}
c_{ab} = \Sigma^{ab}/\sqrt{\Sigma^{aa}\Sigma^{bb}} \,.
\label{correlations}
\end{equation}
We may wish to take into account our prior information on the maximum spin;
we do this in a crude way by assuming
\beq
p^{(0)}(\mbox{\boldmath$\theta$})\propto 
\exp\left[-\frac{1}{2}(\beta/9.4)^2 -\frac{1}{2}(\sigma/2.5)^2 \right]\,.
\eeq
The following derivatives of $\tilde{h}$ will be needed:
\begin{subequations}
\bea
\label{first}
 \f{\p \tilde h}{\p \ln {\cal A}} &=& \tilde h\,, \\
\f{\p \tilde h}{\p t_c} &=& 2\pi \ii f\, \tilde h\,, \\
\f{\p \tilde h}{\p \phi_c} &=& -\ii\, \tilde h\,, \\
\label{derBD}
 \f{\p \tilde h}{\p \varpi}&=&
-\frac{5\ii}{3584} {\cal S}^2 \eta^{2/5}(\pi {\cal M} f)^{-7/3}\,\tilde h\,,\\
\label{derMG}
 \f{\p \tilde h}{\p \beta_{g}}&=& -\frac{\ii}{\pi {\cal M} f}\,\tilde h \,,
\\
\f{\p \tilde h}{\p \ln {\cal M}}&=&-\frac{5\ii}{128}\,(\pi{\cal M}f)^{-5/3}\,
(K_4 v^{-2}+1+A_4v^2+B_4v^3+C_4v^4)\,\tilde h\,,\\
\f{\p \tilde h}{\p \ln \eta}&=&
-\frac{\ii}{96}\,(\pi{\cal M}f)^{-5/3}\,
(K_5v^{-2}+A_5v^2+B_5v^3+C_5v^4)\,\tilde h\,,\\
\f{\p \tilde h}{\p \beta}&=&
-\frac{3\ii}{32}\,\eta^{-3/5}\,(\pi{\cal M}f)^{-2/3}\,\tilde h\,,\\
\f{\p \tilde h}{\p \sigma}&=&
-\frac{15\ii}{64}\,\eta^{-4/5}\,(\pi{\cal M}f)^{-1/3}\,\tilde h\,,
\label{last}
\eea
\label{hderivs}
\end{subequations}
where here we denote $v=(\pi M f)^{1/3}$ and
\begin{subequations}
\bea 
K_4&=&-\frac{{\cal S}^2}{12}\,\varpi\,,\\ 
A_4&=&\frac{4}{3}\left(\frac{743}{336}+\frac{11}{4}\eta\right)
-\frac{128}{5}\beta_g\,\eta^{2/5}\,, \\ 
B_4&=&\frac{8}{5}(\beta-4\pi)\,,\\
C_4&=&2\left(\frac{3058673}{1016064}+\frac{5429}{1008}\eta
+\frac{617}{144}\eta^2-\sigma\right) \,, \\
K_5&=&\frac{3{\cal S}^2}{56}\,\varpi\,, \\
A_5&=&\left(\frac{743}{168}-\frac{33}{4}\eta\right)\,,\\
B_5&=&\frac{27}{5}(\beta-4\pi)\,,\\
C_5&=&18\left(\frac{3058673}{1016064}-\frac{5429}{4032}\eta
-\frac{617}{96}\eta^2-\sigma\right) \,.
\eea
\label{derivcoeffs}
\end{subequations}
For all integrals appearing in the Fisher matrix we will pick the
final frequency, or the upper limit of integration to be $f_{\rm fin}
= {\rm min}(f_{\rm ISCO},f_{\rm end})$. Here $f_{\rm ISCO}$ is twice
the conventional (Schwarzschild) frequency of the innermost stable
circular orbit for a point mass, namely $f_{\rm ISCO} = (6^{3/2}\pi
M)^{-1}$, and $f_{\rm end}=1~$Hz is a conventional upper cutoff on the
{\em LISA} noise curve. The initial frequency $f_{\rm in}$ in the
integrals of the Fisher matrix is determined by assuming that we
observe the inspiral over a time $T_{\rm obs}$ before the ISCO, and by
selecting a cutoff frequency below which the \lisa~ noise curve is not
well characterized.  Our default cutoff is $f_{\rm low}=10^{-5}$ Hz;
in Sec.~\ref{results3} we analyse the effects of increasing this
cutoff frequency to reflect a less optimistic understanding of
{\lisa}'s low frequency noise.  The initial frequency is then given,
in Hz, by the larger of these frequencies,
\beq\label{fin}
f_{\rm in}={\rm max} \biggl \{f_{\rm low}, \, 4.149\times 10^{-5}\left[\f{{\cal M}}{10^6 M_\odot}\right]^{-5/8}
\left(\f{T_{\rm obs}}{1 \rm yr}\right)^{-3/8}
\biggr \} \,.
\eeq
The frequency at a given observation time is calculated using the
quadrupole approximation for radiation damping.
In our calculations we assume that $T_{\rm obs}=1$~yr.  

Since we anticipate setting only lower bounds on $\omega_{BD}$ and
$\lambda_g$, we choose the nominal values $\varpi = 0$ and $\beta_g =
0$ in Eqs. (\ref{derivcoeffs}).  For simplicity, we will also assume
that we are estimating spins in the case where spins are dynamically
small.  This is generally the case for neutron stars (see
\cite{blanchetplus4} for discussion); for black holes, it means that
we are considering only slowly rotating (non extremal) black holes.
Consequently we also choose the nominal values $\beta = \sigma = 0$ in
Eqs.  (\ref{derivcoeffs}).

For a zero--spatial-curvature Universe ($\Omega_\kappa=0,
\Omega_\Lambda + \Omega_M =1$), the luminosity distance is given by
\beq
D_L=\f{1+z}{H_0}\int_0^z \frac{dz'}{\left[\Omega_M(1+z')^3+\Omega_\Lambda\right]^{1/2}}\,.
\label{DsubL}
\eeq
%
The quantity $D$ appearing in Eq.~(\ref{betag}) is defined by ($\Omega_\kappa=0$)
\beq
D=\f{1+z}{H_0}\int_0^z \frac{dz'}{(1+z')^2\left[\Omega_M(1+z')^3+\Omega_\Lambda\right]^{1/2}}\,,
\label{defD}
\eeq
(see Eq. (2.5) of~\cite{willgraviton}).
%
For the Hubble constant we assume $H_0=72$ km~s$^{-1}$~Mpc$^{-1}$,
according to the present observational estimates~\cite{cosmology}.

A useful quantity to characterize the effect of the various terms
(Brans-Dicke, massive graviton, spin couplings and PN corrections) on
the evolution of the GW frequency is the number of GW cycles
accumulated within a certain frequency band. This quantity is defined
as:
\beq
{\cal N}_{\rm GW} \equiv 
\int_{f_{\rm in}}^{f_{\rm fin}}\,
\frac{f}{\dot{f}}\,d f\,.
\label{numcycles}
\eeq
To derive the number of cycles contributed by individual terms in the
phasing, we use an expression for $\dot f$ that includes post-Newtonian GR
terms, plus the Brans-Dicke and graviton-mass contributions, given by
\beq
\frac{df}{dt} = \frac{96}{5 \pi {\cal M}^2}\,(\pi {\cal M}
f)^{11/3}\,\left \{ 1 + \frac{5{\cal S}^2\,{\varpi}}{48}\,
{\eta^{2/5}}\,(\pi {\cal M}\, f)^{-2/3}  + \frac{96\beta_g}{5}\,(\pi {\cal M}
f)^{2/3} + {\rm PN\,corrections} \right \}\,,  
\label{fdot}
\eeq
where the ``PN corrections'' up to 2PN order, including spin terms,
can be found in~\cite{blanchetplus4}.  When we include the
massive-graviton term, the frequency $f$ and time $t$ appearing in
Eq.~(\ref{fdot}) should be considered as the {\it arrival} frequency
and time, respectively (the number of gravitational-wave cycles due to
the massive graviton being an effective number of cycles seen by the
observer at the detector location).

Contributions of individual terms in the integral (\ref{numcycles})
are generally considered significant if they exceed one wave cycle
over the observation time.  For various examples of the two source
targets discussed in this paper, NS inspiral into IMBH, and MBH
binaries, we show the individual contributions to the number of
cycles, along with the initial and final frequencies, in
Tables~\ref{LISAcycBD} and \ref{LISAcycMG}, respectively.

With the restricted post-Newtonian form for $\tilde{h}$ in
Eq. (\ref{hrestrict}), we can
express the SNR $\sqrt{\langle \rho^2\rangle}$
in the form
\begin{widetext}
\beq
\sqrt{\langle \rho^2 \rangle} = 6.245 \times 10^{-23} \,\left (\frac{M}{M_\odot}\right )^{5/6}\,
\eta^{1/2}\,\left (\frac{1 {\rm Gpc}}{D_L}\right )\, 
\sqrt{\int_{f_{\rm in}}^{f_{\rm end}}\frac{3}{4} \frac{f^{-7/3}}{S_h(f)} df}\,,
\eeq
\end{widetext}
where angular braces mean that we are averaging over {\em LISA}
pattern functions.

\subsection{Parameter estimation without averaging over pattern functions}
\label{subsec2}

In this section we consider parameter estimation without averaging
over the relative orientation of the binary with respect to {\em
LISA}.  We assume, as in~\cite{CC}, that two independent Michelson
outputs can be constructed from the readouts of the three {\em LISA}
arms if the noise is totally symmetric.  The signal {\it measured} by
{\em LISA}, $h_\alpha(t)$ with $\alpha = {\rm I,II}$, can be written
as:
\beq
h_\alpha(t) = \frac{\sqrt{3}}{2}\;\frac{2 m_1\,m_2}{r(t)\,D_{\rm L}}\,
\tilde{A}_\alpha(t) \cos \;\biggl( \int_0^t{f(t') \; dt'
+ \varphi_{p,\alpha}(t) + \varphi_D(t) \biggr)}\,, 
\eeq
where $r(t)$ is the relative distance between the two compact bodies,
$\varphi_{p,\alpha}(t)$ is the waveform polarization phase [see
Eq.~(\ref{phip})] and $\varphi_{D}(t)$ the Doppler phase [see
Eq.~(\ref{phiD})].  $\tilde{A}_{\alpha}(t)$ is defined by
\beq
\tilde{A}_{\alpha}(t) = 
\sqrt{[1+ (\hat \vL \cdot \vn)^2]^2\,F_{\alpha}^{+\,2} + 4 
(\hat \vL \cdot \vn)^2\,F_{\alpha}^{\times\,2}}\,,
\eeq
where $\hat \vL$ is the orbital angular momentum unit vector, 
and $\vn$ is a unit vector in the direction of
the source on the sky.  The quantities $F_{\alpha}^{+,\times}$ are the
pattern functions, defined by Eqs.~(\ref{patternI}) and
(\ref{patternII}).  The Fourier transform of the measured signal can
be evaluated in the stationary phase approximation, since
$\tilde{A}_{\alpha}(t)$, $\varphi_{p,\alpha}(t)$ and $\varphi_D(t)$
vary on time scales on the order of 1 year (thus much larger than the
binary orbital period $\sim 2/f$). The result is
\bea
\label{hSPA}
\tilde h_\alpha(f) = \frac{\sqrt{3}}{2}
{\cal A}\, f^{-7/6}\,e^{i \Psi(f)}\, \left \{ \frac{5}{4} \tilde{A}_\alpha(t(f)) \right \}
e^{-\ii\bigl( \varphi_{p,\alpha}(t(f)) +  \varphi_D(t(f))\bigr)}
\,,
\eea
where to 2PN order (including also the Brans-Dicke parameter and the
graviton-mass term) $t(f)$ is given by
\bea
t(f) & = & t_c - \frac{5}{256{\cal M}}(\pi{\cal M}f)^{-8/3} 
\biggl[1 -\frac{{\cal S}^2\,\varpi}{12}\,\eta^{2/5}\,(\pi {\cal M} f)^{-2/3} 
-\frac{4\,\beta_g}{3}\,(\pi {\cal M} f)^{2/3} \nonumber \\
&& + \frac{4}{3}\left(\frac{743}{336} + \frac{11}{4}\,\eta \right)\,\eta^{-2/5}\,
(\pi {\cal M} f)^{2/3} 
- \frac{8}{5}(4\pi -\beta)\,\eta^{-3/5}\,(\pi {\cal M} f) \nonumber \\
&& + 2\left (\frac{3058673}{1016064}+\frac{5429}{1008}\,\eta 
+ \frac{617}{144}\,\eta^2-\sigma \right)\,\eta^{-4/5}\,(\pi {\cal M} f)^{4/3} \biggr] \;. 
\eea
In Appendix \ref{app2}, using equations of Ref.~\cite{CC}, we show how
to express the angular parts of $\tilde h_\alpha (f)$ in terms of the
angles $\bt_S$, $\bph_S$, $\bt_L$, $\bph_L$, which describe the source
location and orbital angular momentum direction in the reference frame
attached to the solar system barycenter.  To evaluate the Fisher
matrix we use the derivatives with respect to the parameters ${\cal
M}$, $\eta$, $\beta$, $\sigma$, $\phi_c$, $t_c$, ${\rm ln}\,{\cal A}$,
$\varpi$ and $\beta_g$, given by Eqs.~(\ref{hderivs}).  We also
determine analytically the angular derivatives with respect to
$\bt_S$, $\bph_S$, $\bt_L$, $\bph_L$ using formulas given in Appendix
\ref{app2}. The final results are lengthy and unenlightening, so we do
not write them down here. We choose to evaluate the angular
derivatives analytically since this is likely to be more accurate than
the numerical finite-differencing adopted in Refs.~\cite{CC,SH}.  As
before, we choose the nominal values $\varpi = \beta_g = \beta =
\sigma = 0$.

The non-averaged SNR is ($\alpha = {\rm I, II}$)
\begin{widetext}
\bea
\rho_\alpha(\bt_S, \bph_S, \bt_L, \bph_L) 
= 7.807 \times 10^{-23} \,\left (\frac{M}{M_\odot}\right )^{5/6}\,
\eta^{1/2}\,\left (\frac{1 {\rm Gpc}}{D_L}\right )\, 
\sqrt{\int_{f_{\rm in}}^{f_{\rm end}} \frac{3}{4} \tilde{A}^2_\alpha(t(f);
\bt_S, \bph_S, \bt_L, \bph_L) 
\frac{f^{-7/3}}{S_h(f)} df}\,.
\eea
\end{widetext}

For some binary-mass configurations we estimate the parameters using
the two \lisa ~ detectors. In this case the Fisher matrix is
\begin{equation}
\Gamma_{ab}^{\rm tot} \equiv \left( \frac{\partial h_{\rm I}}{\partial\theta^a} \mid
\frac{\partial h_{\rm I}}{\partial\theta^b} \right) +
\left( \frac{\partial h_{\rm II}}{\partial\theta^a} \mid
\frac{\partial h_{\rm II}}{\partial\theta^b} \right)\,,
\end{equation}
and the rms error, $\Delta\theta^a$, in measuring the parameter
$\theta^a$ is $\Delta\theta^a = \sqrt{\Sigma^{aa}}$ with $\Sigma =
\left[\Gamma^{\rm tot}\right]^{-1}$ \,.  The total SNR is $\rho_{\rm
tot}=\sqrt{\rho_{\rm I}^2+\rho_{\rm II}^2}$.

We expect that estimates obtained for parameters such as ${\cal M}$,
${\cal \eta}$, $\beta$, $\sigma$, $\varpi$ and $\beta_g$ when we do
not pattern average will not differ qualitatively from those obtained
using pattern averaged templates.  As we will see, those parameters,
which appear in the phasing of the signal, are relatively uncorrelated
with the parameters appearing in the amplitude, such as $\bt_S$,
$\bph_S$, $\bt_L$, $\bph_L$.

\subsection{Noise curve for the \lisa ~instrument}
\label{noisecurves}

\noindent
The {\it non-sky-averaged} noise spectral density of \lisa~ depends on
the relative orientation between the instrument and the source, and it
is very hard to implement in estimating binary parameters.  Generally,
the \lisa~ community has been using the so-called sky-averaged
spectral density $S_h^{\rm SA}$ [see e.g., Ref.~\cite{FT} and the LISA
Pre-Phase A Report].  The sky-averaged spectral density is computed by
a combination of three factors, including: (i) the raw spectral noise
density $S_n$, (ii) the gravitational-wave transfer (response)
function $R$ and (iii) the noise transfer (response) function $R_n$.
They combine together in~\cite{LHH}
\beq
S_h^{\rm SA} = \frac{S_n\,R_n}{R}\,.
\label{SA}
\eeq
In this paper we are also interested in determining binary parameters
{\em without} averaging over the source location, so we are not
allowed, in principle, to use $S_h^{\rm SA}$.  To overcome this
difficulty we evaluate an {\it effective} non-sky-averaged spectral
density which gives the correct result at low frequency, but is only
approximately valid in the high-frequency region. In the low
frequency limit, the GW transfer function used in the {\em LISA}
Sensitivity Curve Generator~\cite{SCG} is $R=4(\sqrt{3}/2)^2 1/5=3/5$,
where the factor $(\sqrt{3}/2)^2$ comes from the LISA arms being at
$60^{o}$, the factor 1/5 is due to the sky-average of the
pattern functions ($\langle F_{+,\times}^2 \rangle = 1/5$) and the
factor 4 depends on the particular read-out variable used. Since our
definition of the GW signal already includes the factor $\sqrt{3}/2$
[see Eqs. (\ref{hnah}) and (\ref{hSPA})], to obtain the effective
non-sky-averaged spectral density we must multiply $S_h^{\rm SA}$ by
$(\sqrt{3}/2)^2/5=3/20$. The final result is:
\beq
S_h^{\rm NSA}(f)=
\left[9.18\times 10^{-52}\left(\f{f}{1~{\rm Hz}}\right)^{-4}
+1.59\times 10^{-41} 
+9.18\times 10^{-38}\left(\f{f}{1~{\rm Hz}}\right)^2\right]~{\rm Hz}^{-1}\,,
\eeq
and has been obtained also in Ref.~\cite{BC}.  We estimate white-dwarf
confusion noise following~\cite{BC}, which uses results
from~\cite{conf1,conf2}: the galactic contribution is approximated as
\beq
S_h^{\rm gal}(f)=
2.1\times 10^{-45}\left(\f{f}{1~{\rm Hz}}\right)^{-7/3}~{\rm Hz}^{-1}\,,
\eeq
and the contribution from extra-galactic white dwarfs as 
\beq
S_h^{\rm ex-gal}(f)=
4.2\times 10^{-47}\left(\f{f}{1~{\rm Hz}}\right)^{-7/3}~{\rm Hz}^{-1}\,.
\eeq
We compute the total (instrumental plus confusion) noise as
\beq
S_h(f)={\rm min}\left\{
S_h^{\rm NSA}(f)/{\rm exp}
\left(-\kappa T^{-1}_{\rm mission} dN/df\right),~
S_h^{\rm NSA}(f)+S_h^{\rm gal}(f)
\right\}+S_h^{\rm ex-gal}(f)\,.
\label{Shtot}
\eeq
Here $dN/df$ is the number density of galactic white-dwarf binaries
per unit gravitational-wave frequency, for which we adopt the estimate
\beq
\f{dN}{df}=2\times 10^{-3}~{\rm Hz}^{-1}
\left(\f{1~{\rm Hz}}{f}\right)^{11/3}\,;
\eeq
$\Delta f=T^{-1}_{\rm mission}$ is the bin size of the discretely
Fourier transformed data for a {\em LISA} mission lasting a time
$T_{\rm mission}$; and $\kappa\simeq 4.5$ is the average number of
frequency bins that are lost when each galactic binary is fitted
out. The factor ${\rm exp}\left(-\kappa T^{-1}_{\rm mission}
dN/df\right)$ thus represents the fraction of ``uncorrupted'' bins
where instrumental noise still dominates. At variance with~\cite{BC},
in our calculations we always assume that the duration of the {\em
LISA} mission $T_{\rm mission}=1$~yr, consistently with the choice we
made for the observation time $T_{\rm obs}$ in Eq.~(\ref{fin}). The
analytic root noise spectral density curve (\ref{Shtot}) used in this
paper is shown in Fig.~\ref{noise} together with the corresponding
root noise spectral density curve from the {\em LISA} Sensitivity
Curve Generator~\cite{SCG}.  The SCG curve shown is obtained using the
nominal values SNR=1, arm length $=5 \times 10^9$ m, telescope
diameter $=0.3$ m, laser wavelength $=1064$ nanometers, laser power
$=1.0$ Watts, optical train efficiency $=0.3$, acceleration noise $=3
\times 10^{-15} \,{\rm m}\,{\rm s}^{-2}\,{\rm Hz}^{-1/2}$, and
position noise budget $=2 \times 10^{-11} \,{\rm m}\,{\rm Hz}^{-1/2}$,
with position noise setting the floor at high frequency.  The data
returned by the SCG is then multiplied by $\sqrt{3/20}$ to obtain the
effective non-sky averaged curve shown in Fig. ~\ref{noise}.

\section{Results of parameter estimation}
\label{results}

\subsection{Estimates using pattern-averaged templates}
\label{results1}

\noindent
We begin with neutron-star inspirals into intermediate mass black
holes.  These are the best sources for bounding scalar-tensor gravity,
for the following reasons.  In scalar-tensor theory, dipole
gravitational radiation is controlled by the difference ${\cal
S}=(\hat \alpha_1-\hat \alpha_2)/2$ in the rescaled scalar charge
$\hat \alpha_i$ between the two bodies. We recall that $\hat
\alpha_i=(1-2 s_i)$ and
\beq
s_i = \left ( \frac{\partial (\ln m_i)}{\partial (\ln G_{\rm eff})} 
\right )_N \,,
\eeq
where $m_i$ is the total mass of the body, $G_{\rm eff}$ is the
effective gravitational constant at the location of the body (which is
related to the value there of the Brans-Dicke scalar field) and the
subscript $N$ denotes that the number of baryons is held fixed.  For
neutron stars, $s_i$ can be substantial, and thus $\hat \alpha_i$ can
differ markedly from unity ($\hat \alpha_i\sim 0.6-0.8$), but because
it is only weakly dependent on the NS equation of state and
mass~\cite{zaglauer}, the {\em difference} ${\cal S}$ for NS binaries
is typically ${\cal S} < 0.05$, so NS-NS binaries do not provide
interesting bounds on Brans-Dicke theory~\cite{willST} (see
however~\cite{damourfarese} for discussion of more general
scalar-tensor theories). Because of the no-hair theorem, for black
holes $\hat \alpha_{\rm BH}\equiv 0$, so BH-BH binaries cannot be used
to put bounds on the Brans-Dicke parameter via dipole
radiation. Therefore, following previous papers on the
subject~\cite{willST,scharrewill,willyunes}, we only consider NS-BH
binaries as sources for this purpose.  Furthermore, as shown earlier
\cite{scharrewill,willyunes}, inspiral into lower-mass black holes
gives the most promising bounds, primarily because more cycles are
observed in a given period of integration in that case.  The event
rate of such inspirals involving intermediate-mass black holes is
uncertain, but is likely to be very small~\cite{miller,willimbh}; only
a lucky detection of such an inspiral will lead to a suitable test.
White dwarf-BH binaries could also be used to test Brans-Dicke theory,
since $\hat \alpha_{WD} \sim 1$ ($s_{WD} \sim 10^{-4}$), so that
${\cal S} \sim 0.5$, except that tidal effects will play a role in the
late stages of the inspiral, depending on the mass of the black hole
(for discussion, see~\cite{scharrewill}).

For concreteness we focus on four NS-BH binaries, setting the NS mass
$M_{\rm NS}=1.4 M_\odot$ and considering black holes of mass $M_{\rm
BH}=400,~1000,~5000$ and $10^4~M_\odot$.  The borderline between
massive and supermassive BHs is hazy, but we choose not to consider
NSs inspiralling into ``supermassive'' BHs with $M = 10^5 \mbox{-}
10^8 M_\odot$.  Even in the context of pure general relativity, our
approximation that the binary orbits be circular is expected to be
unreliable for these high-mass cases: for high mass ratios the binary
is likely to be formed by capture of the smaller body into the larger
one, and the eccentricity will not be washed out by radiation
reaction.  Adding eccentricity complicates the analysis to a level
that is beyond the scope of this paper, and we plan to return to this
problem in the future.

We first consider the inspiral of these four representative binaries
within general relativity (omitting the BD term in the phase). From
the initial and ending frequencies listed in Table~\ref{LISAcycBD} we
see that these binaries sweep through the high frequency part of the
{\em LISA} band, say from $\sim 10^{-2}$ Hz up to $\sim 1$ Hz. In
Table~\ref{LISAnoBD} we list the errors and correlation coefficients
that are obtained when we truncate the phasing formula at various PN
orders and include spin-orbit and spin-spin effects.  For consistency,
at 1PN order we do not include spin effects, at 1.5PN order we include
only spin-orbit effects, and at 2PN order we include both spin-orbit
and spin-spin effects.  All results assume $\rho=10$. In the left
panel of Fig.~\ref{SNR} we show the corresponding luminosity distance
as a function of the black hole mass $M_{\rm BH}$.

From Table~\ref{LISAnoBD} we see that the errors on all parameters
increase considerably when spin effects are taken into account. This
applies in particular to the chirp mass ${\cal M}$ and the parameter
$\eta$. This spin-induced degradation in parameter estimation has long
been known~\cite{poissonwill}: it occurs because (in the absence of
precessional effects) the parameters are highly correlated, so that
adding parameters effectively dilutes the available information.

For technical reasons, when we consider alternative theories of
gravity we only include spin-orbit effects. If in addition we include
spin-spin effects the dimensionality of the Fisher matrix increases,
and the matrix inversion required to obtain the correlation matrix
appears to be unreliable.  This issue is addressed in
Appendix~\ref{app1}.

In Table~\ref{LISAspin} we show errors and correlation coefficients
for NS-BH binaries at 2PN order when we include the BD term.  For
nonspinning binaries the results are similar to Table I in
Ref.~\cite{willyunes}, except that those authors used templates at
1.5PN order and did not take into account the factor $\sqrt{3}/2$
which appears in Eq.~(\ref{hnah}).  The BD term is highly correlated
with ${\cal M}$ and $\eta$ ($c^{{\cal M}\varpi}$ and $c^{\eta \varpi}$
are both quite large). Correspondingly, the error on both ${\cal M}$
and $\eta$ increases by roughly one order of magnitude with respect to
the ``general relativistic'' values listed in Table~\ref{LISAnoBD}.
We also compute the BD bound obtained by inverting only the diagonal
element $\Gamma_{\varpi \varpi}$ of the Fisher matrix.  This
``uncorrelated'' bound $\omega_{\rm BD,unc}$ is always about two
orders of magnitude larger than the actual value we obtain by
inverting the full Fisher matrix in the absence of spins.

Notice also that the BD bound {\em decreases} with increasing black
hole mass. This can be partially understood by the following argument:
the derivative of the GW signal with respect to $\varpi$ is
proportional to $M^{-1}_{\rm NS} M^{-4/3}_{\rm BH}$ [see
Eq.~(\ref{derBD}) and use $M_{\rm BH} \gg M_{\rm NS}$]. Therefore the
derivative decreases as $M_{\rm BH}$ increases; the integration over
the frequency range (which also depends on mass) modifies this
dependence somewhat, but the final conclusion is that the higher the
BH mass, the lower the Brans-Dicke bound.

From Table~\ref{LISAspin} we also see that non-precessional spin
effects reduce considerably the bound on the Brans-Dicke
parameter. For example, for a $(1.4+1000)~M_\odot$ binary the bound
decreases by a factor 10 (from $\sim 2\times 10^5$ to $\sim 2\times
10^4$) when we include the spin-orbit term.  We will see later
(Table~\ref{BDaverage}) that a further reduction of a factor $\sim 2$
comes from inclusion of effects related to the orbital motion of {\em
LISA}.  We also found that including priors, that is, assuming that we
know {\it a priori} from general relativity that the compact objects'
spins are bounded from above (see~\cite{poissonwill} for a discussion)
has completely negligible effects when we include spin-orbit terms.

In setting bounds on the graviton mass, we consider massive and
supermassive binary black holes with $M = 10^4 \mbox{-} 10^7 M_\odot$.
The derivative of the GW signal with respect to $\beta_g$ is inversely
proportional to the chirp mass ${\cal M}$, so for comparable-mass
binaries, the higher the total mass the higher the graviton-mass
bound. However, as seen from the initial and ending frequencies in
Table~\ref{LISAcycMG}, for $M_{\rm BH} > 10^7 M_\odot$, the binary
sweeps through the low-frequency end of the {\em LISA} band below
$10^{-4}$~Hz, where the predicted sensitivity of \lisa~ is not very
robust at present.  The sensitivity in this low-frequency regime
depends on how efficiently the acceleration noise can be reduced. In
Sec.~\ref{subsec2}, we shall investigate the effect on the estimation
of the parameters and on the graviton-mass bound, if the {\em LISA}
noise curve can be trusted {\em only} down to a lower frequency
$f_{\rm low} \sim 10^{-4}$ or $5 \times 10^{-5}$ Hz.

In Tables~\ref{MG1} and~\ref{MG2} we list the errors and correlation
coefficients when binaries with high, comparable masses are detected
using pattern-averaged templates at 2PN order.  Table~\ref{MG1} shows
results for pure general relativity, with spin-orbit and spin-spin
effects included.  Table~\ref{MG2} shows results when a massive
graviton term and a spin-orbit term are included.  As in the BD case,
we do not show results for a massive graviton combined with spin-orbit
and spin-spin effects, because the inversion of the large Fisher
matrix in this case appears to be unreliable (see
Appendix~\ref{app1}).

As in the case of NS-BH binaries, adding new parameters causes a
degradation in the accuracy with which we can estimate parameters.
All the values reported in Tables~\ref{MG1} and ~\ref{MG2} are
obtained for binaries at 3 Gpc. The corresponding SNR for equal mass
BH-BH systems is shown in the right panel of Fig.~\ref{SNR} as a
function of the total mass of the binary.  For $M\gtrsim 10^6 M_\odot$
we observe the appearance of a relative minimum, corresponding to the
range of frequencies in which white-dwarf confusion noise dominates
over instrumental noise.

Although we only report results for the currently favoured values of
the cosmological parameters, we verified that the upper bound on the
graviton wavelength depends only weakly on the underlying cosmological
model. The ``uncorrelated'' bound $\lambda_{g, \rm unc}$ obtained by
inverting only the diagonal element $\Gamma_{\beta_g \beta_g}$ of the
Fisher matrix is about one order of magnitude larger than the result
obtained by inverting the full Fisher matrix (for $\omega_{\rm BD}$
the difference was about {\em two} orders of magnitude). Compared with
the case of scalar-tensor theories, bounds on massive graviton
theories seem to be less sensitive to correlations among different
parameters.

\subsection{Estimates using templates without pattern averaging}
\label{results2}

To assess the effect of pattern-averaging on parameter estimation, and
also to determine how accurately {\em LISA} can measure source
locations and luminosity distances, we adopt the non-averaged
templates of Sec.~\ref{subsec2}, and perform Monte Carlo simulations
using a population of sources across the sky. We consider in detail
two systems: (i) a NS-BH binary with mass of $(1.4+1000) M_\odot$
observed with a single-detector SNR $\rho_I=10$ (a typical target
system used to place bounds on the BD parameter), and (ii) a BH-BH
binary with mass of $(10^6+10^6) M_\odot$ at distance $D_L=3$~Gpc (a
typical system in the context of massive graviton theories).

For each of these systems we distribute $10^4$ sources over sky
position and orientation. We randomly generate the angles $\bph_S$,
$\bph_L$ in the range $[0,2 \pi]$ and $\mu_S = \cos \bt_S$, $\mu_L =
\cos \bt_L$ in the range $[-1,1]$. As in~\cite{SH}, to generate random
numbers we use the routine RAN2~\cite{numrec}. Computing and inverting
the Fisher matrix for $10^4$ binaries typically takes $\sim 8-15$
minutes (depending on the dimensionality of the matrix) on an ordinary
laptop. This is much faster (by a factor $\sim 500$) than previous
Monte Carlo simulations~\cite{SHPC}.  A marginal difference with
previous codes is that we compute angular derivatives {\it
analytically} instead of using finite differencing, but the major
improvement is due to our use of a numerical integrator based on
spectral methods, namely the Gauss-Legendre routine GAULEG
\cite{numrec}.  Numerical experiments show that $\sim 600$ points in
the spectral expansion are sufficient to obtain an accuracy of one
part in $10^4$ in all parameter errors. This is true even when we use
splines to interpolate tabulated data from the {\em LISA} Sensitivity
Curve Generator, instead of adopting the analytical noise curve of
Eq. (\ref{Shtot}). When the waveform contains a large number of highly
correlated parameters, computing the inverse of the Fisher matrix can
be numerically difficult. The method we used to check the robustness
of our results is described in Appendix~\ref{app1}.

Once we have computed the errors for all $10^4$ binaries we group them
into $N_{\rm bins}$ bins depending on the (logarithmic) distribution
of their errors: a binary belongs to the $j$-th bin if the error on
some parameter $X$ satisfies
\beq
\left\{
\ln(X_{\rm min})+
\frac{(j-1)[\ln(X_{\rm max})-\ln(X_{\rm min})]}{N_{\rm bins}}
\right\}<
\ln(X)
\leq 
\left\{
\ln(X_{\rm min})+
\frac{j[\ln(X_{\rm max})-\ln(X_{\rm min})]}{N_{\rm bins}}\,
\right\} \,,
\eeq
for $j=1,\dots,N_{\rm bins}$. In this paper we fix $N_{\rm bins}=50$.
Once we have binned the data, we normalize the binaries in each bin to
the total number of binaries to get a ``probability distribution'' of
the error on the variable $X$.

In Figs.~\ref{histoGR3} and \ref{histoGR4} we show the resulting
histograms for a NS-BH binary of $(1.4 + 10^3) M_\odot$ with $\rho_I =
10$.  The plots contain various histograms for parameter estimations
made when spins are absent, when spin-orbit is included, and when both
spin-orbit and spin-spin are included.  The histograms come in pairs:
in each case the solid-line histogram refers to measurements carried
out with only one data stream from the Michelson interferometer ${\rm
I}$, the dashed histogram refers to measurements made when both data
streams from Michelson interferometers ${\rm I}$ and ${\rm II}$ are
combined.  Not surprisingly, the accuracy is improved with the use of
two outputs, very roughly by a factor of order $\sqrt{2}$ in most
cases.

In Fig. ~\ref{histoGR3} we plot the probability distribution for the
luminosity distance $\Delta D_L/D_L$ and for the angular resolution
$\Delta \Omega_S$, defined as
\beq
\Delta \Omega_S = 
2 \pi |\sin \bar \theta_S|\left\{ \Sigma_{\bar \theta_S \bar \theta_S} 
\Sigma_{\bph_S \bph_S}-\Sigma^2_{\bar \theta_S \bph_S} \right\}^{1/2}\,.
\eeq
In Fig.  ~\ref{histoGR4} we plot the distributions for the chirp mass
$\Delta {\cal M}/{\cal M}$, the reduced mass $\Delta \mu/\mu$, the
spin parameters $\Delta \beta$ and $\Delta \sigma$, and the bound on
the BD parameter $\omega_{\rm BD}$.  The distribution for $\Delta
\mu/\mu$ can be obtained from the errors on $\eta$ and ${\cal M}$ by
error propagation, taking into account that the correlation between
the two mass parameters, as defined in Eq.~(\ref{correlations}), can
be large:
\beq
\frac{\Delta \mu}{\mu}=
\left[
\left(\frac{\Delta {\cal M}}{{\cal M}}\right)^2+
\left(\frac{2}{5} \frac{\Delta \eta}{\eta}\right)^2+
\frac{4}{5}
\left(\frac{\Delta {\cal M}}{{\cal M}}\right)
\left(\frac{\Delta \eta}{\eta}\right)
c^{{\cal M}\eta}
\right]^{1/2}\,.
\eeq
The plots contain histograms evaluated when spins are absent, when
spin-orbit alone is considered and when both spin-orbit and spin-spin
are included. We complement these plots by Table~\ref{BDaverage},
which shows the average errors obtained summing over all binaries,
both with and without the BD term.  Notice that this procedure is {\em
different} from averaging over the sky without taking into account the
orbital motion of {\em LISA}.  For each model, the first line in
Table~\ref{BDaverage} refers to errors obtained averaging over all
binaries and using only detector I, the second line refers to an
average over all binaries using both detectors, while the third line
reproduces, for comparison, the corresponding pattern-averaged results
from Tables~\ref{LISAnoBD} and \ref{LISAspin}.  In general, the
pattern-averaged procedure gives good qualitative, albeit
systematically low estimates of the measurement errors, compared to
the Monte Carlo results.
For stellar mass inspirals into intermediate-mass black holes, both
the angular resolution and the distance determination accuracy are
poor.  The Monte Carlo simulation gives rather broad probability
distributions, shown in Fig.~\ref{histoGR3}, with minimum error in
distance around 0.1, but with a tail extending up to $\Delta
D_L/D_L\sim 10$ or so.
The values for $\Delta \Omega_S$ in steradians look rather small, but
when expressed in arcminutes, with $\delta \theta_S \approx [\Delta
\Omega_S \times (3283/{\rm str})]^{1/2} \times 60$ arcmin, they are
substantial.  For comparison, the angular diameter of the Moon (and
the Sun) as seen from the Earth is $\simeq 30$ arcmins.  The angular
resolution is degraded when spin terms are included, as is apparent in
the top panels of Fig.~\ref{histoGR3}, while the distance
determination is relatively insensitive to the inclusion of spin
terms.  As first noticed by Cutler~\cite{CC}, both the angular
distribution and the distance determination improve when we use both
detectors (dashed lines) instead of a single detector (solid lines).

A noteworthy feature of the histograms in Fig.~\ref{histoGR4} is that
the errors on ${\cal M}$, $\mu$, $\beta$, $\sigma$ and $\omega_{\rm
BD}$ show a peculiar ``double-peak'' structure which is absent for
the high-mass BH-BH binaries.  We see this double-peak structure for
the first time because our fast spectral integrator allows us to
simulate a sufficiently high number of binaries, but we have no
analytical understanding of this behavior.
The inclusion of each spin-coupling term degrades the determination of
both ${\cal M}$ and $\mu$ by roughly one order of magnitude.  The
large reduction on the Brans-Dicke bound caused by the inclusion of
non-precessing spins is one of the main results of this work (bottom
panel of Fig.~\ref{histoGR4}).

In Figs.~\ref{histoGR1} and~\ref{histoGR2} we show similar histograms
for a binary of total observed mass $(10^6 + 10^6) M_\odot$ at a fixed
distance of 3 Gpc, and in Table~\ref{MGaverage} we display average
errors obtained by summing over all binaries.  Supermassive BH-BH
binaries can be observed at much higher redshifts than NS-BH
binaries. {\em LISA}'s accuracy in measuring the luminosity distance
$D_L$ can thus be exploited to infer the redshift $z$ of the source,
disentangling the mass-redshift degeneracy of the waveforms and
allowing the determination of the masses in the source rest
frame. Indeed, Hughes proposed to use gravitational wave observations
in this way to map the merger history of supermassive black holes
\cite{SH}.  Once we have $\Delta D_L/D_L$ we can compute the error on
the binary's redshift by the following procedure. If we assume that
the cosmological parameters $\Omega_\Lambda$ and $H_0$ are known with
an accuracy $\simeq 10 \%$, we can use error propagation to get
$\Delta z$ from $\Delta D_L$ following Hughes~\cite{SH}:
\beq
\Delta z=\left(\frac{\p D_L}{\p z}\right)^{-1}
\left[\Delta D_L^2
+\left(\frac{\p D_L}{\p \Omega_\Lambda}\right)^{2}\Delta \Omega_\Lambda^2
+\left(\frac{\p D_L}{\p H_0}\right)^{2}\Delta H_0^2
\right]^{1/2}\,.
\label{dz}
\eeq
%
From Eq. (\ref{DsubL}), and assuming that $\Omega_M + \Omega_\Lambda =1$, we
find the derivatives 
\bes
\bea
\frac{\p D_L}{\p z} &=& \frac{D_L}{1+z} 
+ \frac{1+z}{H_0 \sqrt{(1-\Omega_\Lambda)\,(1+z)^3 + \Omega_\Lambda}}\,,
\label{dDdz}\\
\frac{\p D_L}{\p H_0} &=& - \frac{D_L}{H_0} \,, 
\\
{\frac{\p D_L}{\p \Omega_\Lambda}} &=& \frac{1+z}{2H_0}
\int_0^z \frac{[(1+z^\prime)^3-1]\,dz^\prime}{\left[(1-\Omega_\Lambda)\,(1+z^\prime)^3 + \Omega_\Lambda\right]^{3/2}}\,.
\eea
\label{Dderivs}
\ees
Then Eq.~(\ref{dz}) can be re-cast in the form
\beq
\Delta z = \left({\frac{\p D_L}{\p z}}\right)^{-1}\,
\sqrt{
\left[\frac{\Delta D_L^2}{D_L^2}\,+ \frac{\Delta
    H_0^2}{H_0^2}\right]\,{D}_L^2 
+ \Delta \Omega_\Lambda^2\, \left ({\frac{\p D_L}{\p \Omega_\Lambda}}\right )^2}\,.
\label{deltaz}
\eeq
Thus $\Delta z$ is completely determined once we fix $z$, $\Delta
D_L/D_L$, $\Delta H_0/H_0$, $\Omega_\Lambda$, $\Delta \Omega_\Lambda$.
We also compute the best possible redshift determination $(\Delta
z)_{\rm best}$ that {\em LISA} could achieve assuming (perhaps not too
optimistically) that by the time {\em LISA} flies the cosmological
parameters are known to much better precision than {\em LISA}'s
distance determinations ($\Delta \Omega_\Lambda \approx 0\,, \,\Delta
H_0 \approx 0$ in Eq.~(\ref{dz})).

Figure~\ref{histoGR1} shows the resulting probability distributions
for the SNR, the luminosity distance $\Delta D_L/D_L$, the redshift
determinations $\Delta z/z$ and $(\Delta z/z)_{\rm best}$, and the
angular resolution $\Delta \Omega_S$ in steradians.  Unlike the NS-BH
systems considered in Fig. \ref{histoGR3}, the probability
distribution of all these quantities (at fixed distance) depends very
weakly on whether we include or omit spins, so we display only the
spinless results.  The SNR distribution has the same shape and average
value for the two Michelson detectors; it increases (on average) by a
factor $\simeq \sqrt{2}$ when we use both detectors.  Not
surprisingly, the distribution of $(\Delta z/z)_{\rm best}$ and
$\Delta D_L/D_L$ are identical, apart from the scale factor from $\p
D_L/\p z$ in Eq. (\ref{dDdz}).  The shape of the distribution of
$\Delta D_L/D_L$ is different from that shown in Fig. \ref{histoGR3}.
The distribution of $\Delta z/z$ is dominated by the 10 per cent
errors assumed for the cosmological parameters, and shows only small
effects of the distribution of $\Delta D_L/D_L$.  At this relatively
small redshift, the present uncertainty on cosmological parameters
dominates over the accuracy of \lisa~\cite{SH}.  For larger values of
the redshift {\em LISA} distance and redshift measurements become less
accurate, eventually dominating the error at some critical value of
$z$ that depends on the MBH masses (see Sec.~\ref{results2b}).

Figure~\ref{histoGR2} shows distributions for the chirp mass $\Delta
{\cal M}/{\cal M}$, the reduced mass $\Delta \mu/\mu$, the spin
parameters, $\Delta \beta$ and $\Delta \sigma$, in general relativity,
and finally the bound on the graviton Compton wavelength when that
term is included.  The corresponding errors are listed in
Table~\ref{MGaverage}.  It turns out in this case that the
``pattern-averaged'' approach of Sec.~\ref{subsec1} provides very
good estimates of the errors on the parameters $\cal M$, $\eta$,
$\lambda_g$, $\beta$ and $\sigma$ (in fact, these estimates are almost
identical to those obtained from the Monte Carlo simulations using
both detectors, as can be seen by comparing the relevant values in
Tables~\ref{MG1} and \ref{MG2} with Table~\ref{MGaverage}).  Quite
remarkably, the errors on $D_L$, ${\cal M}$, $\Omega_S$ and $z$ in the
case GR+SO+SS are exactly the same as the corresponding errors for the
case MG+SO, while the errors on $\eta$ and $\beta$ differ (a similar
consideration applies to the angle-average calculation of
Sec.~\ref{results1}).

The determination of ${\cal M}$ and $\mu$ is affected by the use of
one or two detectors and by the inclusion of the spin couplings in the
same way.  Notice however that $\Delta \mu/\mu$ is always about two
orders of magnitude larger than $\Delta {\cal M}/{\cal M}$.  The
determination of the spin-spin parameter $\sigma$ without large errors
is only made possible by the use of two detectors.  

When we consider alternative theories of gravity (either with a BD or
a massive graviton term) we do not include spin-spin effects because
the Fisher matrix becomes ill-conditioned and non-invertible (see
Appendix~\ref{app1}). The non-invertibility of the Fisher matrix in
the case MG+SO+SS is probably related to the ``degeneracy'' of the
errors in the cases GR+SO+SS and MG+SO.

\subsection{Redshift dependence of parameter estimation}
\label{results2b}

Inspiralling MBH can be observed by {\em LISA} out to enormous
distances. If their masses and luminosity distances are determined
with sufficient accuracy, {\em LISA} can be a source of information on
the growth of structures at high redshift. In this context the
redshift $z$ can be large, and the distinction between observed masses
and masses as measured in the source rest frame -- given by the simple
rescaling (\ref{source}) -- is important.  {\em LISA} can only measure
redshifted combinations of the intrinsic source parameters (masses and
spins), so it cannot measure the redshift $z$. If cosmological
parameters are known, Eq.  (\ref{DsubL}) can be inverted to yield $z$
as a function of $D_L$~\cite{SH}, and {\em LISA} measurements of the
luminosity distance can be used to obtain black hole masses as a
function of redshift, thus constraining hierarchical merger
scenarios. Alternatively: if we can obtain the binary's redshift by
some other means, e.g., from an electromagnetic counterpart, then {\em
LISA} measurements of $D_L$ can be used to improve our knowledge of
the cosmological parameters~\cite{schutz,markovic,HH2}.

These exciting applications depend, of course, on {\em LISA}'s
measurement accuracy at large redshifts. In the following we look at
the redshift dependence of measurement errors for two representative
MBH binaries having masses $(10^6+10^6) M_\odot$ and $(10^7+10^7)
M_\odot$ {\em as measured in the source rest frame}. (This choice is
at variance with the rest of the paper, where we fix instead the
values of the measured masses {\em at the detector}.) We consider a
zero--curvature--universe with ($\Omega_M=0.3$, $\Omega_\Lambda=0.7$)
and $H_0=72$ km~s$^{-1}$~Mpc$^{-1}$, according to the present
observational estimates. We also assume that the {\em LISA} noise can
be extrapolated down to $f_{\rm low}=10^{-5}$~Hz; more conservative
assumptions on $f_{\rm low}$ could significantly affect our
conclusions (see~\cite{HH,Baker} and Sec.~\ref{results3}). We compute
the errors, as in Sec.~\ref{results2}, performing Monte Carlo
simulations of $10^4$ binaries for different values of the redshift
and then averaging over all binaries.

Fig. \ref{zeta} shows the redshift dependence of the average errors on
various quantities (all errors are computed using two detectors). The
left panel corresponds to the general relativistic inspiral of a
nonspinning binary. In the right panel we include, in addition, SO and
SS terms. Solid (dashed) lines refer to MBH binaries having mass
$(10^6+10^6) M_\odot$ [$(10^7+10^7) M_\odot$, respectively] as
measured in the source rest frame.

As expected from the discussion in Sec.~\ref{results2}, distance
determination and angular resolution are essentially independent of
the inclusion of spin terms. The relative error on $D_L$ for the
lower-mass binary system is $\sim 2 \%$ at $z=1$, $\sim 5 \%$ at $z=2$
and $\sim 11 \%$ at $z=4$.  This reduction in accuracy is due to the
fact that the signal spends less and less time in band as the redshift
is increased. We only consider values of the redshift such that the
binary spends at least one month in band before coalescing: for the
case $(10^7+10^7) M_\odot$, this corresponds to $z\sim 4$. The
distance determination error for this high-mass binary grows quite
rapidly, being $\sim 2 \%$ at $z=1$, $\sim 6 \%$ at $z=2$ and $\sim 21
\%$ at $z=4$. {\em LISA}'s angular resolution is rather poor even at
small redshifts, and it rapidly degrades for sources located farther
away, the degradation being more pronounced for higher-mass
binaries. Better distance determinations can be obtained if we are
lucky enough to locate the source in the sky by some other means: for
example, associating the gravitational wave event with an
electromagnetic counterpart. In this case angles and distance would be
decorrelated, allowing order of magnitude improvements in the
determination of $D_L$~\cite{HH}. We should also recall that in our
discussion we are quoting {\it average} errors. Since the logarithmic
distribution of $\Delta D_L/D_L$ has a rather long tail at large
values of the error (Fig.~\ref{histoGR1}), distance errors in a
specific detection could actually be much smaller than the average.

Unimportant as they are for distance determination and angular
resolution, spin effects have a dramatic impact on mass measurement
accuracy. For our low-mass system, in the absence of spins the chirp
mass can be measured with fantastic accuracy up to $z=10$, the largest
error being $\sim 0.06 \%$. Even including SO and SS effects and
ignoring precession, the error on ${\cal M}$ is only $\sim 2.5 \%$ at
$z=10$. Errors are predictably larger for the $(10^7+10^7) M_\odot$
binary. When we omit spin effects ${\cal M}$ can still be measured
with an accuracy better than a percent out to $z=4$, but when we
include SO and SS terms the error at $z=2$ is already $\sim 6 \%$.
Our ability to measure the mass of {\em both} black holes is severely
limited by the error on the reduced mass $\mu$, which is always about
two orders of magnitude larger than the error on ${\cal M}$.  Errors
on $\mu$ for nonspinning binaries of $(10^6+10^6)~M_\odot$ are
remarkably small if we ignore spin effects: at $z=10$ the error is
only $0.6 \%$. Including both SO and SS terms things get much worse,
and even at $z=1$ the reduced mass error is $\sim 6 \%$.

For $(10^6+10^6) M_\odot$ general relativistic nonspinning binaries, a
least-square fit of mass and distance errors in the interval $z\in
[1,10]$ yields:
\bea
\Delta {\cal M}/{\cal M}&=&
(-1.1476+7.2356 z+5.7376 z^2)\times 10^{-6}\,, \nn\\
\Delta \mu/\mu&=&
(-0.61431+1.9018 z+0.43721 z^2)\times 10^{-4}\,, \nn\\
\Delta D_L/D_L&=&
(-0.65651+2.6935 z+0.061595 z^2)\times 10^{-2}\,. \nn
\eea

It is important to remark here that in our study we are ignoring
precessional effects. These effects induce modulations in the
waveform, possibly improving the mass measurements in a significant
way~\cite{vecchio}. The study of precession is therefore crucial to
assess {\em LISA}'s ability to measure MBH masses in galactic
mergers. Such a study is beyond the scope of this paper.

\subsection{Effect of {\em LISA}'s low frequency sensitivity on parameter
estimation}
\label{results3}

High-mass binaries sweep through the low-frequency region of the {\em
LISA} band, where the {\em LISA} sensitivity will ultimately depend on
design choices for the acceleration noise.  To explore the possible
consequences of such design choices, we studied the effects on the
accuracy of parameter estimation if {\em LISA} were completely blind
below some cutoff frequency $f_{\rm low}$.  In all our analyses to
this point, we chose the default value for $f_{\rm low}$ to be
$10^{-5}$ Hz, but we now consider higher cutoff frequencies $5\times
10^{-5}$ ~Hz and $10^{-4}$~Hz.

Table~\ref{MGaverage} shows that if $f_{\rm low}=5\times 10^{-5}$~Hz ,
parameter estimation for binaries of $(10^6+10^6)~M_\odot$ is
essentially unaffected.  But if $f_{\rm low}=10^{-4}$~Hz, the
accuracies of estimating all quantities are degraded by factors
between two and six, both for GR and for massive graviton theories.

Since binaries of larger mass sweep through a lower frequency band, we
expect the degradation to be even worse for such binaries.  To be more
quantitative, we analysed what happens when we increase the {\em LISA}
cutoff frequency $f_{\rm low}$ from $10^{-5}$~Hz to $10^{-4}$~Hz for
massive black hole binaries having total mass larger than $\sim
10^5~M_\odot$. Similar preliminary investigations of the effect of
low-frequency {\em LISA} noise on the distance determination of
massive black holes can be found in~\cite{HH}. In this subsection we
consider for concreteness nonspinning, equal-mass binaries. All quoted
values for the average errors have been obtained using Monte Carlo
simulations with two detectors.

The results are summarized in Fig.~\ref{errGRMG}, where we show errors
computed both in general relativity (left panel) and in massive
graviton theories (right panel).  In the plots we display results for
cutoff frequencies $f_{\rm low}=10^{-5}$~Hz and $f_{\rm
low}=10^{-4}$~Hz.

Consider first the distance determination accuracy $\Delta D_L/D_L$
one could achieve.  For general relativistic binaries, as we move the
low-frequency cutoff $f_{\rm low}$, $\Delta D_L/D_L$ grows from
$1.1~\%$ to $1.8~\%$ for a binary of $(10^6+10^6)~M_\odot$.  If we
double the mass of each black hole, the corresponding increase is
roughly twice as large -- from $1.0~\%$ to $2.6~\%$. For larger values
of the mass and $f_{\rm low}=10^{-4}$~Hz the binary does not spend
much time in band, and the inversion of the Fisher matrix becomes
problematic (see Appendix~\ref{app1}).
The angular resolution is even more sensitive to the low-frequency
cutoff. As the cutoff $f_{\rm low}$ goes from $10^{-5}$~Hz to
$10^{-4}$~Hz, $\Delta \Omega_S$ (in steradians) goes from $9.7\times
10^{-5}$ to $4.8\times 10^{-4}$ for a binary of
$(10^6+10^6)~M_\odot$. If we double the mass of each black hole
$\Delta \Omega_S$ correspondingly goes from $8.4 \times 10^{-5}$ to
$1.8\times 10^{-3}$, increasing by a factor $\sim 10^2$.
We note another interesting feature emerging from Fig.~\ref{errGRMG}.
If present design choices allow {\em LISA} to be sensitive down to
$f_{\rm low}=10^{-5}$~Hz, the distance error and angular resolution
(at least for $D_L=3$~Gpc) will be decreasing functions of $M$ in the
supermassive black hole mass range $M\in[10^6-10^7]~M_\odot$. This
feature could be used to study the merger history of black holes at
galactic centers, and to map structure formation in the early universe.

Mass determinations are also strongly affected by low-frequency
sensitivity, as observed in Refs.~\cite{HH,Baker}.  For a
$(10^6+10^6)~M_\odot$ binary in general relativity, a lower cutoff at
$f_{\rm low}=10^{-4}$~Hz increases the error on the chirp mass by a
factor 3 for a $(10^6+10^6)~M_\odot$ binary, while the error in $\mu$
correspondingly increases from $0.0066~\%$ to $0.012~\%$.  An error of
the order of a percent on mass determination during coalescence could
be a problem for the identification of ``golden binaries'' (binaries
for which we can measure the total mass-energy lost to GWs~\cite{HM})
with {\em LISA}.  If in addition to a cutoff at $f_{\rm
low}=10^{-4}$~Hz we also include a massive graviton term, the errors
for binaries of total mass $10^7~M_\odot$ become unacceptably large:
$\sim 0.6~\%$ for the chirp mass and $\sim 4~\%$ for the reduced mass.

Next we consider how the low-frequency cutoff affects bounds on the
graviton mass.  The results are shown in Fig.~\ref{lambda}. It turns
out that the bound on the graviton Compton wavelength is affected by
low frequency noise in the same way as the accuracy in distance and
redshift determination .  As we increase $f_{\rm low}$ from
$10^{-5}$~Hz to $10^{-4}$~Hz, the bound on $\lambda_g$ (in units of
$10^{15}~$km) drops from $49.4$ to $29.5$ for a binary of
$(10^6+10^6)~M_\odot$. For an equal-mass BH-BH binary of total mass
$10^7~M_\odot$ the corresponding reduction is from $67.9$ to
$10.3$. Notice also that when we pick $f_{\rm low}=10^{-4}$~Hz the
bound on $\lambda_g$ has a maximum for equal-mass binaries of total
mass $\sim 10^6~M_\odot$. Being sensitive below $10^{-4}$~Hz is
therefore important to put bounds on the graviton mass through
observations of binaries more massive than this.

In conclusion, design choices for the {\em LISA} low-frequency
acceleration noise will have a dramatic impact on our ability to i)
locate massive black hole binaries in the sky, ii) measure their
masses, iii) use them as standard cosmological candles, and iv) bound
the mass of the graviton.

\section{Conclusions}
\label{conclusions}

In this paper we analysed how the inclusion of spin couplings affects
parameter estimation in observations of binary coalescence, and
the bounds that can be placed on alternative theories of gravity, such
as the Brans-Dicke and massive graviton theories.  Extending 
previous investigations~
\cite{willST,willgraviton,scharrewill,willyunes}, we also took into
account the dependence on the four angles describing the source
location and the direction of the (orbital) angular momentum,
performing large-scale Monte Carlo simulation of $10^4$ binaries.

We found that the bound on the Brans-Dicke parameter (and therefore
also the bound on parameters describing more general scalar-tensor
theories, such as those considered in Ref.~\cite{damourfarese}) is
significantly reduced by spin-orbit and spin-spin couplings, while the
bound on the graviton Compton wavelength is only marginally reduced.
As expected, we found that the inclusion of the four orientation
angles does not alter the estimation of the binary masses and
spins. The reason is that the orientation angles are rather
uncorrelated with those parameters, appearing only in the GW amplitude
and not in the phase. For the same reason, we found that spin-orbit
and spin-spin couplings have little effect on the angular resolution,
distance determination and hence on the redshift determination for
massive black-hole binaries.  For NS-IMBH binaries, these extrinsic
parameters are determined rather poorly with or without spin effects.
For massive black-hole binaries, Monte Carlo simulations show that
{\em LISA} can provide reasonably accurate distance determinations out
to $z\sim 2-4$ for black hole masses $\lesssim 10^7 M_\odot$.

The cosmological reach of {\em LISA} will ultimately depend on design
choices for the acceleration noise. The reason is that for massive
binaries the GW signal sweeps through the low-frequency \lisa~ band.
By default we made a rather optimistic assumption that the \lisa~
noise curve can be extrapolated down to a lower frequency of $10^{-5}$
Hz.  We then carried out an explorative survey to see how a higher
(more conservative) low-frequency cutoff affects the determination of
the binary parameters for high-mass configurations.  We found that the
cutoff will have a dramatic impact on our ability to i) locate black
hole binaries of mass $\gtrsim 10^6~M_\odot$ in the sky, ii) measure
their masses, iii) use them as standard cosmological candles, and iv)
bound the mass of the graviton. Our results are compatible with
similar investigations which have appeared in the
literature~\cite{HH,Baker}.
 
Our analysis was limited to non-precessing binaries.
Vecchio~\cite{vecchio} has shown that for comparable high-mass BHs
[e.g., $(10^6 + 10^6) M_\odot$] modulational effects can decorrelate
some of the binary parameters, allowing a better estimation of masses
and distances with respect to the case when spins are aligned or
antialigned with the (orbital) angular momentum.  At this stage it is
not clear if modulational effects can improve the accuracy in
estimating binary parameters also for small mass-ratio binaries (e.g.,
a NS and an IMBH), and allow one to put more stringent bounds on
alternative theories of gravity. Only a direct calculation can clarify
this point, and we plan to tackle it in the near future.

When describing stellar mass objects inspiralling into
intermediate-mass black holes, we only considered circular
orbits. This assumption is barely justified, especially for high mass
ratios. In the future we plan to investigate how the results change
when eccentricity is included.

Finally, in this paper we have focused on statistical errors,
implicitly assuming that the waveform is known with high enough
accuracy to disregard systematic errors.  However, massive binaries
are likely to be detected with rather high SNR, on the order of
100. If this is the case, it might well be that spinning waveforms at
2PN order are not sufficiently accurate to permit one to neglect
systematic over statistical errors.

\acknowledgments 
We thank Luc Blanchet, Curt Cutler, Scott Hughes, Shane Larson, Eric
Poisson, Michele Vallisneri and Alberto Vecchio for useful discussions
and Scott Hughes for providing important comments on the manuscript.
This work was supported in part by the National Science Foundation
under grant PHY 03-53180.

\appendix

\section{Essential formulae for the 
{\em LISA} response to non-precessing spinning binaries}
\label{app2}

In this Appendix we write down the essential formulae of
Ref.~\cite{CC} which we use in Sec.~\ref{subsec2}. (We refer the
reader to Ref.~\cite{CC} for further details and notation.)

Unbarred quantities refer to the rotating {\em LISA}-based coordinate
system, while barred quantities refer to the fixed solar-system-based
coordinate system. Assuming as in Ref.~\cite{CC} that the noise is
symmetric in each pair of {\em LISA} arms, we can reduce {\em LISA} to
two independent Michelson interferometers with equilateral triangle
shape. In this approximation the {\em LISA} beam-pattern functions for
the two Michelson outputs are the same as for a single detector,
except for the factor $\sqrt{3}/2$ which already appears in
Eq.~(\ref{hSPA}), and are given by
\bea
F_{\rm I}^+(\theta_S,\phi_S,\psi_S) &=& \frac{1}{2}(1 + \cos^2 \theta_S) \cos 2\phi_S \cos
2\psi_S \nonumber \\
&& -\cos \theta_S \sin 2\phi_S \sin 2\psi_S\;, \nn\\
F_{\rm I}^{\times}(\theta_S,\phi_S, \psi_S) &=& \frac{1}{2}(1 + \cos^2 \theta_S) \cos 
2\phi_S \sin 2\psi_S \nonumber \\
&& + \cos \theta_S \sin 2\phi_S \cos 2\psi_S\;,
\label{patternI}
\eea
and 
\bea
F_{\rm II}^+(\theta_S,\phi_S,\psi_S) &=& F_{\rm I}^+(\theta_S,\phi_S -\frac{\pi}{4},\psi_S)\,, \nn\\ 
F_{\rm II}^{\times}(\theta_S,\phi_S, \psi_S) &=& 
F_{\rm I}^+(\theta_S,\phi_S -\frac{\pi}{4},\psi_S)\,. 
\label{patternII}
\eea
In the above equations we have denoted by $(\theta_S,\phi_S)$ the
source location and by $\psi_S$ the polarization angle defined as
\beq
\tan \psi_S(t) =  \frac{\hat \vL \cdot \vz -   
(\hat \vL \cdot \vn) (\vz \cdot \vn)}
{\vn \cdot ( \hat \vL \times \vz)}\;,   
\eeq
$\hat \vL$, $\vz$ and $-\vn$ being the unit vectors along the orbital
angular momentum, the unit normal to {\em LISA}'s plane and the GW
direction of propagation, respectively.

The waveform polarization and Doppler phases entering the GW signal
(\ref{hSPA}) are ($\alpha = {\rm I,II}$):
\bes
\bea
\label{phip}
\varphi_{p,\alpha}(t) &=& \tan^{-1}\left[\frac{2 (\hat \vL \cdot  \vn) F_\alpha^\times(t)}
{(1+(\hat \vL \cdot \vn)^2) F_\alpha^+(t)}\right]\;,\\
\label{phiD}
\varphi_D(t) &=& \frac{2\pi f}{c}\,  R \, {\rm sin}\,\bt_S \ 
{\rm cos}(\bph(t) - \bph_S)  \;,
\eea
\ees
with $R = 1 {\,\rm AU}$ and 
$\bph(t) = \bph_0 + 2\pi {t}/{T}$.
Here $T=1$ year is the orbital period of {\em LISA}, and $\bph_0$ is a
constant that specifies the detector's location at time $t=0$.  In
this paper we always assume that there is no precession, so $\hat L^a$
points in a fixed direction $(\bt_L,\bph_L)$.

To express the angles $(\theta_S,\phi_S,\psi_S)$ evaluated with respect to 
the rotating detector-based coordinate system as function of the angles 
$(\bt_S,\bph_S,\bt_L,\bph_L)$ evaluated with respect to 
the fixed solar-system based coordinate system, 
we use the following relations~\cite{CC}:
\bes
\bea
\cos\theta_S(t) &=& \frac{1}{2} \cos\bt_S - \frac{\sqrt{3}}{2}\sin\bt_S
\cos(\bph(t)-\bph_S) \;,
\\
 \phi_S(t) &=& \alpha_0 + \frac{2\pi t}{T} + 
 \tan^{-1} \biggl[\frac{\sqrt{3}\cos\bt_S \ + \ 
\sin\bt_S \cos(\bph(t)-\bph_S)}{2\sin\bt_S\sin(\bph(t)-\bph_S)} \biggr]\;,
\eea
\ees
where $\alpha_0$ is a constant specifying the orientation of
the arms at $t=0$.  Following Cutler~\cite{CC}, we take $\alpha_0=0$
and  $\bar \phi_0=0$, corresponding to a specific choice of the
initial position and orientation of the detector.
In addition,
\bes
\bea
\vz \cdot \vn &=& \cos \theta_S\,,
\\
\hat \vL \cdot \vz &=& \frac{1}{2} {\rm cos}\,\bt_L - \frac{\sqrt{3}}{2} 
{\rm sin}\; \bt_L  {\rm cos}\,\bigl(\bph(t) - \bph_L \bigr)\;,
\\
\hat \vL \cdot \vn &=& {\rm cos}\; \bt_L {\rm cos}\;\bt_S +
{\rm sin}\,\bt_L  \; {\rm sin}\,\bt_S \; 
{\rm cos}\,\bigl(\bph_L - \bph_S \bigr) \;,
\\
\vn \cdot (\hat \vL \times \vz) &=& \frac{1}{2} \,{\rm sin}\,\bt_L \; 
{\rm sin}\,\bt_S\; 
{\rm sin}\,\bigl(\bph_L - \bph_S \bigr) \nonumber \\ 
&&- \frac{\sqrt{3}}{2} {\rm cos}\bph(t)\biggl(
 {\rm cos}\,\bt_L \; {\rm sin}\,\bt_S \; {\rm sin}\,\bph_S - 
 {\rm cos}\,\bt_S \; {\rm sin}\,\bt_L \;{\rm sin}\,\bph_L \biggr)
\nonumber\\
&&- \frac{\sqrt{3}}{2} {\rm sin}\bph(t)\biggl(
 {\rm cos}\,\bt_S \; {\rm sin}\,\bt_L \; {\rm cos}\,\bph_L - 
 {\rm cos}\,\bt_L \; {\rm sin}\,\bt_S \; {\rm cos}\,\bph_S \biggr)\;.
\eea
\ees

\section{Subtleties in the inversion of the Fisher matrix}
\label{app1}

In the paper we evaluate the Fisher matrix using both a Mathematica
and a Fortran code.  In the Fortran code we normally perform the
numerical inversion of the Fisher matrix using the LU decomposition,
which expresses the Fisher matrix as the product of a Lower-triangular
and an Upper-triangular matrix (cf. Sec. 2.3 of~\cite{numrec}). To
check the result we simply multiply the inverse by the original
matrix. In this way we obtain a numerical ``identity matrix'' whose
elements $I^{\rm num}_{ij}$ will be slightly different from the
Kronecker symbol $\delta_{ij}$. We can measure this deviation from the
``true'' identity matrix defining a small quantity
\beq
\epsilon_{\rm inv}\equiv {\rm max}_{i,j}
\left|
I^{\rm num}_{ij}-\delta_{ij}
\right|\,.
\eeq
We found that extreme mass ratio inspirals are more likely to yield an
ill-conditioned Fisher matrix. Therefore, as a rule of thumb, we
consider the inversion successful if the parameter $\epsilon_{\rm
inv}<10^{-3}$ (for NS-BH binaries) and if $\epsilon_{\rm inv}<10^{-4}$
(for massive BH-BH binaries). In Mathematica we perform a similar
check using the built-in matrix inversion routine.

Matrix inversion generally becomes more difficult as the number of
elements of the Fisher matrix increases. In particular, the LU
decomposition and the Mathematica inversion routine fail when we
consider alternative theories of gravity including both spin-orbit and
spin-spin terms.  To understand the reason for this failure we can use
a principal component analysis~\cite{sathya}, also known as
singular-value decomposition (see eg. Sec. 2.6 of~\cite{numrec}). We
decompose the Fisher matrix $\mathbf{F}$ as
\beq
\mathbf{F}=\mathbf{U}\mathbf{W}\mathbf{V^T}\,,
\eeq
$\mathbf{U}$ and $\mathbf{V}$ being orthogonal matrices, and
$\mathbf{V^T}$ being the transpose of $\mathbf{V}$. The matrix
$\mathbf{W}$ is diagonal with positive or zero elements $w_j$ (the
singular values). The inverse of $\mathbf{F}$ is then given by
\beq
\label{SVDinv}
\mathbf{F}=\mathbf{V}\mathbf{W^{-1}}\mathbf{U^T}\,,
\eeq
where $\mathbf{W^{-1}}$ is a diagonal matrix whose elements are the
reciprocals $1/w_j$ of the singular values. Numerically speaking, a
matrix is not invertible when the reciprocal of its {\it condition
number} (defined as the ratio of the largest singular value to the
smallest singular value) approaches the machine's floating-point
precision.  When we consider alternative theories including both
spin-orbit and spin-spin terms, the matrix becomes non-invertible in
this sense: our numerical experiments show that one of the singular
values approaches zero. In principle, even in this case we can still
obtain a ``pseudo-inverse'': the matrix which is closest to the
``real'' inverse in a least-square sense~\cite{numrec}. To do this it
suffices to replace $1/w_j$ by zero whenever $w_j$ is zero in
Eq. (\ref{SVDinv}). However, in this paper we decided not to quote
results obtained in this way. We estimate the binary parameters {\it
only} when spin-orbit and spin-spin terms (Tables~\ref{LISAnoBD}
and~\ref{MG1}), Brans-Dicke and spin-orbit terms
(Table~\ref{LISAspin}) or massive graviton and spin-orbit terms
(Table~\ref{MG2}) are included.


\begin{widetext}

\begin{table}[hbt]
\centering
\caption{Number of GW inspiral cycles contributed by different PN
orders for different NS-BH binaries.  We assume ${\cal S}=0.3$ and an
observation time $T_{\rm obs}=1$ yr.  In the bottom section of the
table, we normalize the number of cycles associated with the
Brans-Dicke parameter to $\varpi$ (first row) and to the Cassini bound
$\omega_{\rm BD} > \omega_{\rm Cassini}=4 \times 10^4$ (second
row). We also show the initial and final GW frequencies, assuming an
upper cutoff of 1.0 Hz for the \lisa~ noise curve.}
\vskip 12pt
\begin{tabular}{@{}|l|c|c|c|c|@{}}
\hline
\hline
PN order	&$(1.4+400) M_\odot$ &$(1.4+1000)M_\odot$ &$(1.4+5000)M_\odot$ &$(1.4+10^4) M_\odot$ \\
\hline
\hline
$f_{\rm in} {\rm (Hz)}$   &$4.601\times 10^{-2}$   &$3.658\times 10^{-2}$  &$2.446\times 10^{-2}$  &$2.057\times 10^{-2}$ \\
$f_{\rm fin}{\rm (Hz)}$   &1.000  &1.000  &0.8792  &0.4397 \\
\hline
\hline
Newtonian	&2,294,904          &1,828,036           &1,224,122         &1,025,711              \\       
1PN             &35,366             &44,712              &67,309            &78,460                 \\
Tail            &-18,064            &-29,081             &-66,278           &-89,793                \\
Spin orbit      &1,437$\beta$       &2,314$\beta$        &5,274$\beta$      &7,145$\beta$           \\
2PN             &422                &868                 &3,016             &4,653                  \\
Spin spin	&-139$\sigma$       &-288$\sigma$        &-1,001$\sigma$    &-1,545$\sigma$         \\
\hline
Brans-Dicke	&-3,560,569$\varpi$ &-1,793,782$\varpi$  &-536,954 $\varpi$ &-319,126$\varpi$       \\
Brans-Dicke &-89~$\omega_{\rm Cassini}/\omega_{\rm BD}$ &-45~$\omega_{\rm Cassini}/\omega_{\rm BD}$  
&-13~$\omega_{\rm Cassini}/\omega_{\rm BD}$ &-8.0~$\omega_{\rm Cassini}/\omega_{\rm BD}$ \\ 
\hline	     	   	       	            	 	     
\hline								     
\end{tabular}
\label{LISAcycBD}
\end{table}

\begin{table}[hbt]
\centering
\caption{Number of GW inspiral cycles contributed by different PN
orders for high-mass BH binaries. We assume an observation time
$T_{\rm obs}=1$ yr.  In the bottom section of the table, we normalize
the number of cycles associated with the graviton-mass term to
$\beta_g$ (first row), and to the Compton wavelength $\lambda_g$,
using Eq.~(\ref{betag}) (second row).  We assume a luminosity distance
$D_L=3$ Gpc, $\Omega_M = 0.3$ and $\Omega_{\Lambda} = 0.7$.}
\vskip 12pt
\begin{tabular}{@{}|l|c|c|c|c|c|@{}}
\hline
\hline
PN order	&$(10^7+10^7) M_\odot$ &$(10^7+10^6)M_\odot$ &$(10^6+10^6)M_\odot$ &$(10^5+10^4)M_\odot$ &$(10^4+10^4) M_\odot$ 
\\
\hline
\hline
$f_{\rm in} {\rm (Hz)}$   &$1.073\times 10^{-5}$ &$2.361\times 10^{-5}$ &$4.525\times 10^{-5}$ &$4.199\times 10^{-4}$ &$8.046\times 10^{-4}$\\
$f_{\rm fin}{\rm (Hz)}$   &$2.199\times 10^{-4}$ &$3.997\times 10^{-4}$ &$2.199\times 10^{-3}$ &$3.997\times 10^{-2}$ &$0.2199$\\
\hline
\hline
Newtonian	&535           &1,174         &2267          &21,058          &40,369\\
1PN             &55            &115           &134           &677             &769\\
Tail            &-48           &-127          &-92           &-450            &-308\\
Spin orbit      &4$\beta$      &10   $\beta$  &7$\beta$      &36   $\beta$    &25$\beta$\\
2PN             &4             &8             &6             &18              &12\\
Spin spin	&-1$\sigma$    &-2   $\sigma$ &-1$\sigma$    &-5   $\sigma$   &-3$\sigma$\\
\hline massive graviton &-209$\beta_g$ &-333$\beta_g$ &-512$\beta_g$ &-1967$\beta_g$ &-2926$\beta_g$\\
massive graviton 
& -1063 $\left (10^{15} {\rm km}/\lambda_g\right )^2$ 
& -478  $\left (10^{15} {\rm km}/\lambda_g\right )^2$ 
& -260  $\left (10^{15} {\rm km}/\lambda_g\right )^2$ 
& -28   $\left (10^{15} {\rm km}/\lambda_g\right )^2$ 
& -15   $\left (10^{15} {\rm km}/\lambda_g\right )^2$ \\
\hline	     	   	       	            	 	     
\hline								     
\end{tabular}
\label{LISAcycMG}
\end{table}

\begin{table}[hbt]
\centering
\caption{ Errors and correlation coefficients for different NS-BH
binaries at different PN orders in general relativity (no Brans-Dicke
term) with and without spin-orbit and spin-spin terms.  We consider
one detector and set $\rho=10$.}
\vskip 12pt
\begin{tabular}{@{}ccccccccccccc@{}}
\hline
\hline
PN order &$\Delta t_c$ &$\Delta \phi_c$ &$\Delta {\cal M}/{\cal M}$ &$\Delta \eta/\eta$ &$\Delta \beta$ &$\Delta \sigma$ 
&$c^{{\cal M}\eta}$ &$c^{{\cal M}\beta}$ &$c^{\eta \beta}$ &$c^{{\cal M}\sigma}$ &$c^{\eta \sigma}$ &$c^{\beta \sigma}$\\
&(s)&&(\%)&(\%)&\\
\hline
\multicolumn{13}{l}{$(1.4+400)M_\odot$}\\
1    &1.59 &3.61 &0.0000710 &0.0206  &-       &-     &-0.995 &-         &-         &-         &-         &-        \\	
1.5  &3.00 &15.6 &0.000148  &0.149   &-       &-     &-0.999 &-         &-         &-         &-         &-        \\	
1.5  &4.07 &28.1 &0.000478  &0.375   &0.00346 &-     &-0.996 &0.951 &-0.918 &-         &-         &-        \\	
2    &3.33 &17.6 &0.000220  &0.208   &-       &-     &-0.999 &-         &-         &-         &-         &-        \\	
2    &4.00 &25.7 &0.000491  &0.393   &0.00260 &-     &-0.996 &0.893 &-0.849 &-         &-         &-        \\	
2    &16.2 &508  &0.00266   &3.85    &0.206   &1.91  &-0.996 &-0.981 &0.994 &-0.983 &0.995 &1.000\\                    
\hline											  	      		  	      		  
\multicolumn{13}{l}{$(1.4+1000)M_\odot$}\\
1    &1.86 &3.41 &0.0000861 &0.0157  &-       &-     &-0.995 &-         &-         &-         &-         &-        \\	
1.5  &2.52 &8.08 &0.0000233 &0.0369  &-       &-     &-0.922 &-         &-         &-         &-         &-        \\	
1.5  &4.45 &25.2 &0.000560  &0.275   &0.00618 &-     &-0.995 &0.999 &-0.991 &-         &-         &-        \\	
2    &2.55 &7.89 &0.0000341 &0.0439  &-       &-     &-0.965 &-         &-         &-         &-         &-        \\	
2    &4.33 &21.8 &0.000584  &0.297   &0.00560 &-     &-0.996 &0.998 &-0.989 &-         &-         &-        \\	
2    &16.1 &425  &0.00296   &2.65    &0.0705  &0.793 &-0.996 &-0.962 &0.982 &-0.980 &0.994 &0.997\\                   
\hline											  	      		  	      		  
\multicolumn{13}{l}{$(1.4+5000)M_\odot$}\\
1    &2.55 &3.19 &0.000123  &0.0101  &-       &-     &-0.995 &-         &-         &-         &-         &-        \\
1.5  &3.03 &5.61 &0.0000386 &0.00870 &-       &-     &0.941 &-         &-         &-         &-         &-        \\ 
1.5  &5.68 &22.4 &0.000771  &0.168   &0.00673 &-     &-0.995 &0.999 &-0.999 &-         &-         &-        \\       
2    &3.00 &5.18 &0.0000360 &0.0101  &-       &-     &0.932 &-         &-         &-         &-         &-        \\ 
2    &5.34 &14.7 &0.000857  &0.202   &0.00700 &-     &-0.996 &0.999 &-0.999 &-         &-         &-        \\       
2    &18.5 &349  &0.00386   &1.52    &0.0137  &0.178 &-0.996 &0.951 &-0.920 &-0.975 &0.991 &-0.860\\                 
\hline											  	      		  	      		  
\multicolumn{13}{l}{$(1.4+10^4)M_\odot$}\\
1    &3.33 &3.43 &0.000154  &0.00914 &-       &-     &-0.995 &-         &-         &-         &-         &-        \\
1.5  &3.96 &5.82 &0.0000607 &0.00582 &-       &-     &0.965 &-         &-         &-         &-         &-        \\ 
1.5  &8.60 &26.7 &0.00103   &0.163   &0.00737 &-     &-0.996 &0.998 &-0.999 &-         &-         &-        \\       
2    &3.91 &5.26 &0.0000592 &0.00678 &-       &-     &0.964 &-         &-         &-         &-         &-        \\ 
2    &7.72 &13.0 &0.00120   &0.211   &0.00822 &-     &-0.997 &0.999 &-0.999 &-         &-         &-        \\       
2    &34.6 &487  &0.00576   &1.66    &0.0299  &0.133 &-0.997 &0.998 &-0.989 &-0.978 &0.992 &-0.961\\                 
\hline
\hline
\end{tabular}
\label{LISAnoBD}
\end{table}

\begin{table}[hbt]
\centering
\caption{Errors and correlation coefficients in Brans-Dicke theory
using 2PN templates, with and without the spin-orbit term. We consider
one detector and set $\rho=10$.  In the first row we do not consider
spin terms; in the second row we also include spin-orbit effects.
When we include the spin-orbit term priors do not have an appreciable
effect on parameter estimation. For each binary we also give the bound
$\omega_{\rm BD,unc}$ that could be obtained (in principle) if all the
binary parameters were known and not correlated with the BD term.}
\vskip 12pt
\begin{tabular}{@{}lccccccccccc@{}}
\hline
\hline
$\Delta t_c$ &$\Delta \phi_c$ &$\Delta {\cal M}/{\cal M}$ &$\Delta \eta/\eta$ 
&$\omega_{\rm BD}$
&$\Delta \beta$ 
&$c^{{\cal M}\eta}$ &$c^{{\cal M}\varpi}$ &$c^{{\cal M}\beta}$ 
&$c^{\eta \varpi}$  &$c^{\eta \beta}$
&$c^{\varpi \beta}$\\
(s)&&(\%)&(\%)&&&&&&&&\\
\hline
\multicolumn{12}{l}{$(1.4+400) M_\odot$,~$\omega_{\rm BD,unc}=43,057,645$}\\
3.82 &23.2 &0.000243 &0.293  &765,014 &-      &-0.939 &0.421  &-      &-0.705 &- &- \\
7.95 &76.7 &0.00657  &2.50   &39,190  &0.0508 &-0.997 &-0.997 &0.999     &0.988 &-0.993     &-0.999        \\  
\hline
\multicolumn{12}{l}{$(1.4+1000) M_\odot$,~$\omega_{\rm BD,unc}=21,602,414$}\\
3.79 &16.7 &0.000189 &0.116  &211,389 &-      &0.845  &-0.984 &- &-0.926 &- &- \\
7.99 &58.4 &0.00764  &1.86   &21,257  &0.0557 &-0.996 &-0.997 &1.000     &0.987 &-0.998     &-0.995        \\
\hline
\multicolumn{12}{l}{$(1.4+5000) M_\odot$,~$\omega_{\rm BD,unc}=6,388,639$}\\
4.60 &12.5 &0.000600 &0.0342 &50,925  &-      &0.970 &-0.998 &- &-0.955 &- &- \\
8.79 &23.4 &0.0114   &1.33   &6,486   &0.0550 &-0.997 &-0.997 &0.999     &0.988 &-1.000     &-0.992        \\ 
\hline
\multicolumn{12}{l}{$(1.4+10^4) M_\odot$,~$\omega_{\rm BD,unc}=3,768,347$}\\
6.59 &13.8 &0.000877 &0.0253 &26,426  &-      &0.979  &-0.998 &- &-0.963 &- &- \\
13.6 &15.5 &0.0178   &1.61   &3,076   &0.0706 &-0.998 &-0.998 &0.999     &0.991 &-1.000     &-0.993        \\ 
\hline	     	   	       	            	 	     
\hline								     
\end{tabular}
\label{LISAspin}
\end{table}

\begin{table}[hbt]
\centering
\caption{Errors and correlation coefficients for various high mass BH
binaries in general relativity using one detector, with and without
spin-orbit and spin-spin terms. We set $D_L=3$ Gpc and assume $H_0=72$
km~s$^{-1}$~Mpc$^{-1}$. The effect of adding priors is practically
negligible in all cases.}
\vskip 12pt
\begin{tabular}{@{}cccccccccccc@{}}
\hline
\hline
$\Delta t_c$ &$\Delta \phi_c$ &$\Delta {\cal M}/{\cal M}$ &$\Delta \eta/\eta$ &$\Delta \beta$ &$\Delta \sigma$ 
&$c^{{\cal M}\eta}$ &$c^{{\cal M}\beta}$ &$c^{\eta \beta}$ &$c^{{\cal M}\sigma}$ &$c^{\eta \sigma}$ &$c^{\beta \sigma}$\\
(s)&&(\%)&(\%)&\\
\hline
\multicolumn{12}{l}{$(10^7+10^7)M_\odot$, SNR=2063}\\
5.27   &0.0108  &0.00189   &0.0401  &-       &-      &0.930 &-         &-         &-         &-         &-        \\
7.26   &0.0717  &0.0224    &5.56    &0.315   &-      &-0.996 &0.996 &-1.000 &-         &-         &-        \\
79.6   &1.12    &0.0703    &25.4     &1.29    &0.438  &-0.994 &0.997 &-1.000 &-0.948 &0.976 &-0.970\\                 
\hline
\multicolumn{12}{l}{$(10^7+10^6)M_\odot$, SNR=1204}\\
5.11   &0.0187  &0.00104   &0.0267  &-       &-      &0.934 &-         &-         &-         &-         &-        \\
9.50   &0.0589  &0.0132    &1.86    &0.0891  &-      &-0.996 &0.997 &-1.000 &-         &-         &-        \\
83.4   &2.18    &0.0499    &10.8     &0.402   &0.340  &-0.995 &0.999 &-0.999 &-0.964 &0.985 &-0.975\\                 
\hline
\multicolumn{12}{l}{$(10^6+10^6)M_\odot$, SNR=2143}\\
0.307  &0.00551 &0.000369  &0.0157  &-       &-      &0.887 &-         &-         &-         &-         &-        \\
0.496  &0.00819 &0.00303   &1.02    &0.0566  &-      &-0.991 &0.993 &-1.000 &-         &-         &-        \\
3.02   &0.317   &0.00872   &4.60    &0.213   &0.140  &-0.991 &0.996 &-0.999 &-0.938 &0.975 &-0.964\\                
\hline
\multicolumn{12}{l}{$(10^6+10^5)M_\odot$, SNR=2378}\\
0.214  &0.00734 &0.000202  &0.00976 &-       &-      &0.871 &-         &-         &-         &-         &-        \\
0.380  &0.00773 &0.00188   &0.352   &0.0162  &-      &-0.991 &0.994 &-1.000 &-         &-         &-        \\
2.46   &0.521   &0.00587   &1.79    &0.0560  &0.0858 &-0.991 &0.999 &-0.995 &-0.948 &0.980 &-0.957\\                
\hline
\multicolumn{12}{l}{$(10^5+10^5)M_\odot$, SNR=1710}\\
0.0521 &0.00539 &0.000114  &0.0113  &-       &-      &0.799 &-         &-         &-         &-         &-        \\
0.0746 &0.00614 &0.000766  &0.366   &0.0196  &-      &-0.985 &0.989 &-1.000 &-         &-         &-        \\
0.300  &0.201   &0.00210   &1.70    &0.0685  &0.0892 &-0.987 &0.995 &-0.997 &-0.931 &0.977 &-0.958\\                
\hline
\multicolumn{12}{l}{$(10^5+10^4)M_\odot$, SNR=601}\\
0.122  &0.0165  &0.0000810 &0.0111  &-       &-      &0.732 &-         &-         &-         &-         &-        \\
0.190  &0.0301  &0.000767  &0.225   &0.00934 &-      &-0.989 &0.994 &-0.999 &-         &-         &-        \\
0.643  &0.663   &0.00219   &1.08    &0.0213  &0.101  &-0.988 &0.995 &-0.970 &-0.937 &0.978 &-0.899\\                
\hline
\multicolumn{12}{l}{$(10^4+10^4)M_\odot$, SNR=252}\\
0.222  &0.0419  &0.0000822 &0.0308  &-       &-      &0.846 &-         &-         &-         &-         &-        \\
0.327  &0.103   &0.000622  &0.612   &0.0275  &-      &-0.984 &0.991 &-0.999 &-         &-         &-        \\
0.758  &1.39    &0.00200   &3.39    &0.0786  &0.480  &-0.990 &0.998 &-0.984 &-0.951 &0.984 &-0.937\\                
\hline
\hline
\end{tabular}
\label{MG1}
\end{table}

\begin{table}[hbt]
\centering
\caption{Errors and correlation coefficients for various high mass BH
binaries including the massive graviton and spin-orbit
terms.  We use one detector, set $D_L=3$ Gpc and assume $H_0=72$
km~s$^{-1}$~Mpc$^{-1}$.  We show the massive graviton bound obtained
assuming ($\Omega_M=0.3$, $\Omega_\Lambda=0.7$). The effect of adding
priors is practically negligible in all cases.  For each binary we
also give the bound $\lambda_{g, \rm unc}$ that could be obtained (in
principle) if all the binary parameters were known and not correlated
with the massive graviton term.}
\vskip 12pt
\begin{tabular}{@{}cccccccccccc@{}}
\hline
\hline
$\Delta t_c$ &$\Delta \phi_c$ &$\Delta {\cal M}/{\cal M}$ &$\Delta \eta/\eta$ 
&$\lambda_g$ &$\Delta \beta$ 
&$c^{{\cal M}\eta}$ &$c^{{\cal M}\beta_g}$ &$c^{\eta \beta_g}$ &$c^{{\cal M}\beta}$ &$c^{\eta \beta}$ &$c^{\beta_g \beta}$\\
(s)&&(\%)&(\%)&$(10^{15}\,{\rm km})$&&&&&&&\\
\hline
\multicolumn{12}{l}{$(10^7+10^7)M_\odot$, SNR=2063, $\lambda_{g, \rm unc}/(10^{15}\,{\rm km})=880$}\\
5.27   &0.0108  &0.00189   &0.0401  &-             &-     &0.930 &-         &-         &-         &-         &-        \\
14.1   &0.0448  &0.0155    &0.534   &69.4    &-     &-0.981 &-0.993 &0.997 &-         &-         &-        \\
79.6   &1.12    &0.0703    &49.2     &22.2    &3.08  &-0.978 &0.948 &-0.994 &0.975 &-1.000 &0.995\\
\hline
\multicolumn{12}{l}{$(10^7+10^6)M_\odot$, SNR=1204, $\lambda_{g, \rm unc}/(10^{15}\,{\rm km})=527$}\\
5.11   &0.0187  &0.00104   &0.0267  &-             &-     &0.934 &-         &-         &-         &-         &-        \\
13.8   &0.0749  &0.0104    &0.352   &39.5    &-     &-0.985 &-0.995 &0.997 &-         &-         &-        \\
83.4   &2.18    &0.0499    &25.6     &9.57    &1.52  &-0.981 &0.964 &-0.997 &0.978 &-1.000 &0.998\\
\hline
\multicolumn{12}{l}{$(10^6+10^6)M_\odot$, SNR=2143, $\lambda_{g, \rm unc}/(10^{15}\,{\rm km})=351$}\\
0.307  &0.00551 &0.000369  &0.0157  &-             &-     &0.887 &-         &-         &-         &-         &-        \\
0.675  &0.0175  &0.00244   &0.146   &46.3   &-     &-0.968 &-0.988 &0.994 &-         &-         &-        \\
3.02   &0.317   &0.00872   &12.2     &12.4   &0.790 &-0.963 &0.938 &-0.997 &0.960 &-1.000 &0.997\\
\hline
\multicolumn{12}{l}{$(10^6+10^5)M_\odot$, SNR=2378, $\lambda_{g, \rm unc}/(10^{15}\,{\rm km})=215$}\\
0.214  &0.00734 &0.000202  &0.00976 &-             &-     &0.871 &-         &-         &-         &-         &-        \\
0.481  &0.0238  &0.00161   &0.0865  &27.4    &-     &-0.973 &-0.992 &0.994 &-         &-         &-        \\
2.46   &0.521   &0.00587   &5.52    &6.02    &0.337 &-0.966 &0.948 &-0.998 &0.962 &-1.000 &0.999\\
\hline
\multicolumn{12}{l}{$(10^5+10^5)M_\odot$, SNR=1710, $\lambda_{g, \rm unc}/(10^{15}\,{\rm km})=139$}\\
0.0521 &0.00539 &0.000114  &0.0113  &-             &-     &0.799 &-         &-         &-         &-         &-        \\
0.0843 &0.0131  &0.000678  &0.0713  &23.2   &-     &-0.952 &-0.986 &0.987 &-         &-         &-        \\
0.300  &0.201   &0.00210   &6.58    &4.92   &0.436 &-0.950 &0.931 &-0.998 &0.947 &-1.000 &0.999\\
\hline
\multicolumn{12}{l}{$(10^5+10^4)M_\odot$, SNR=601, $\lambda_{g, \rm unc}/(10^{15}\,{\rm km})=77.4$}\\
0.122  &0.0165  &0.0000810 &0.0111  &-             &-     &0.732 &-         &-         &-         &-         &-        \\
0.203  &0.0458  &0.000712  &0.0804  &10.8   &-     &-0.973 &-0.994 &0.990 &-         &-         &-        \\
0.643  &0.663   &0.00219   &5.48    &1.76   &0.352 &-0.950 &0.937 &-0.999 &0.946 &-1.000 &1.000\\
\hline
\multicolumn{12}{l}{$(10^4+10^4)M_\odot$, SNR=252, $\lambda_{g, \rm unc}/(10^{15}\,{\rm km})=48.5$}\\
0.222  &0.0419  &0.0000822 &0.0308  &-             &-     &0.846 &-         &-         &-         &-         &-        \\
0.333  &0.120   &0.000597  &0.217   &5.80    &-     &-0.964 &-0.990 &0.990 &-         &-         &-        \\
0.758  &1.39    &0.00200   &29.7     &0.670 &2.06  &-0.957 &0.951 &-1.000 &0.955 &-1.000 &1.000\\
\hline
\hline
\end{tabular}
\label{MG2}
\end{table}

\begin{table}[hbt]
\centering
\caption{Average errors for a Monte Carlo simulation of $10^4$ NS-BH
binaries randomly located and oriented on the sky with mass of
$(1.4+1000) M_\odot$. We first consider general relativistic waveforms
(GR) and add spin-orbit (SO) and spin-spin (SS) couplings, fixing the
single-detector SNR $\rho_I=10$. Then we do the same including also a
Brans-Dicke (BD) term. In each case, the first line refers to the
errors obtained using only the first detector; the second line gives
the errors obtained using both detectors; the third line gives
pattern-averaged results from the relevant entries of Tables
~\ref{LISAnoBD} and \ref{LISAspin}.  }
\vskip 12pt
\begin{tabular}{@{}lcccccccc@{}}
\hline
\hline
Case
&$\Delta D_L/D_L$ &$\Delta {\cal M}/{\cal M}$ 
&$\Delta \eta/\eta$
&$\Delta \Omega_S$ &$\delta \theta_S$
&$\omega_{\rm BD}$ 
&$\Delta \beta$ &$\Delta \sigma$\\
&&(\%)&(\%)&($10^{-5}$~str)&(arcmin)&\\
\hline
\hline
GR                       & 0.782     &0.0000633 &0.103  &15.3  &42.5     &-     &-     &-          \\
                         & 0.376     &0.0000378 &0.0617 &5.95  &26.5     &-     &-     &-          \\
                         &-           &0.0000341  &0.0439  &-      &-     &-     &-     &-    \\
\hline		  						   	  	      	  				  
\hline								   	  	      	  						  
GR+SO                    & 0.797     &0.00178   &0.816  &35.3  &64.6     &-     &0.0179    &-          \\
                         & 0.374     &0.00111   &0.507  &13.8  &40.4     &-     &0.0111    &-          \\
                         &-           &0.000584   &0.297   &-      &-     &-     &0.00560    &-    \\
\hline		  						   	  	     	      				  
GR+SO+SS                 & 0.900     &0.00869   &6.96   &60.2  &84.3     &-     &0.143     &1.82  \\
                         & 0.420     &0.00562   &4.51   &23.9  &53.1     &-     &0.0930    &1.18  \\
                         &-           &0.00296    &2.65    &-      &-     &-     &0.0705     &0.793\\ 
\hline								   			     				  
BD                       & 0.764     &0.000789  &0.283  &40.8  &69.4     &62,561 &-               \\
                         & 0.359     &0.000488  &0.174  &15.8  &43.2     &96,719 &-               \\
                         &-           &0.000189   &0.116   &-      &-     &211,389&-               \\
\hline								   	     		     
BD+SO                    & 0.898     &0.0225    &4.87   &86.8  &101.3     &7,209  &0.157          \\
                         & 0.413     &0.0143    &3.10   &34.1  &63.5     &10,799 &0.100          \\
                         &-           &0.00764    &1.86    &-      &-     &21,257 &0.0557          \\
\hline
\hline
\end{tabular}
\label{BDaverage}
\end{table}

\begin{table}[hbt]
\centering
\caption{Average SNRs and errors for a Monte Carlo simulation of
$10^4$ BH-BH binaries randomly located and oriented in the sky with
mass $(10^6+10^6) M_\odot$. We fix $D_L=3$ Gpc and, where indicated,
include a massive-graviton (MG) term, spin-orbit (SO) and spin-spin
(SS) couplings. We also illustrate the deterioration in parameter
estimation when we assume that {\em LISA} is blind below some cutoff
frequency $f_{\rm low}$.  The default value for $f_{\rm low}$ is
$10^{-5}$ Hz.  In each case, the first (second) line refers to the
errors of a $(10^6+10^6) M_\odot$ binary using one (two) detectors. We
assume a cosmological model with $\Omega_M=0.3$ and
$\Omega_\Lambda=0.7$.}
\vskip 12pt
\begin{tabular}{@{}lcccccccccc@{}}
\hline
\hline
Case &SNR
&$\Delta D_L/D_L$ &$\Delta {\cal M}/{\cal M}$ 
&$\Delta \eta/\eta$
&$\Delta \Omega_S$
&$(\Delta z/z)$ &$(\Delta z/z)_{\rm best}$
&$\lambda_{g}$ &$\Delta \beta$ &$\Delta \sigma$\\
&&&(\%)&(\%)&($10^{-5}$ str)&&&($10^{15}$~km)&&\\
\hline
GR                           &1861 &0.0458 &0.000614 &0.0304  &59.9 &0.101   &0.0380  &-     &-       &-         \\
                             &2693 &0.0106 &0.000349 &0.0155  &9.76 &0.0873  &0.00880 &-     &-       &-         \\
\hline			     								 									     
GR+cutoff                    &1823 &0.341  &0.00966  &0.109   &5894  &0.306   &0.283   &-     &-       &-         \\
$f_{\rm low}=10^{-4}$~Hz     &2640 &0.0164 &0.00121  &0.0289  &48.9 &0.0881  &0.0136  &-     &-       &-         \\
\hline			     								 									     
GR+SO                        &1863 &0.0492 &0.00558  &1.93    &64.4 &0.103   &0.0408  &-     &0.107  &-         \\
                             &2696 &0.0107 &0.00295  &1.01    &10.2 &0.0873  &0.00891 &-     &0.0558 &-         \\
\hline			     								 									     
GR+SO+SS                     &1862 &0.0504 &0.0152   &8.05    &67.1 &0.104   &0.0418  &-     &0.374  &0.248    \\
                             &2695 &0.0109 &0.00852  &4.52    &10.4 &0.0873  &0.00902 &-     &0.209  &0.139    \\
\hline			     								 									     
MG                           &1861 &0.0486 &0.00447  &0.273   &64.1 &0.103   &0.0403  &37.4  &-       &-         \\
                             &2693 &0.0107 &0.00237  &0.145   &10.2 &0.0873  &0.00889 &49.5  &-       &-         \\
\hline			     								 									     
MG+cutoff                    &1787 &0.442  &0.0443   &1.18    &13290 &0.386   &0.367   &16.2  &-       &-         \\
$f_{\rm low}=10^{-4}$~Hz     &2592 &0.0159 &0.00921  &0.362   &51.2 &0.0878  &0.0132  &29.5  &-       &-         \\
\hline			     								 									     
MG+cutoff                        &1859 &0.0511 &0.00522  &0.301   &71.8 &0.104   &0.0424  &35.4  &-       &-         \\
$f_{\rm low}=5\times 10^{-5}$~Hz &2691 &0.0108 &0.00274  &0.158   &10.6 &0.0873  &0.00895 &46.7  &-       &-         \\
\hline			     								 									     
MG+SO                        &1861 &0.0495 &0.0152   &21.5    &67.2 &0.103   &0.0411  &10.6  &1.39   &-         \\
                             &2693 &0.0108 &0.00852  &12.1    &10.4 &0.0873  &0.00896 &13.3  &0.780  &-         \\
\hline
\hline
\end{tabular}
\label{MGaverage}
\end{table}

\clearpage

\begin{figure*}
\epsfig{file=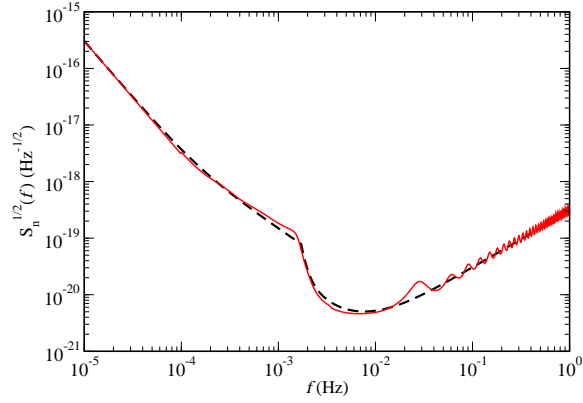,width=0.75\sizeonefig,angle=-90}
\caption{Analytic approximation to the \lisa~ root noise spectral
density curve used in this paper and in Ref.~\cite{BC} (dashed line)
and the curve produced using the {\em LISA} Sensitivity Curve
Generator~\cite{SCG} (solid line).  The SCG curve has been multiplied
by a factor of $\sqrt{3/20}$ to obtain an {\em effective} non-sky
averaged noise spectral density (see Sec. \ref{noisecurves}).  The SCG
noise curve does not include the extragalactic white dwarf confusion
noise while the analytical approximation curve does.
\label{noise}}
\end{figure*}

\begin{figure*}
\begin{center}
\begin{tabular}{cc}
\epsfig{file=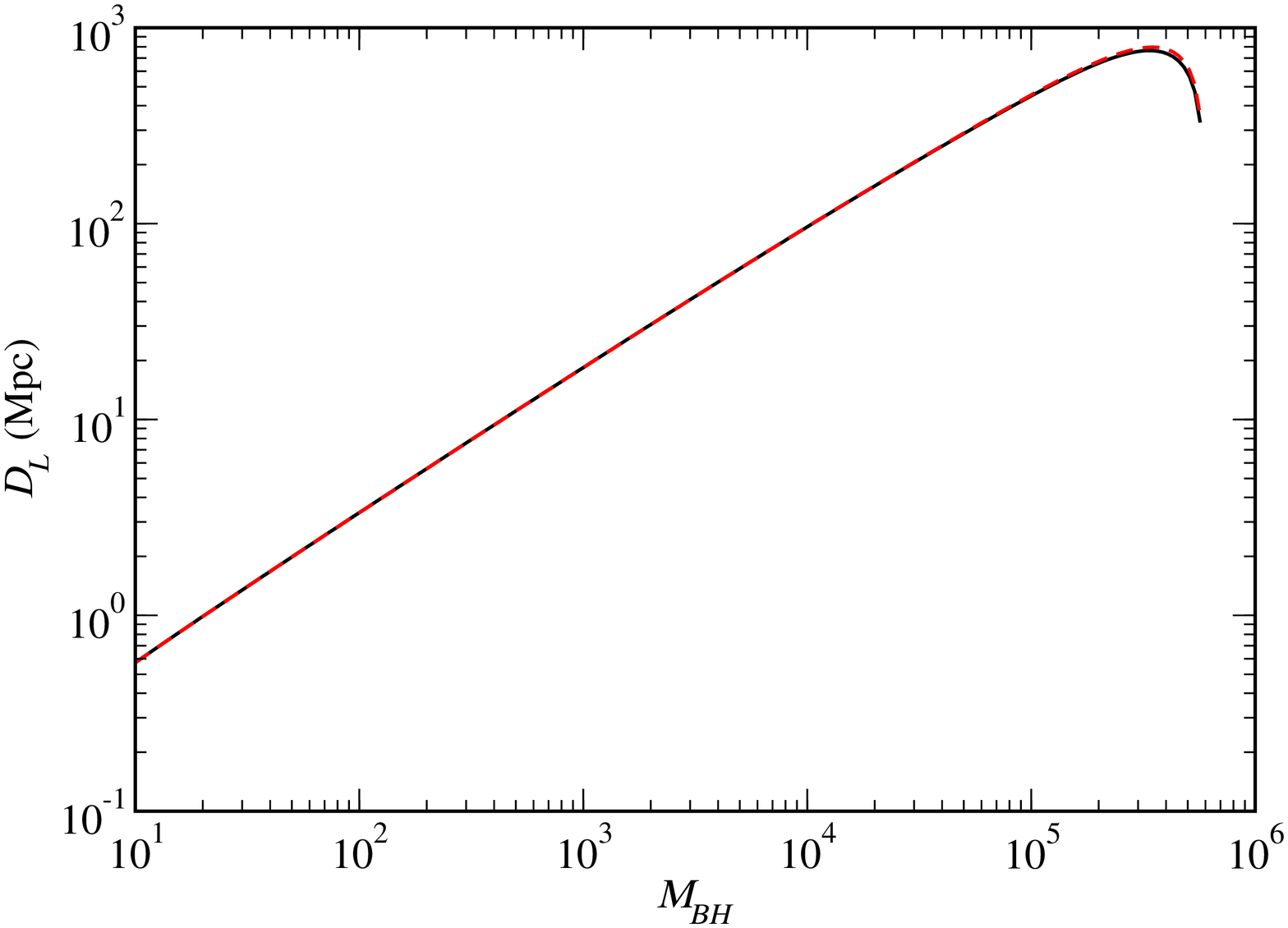,width=0.75\sizeonefig,angle=-90}&
\epsfig{file=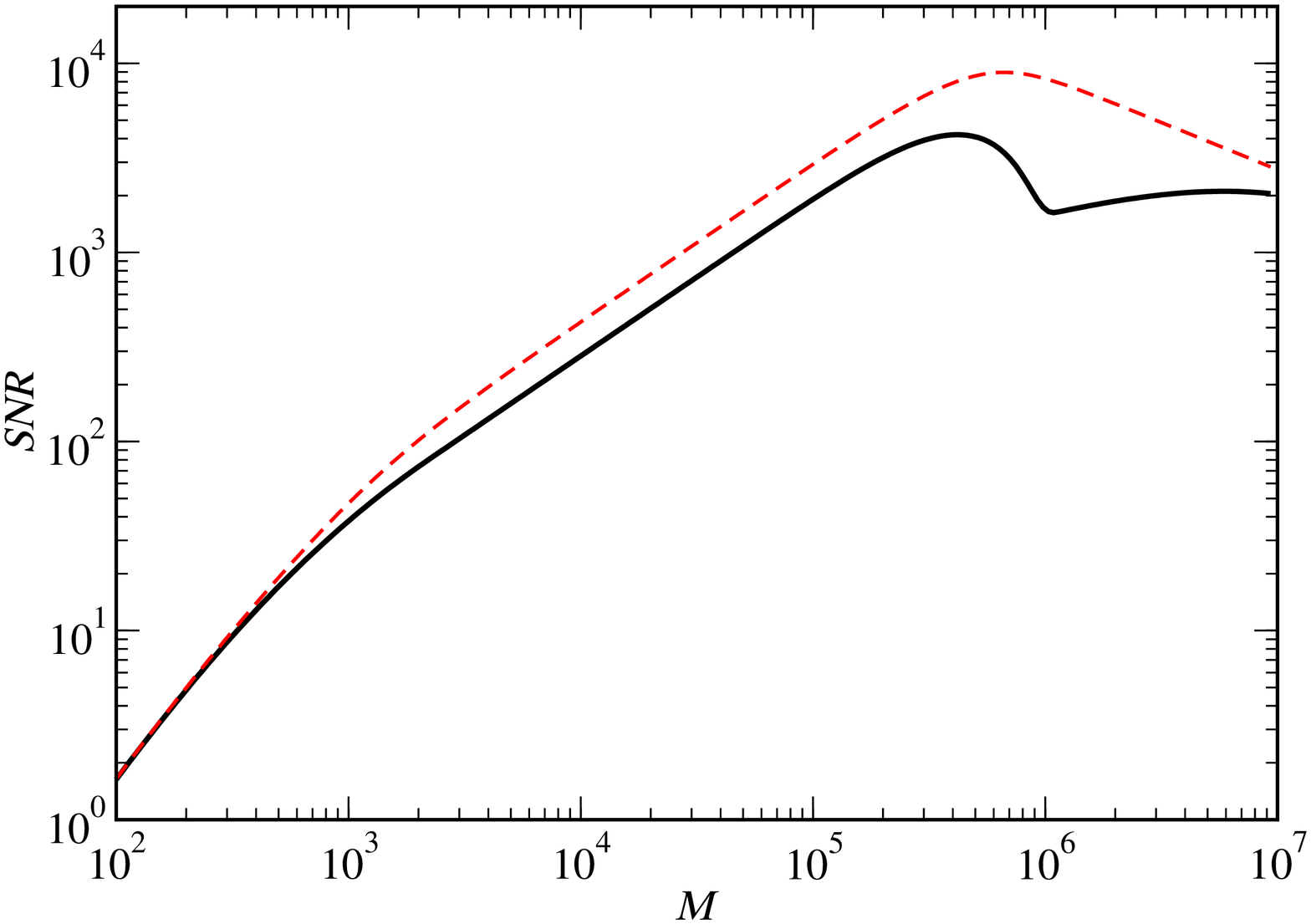,width=0.75\sizeonefig,angle=-90}
\end{tabular}
\caption{Left: luminosity distances $D_L$ of NS-BH binaries observed
with SNR=10 as a function of the black hole mass.  We assume that the
NS mass $M_{\rm NS}=1.4 M_\odot$.  Right: SNR for equal mass BH-BH
binaries at $D_L=3$ Gpc as a function of the total mass.  Solid lines
refer to the {\em LISA} noise curve (\ref{Shtot}) used in this paper;
dashed lines refer to the same noise curve without including the
white-dwarf confusion noise.  The ``bump'' in the noise curve due to
white-dwarf confusion noise is responsible for the dip in the SNR for
MBH binaries of masses $\sim 10^6 M_\odot$.
\label{SNR}}
\end{center}
\end{figure*}

\begin{figure*}
\begin{center}
\begin{tabular}{cc}
\epsfig{file=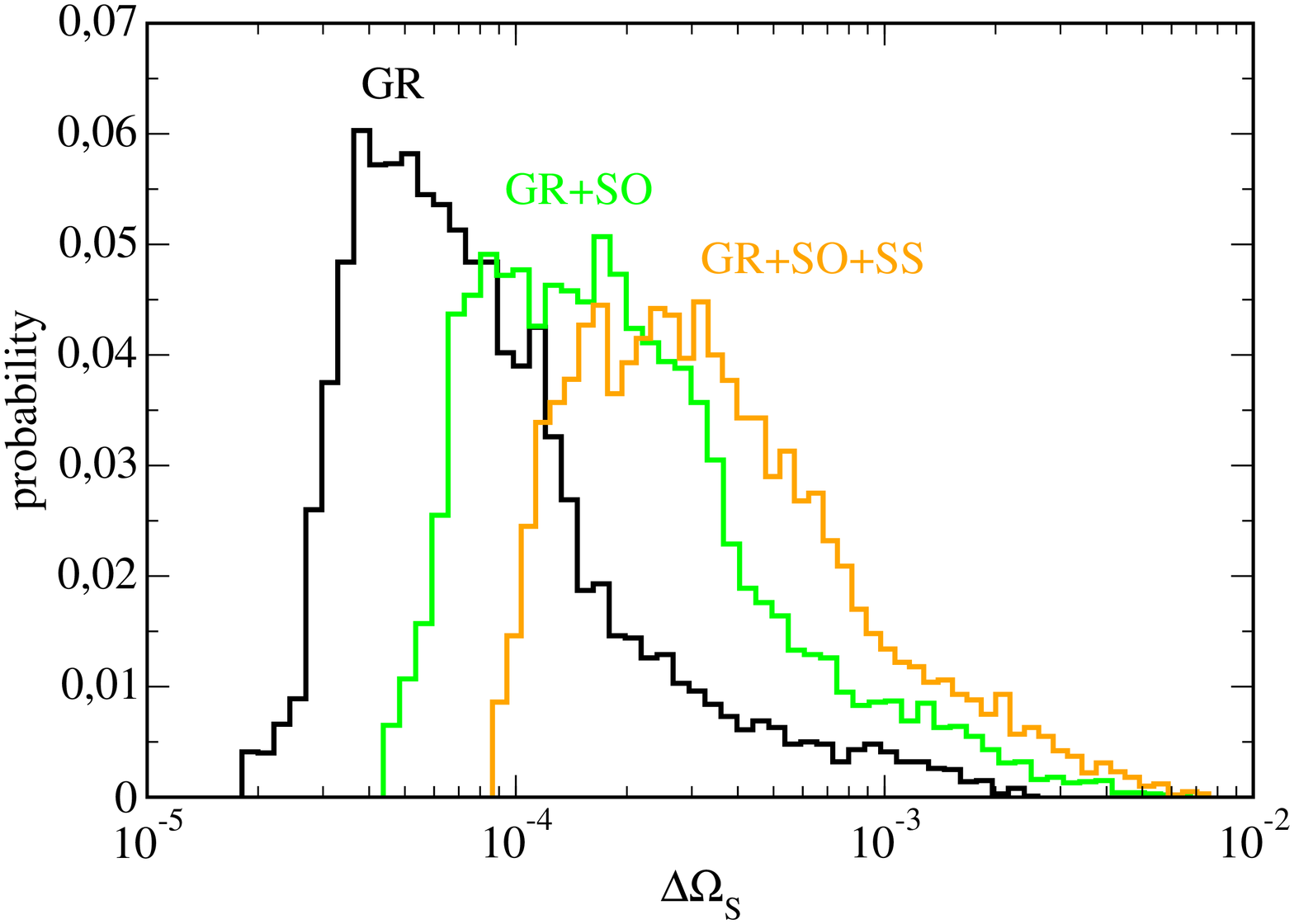,width=0.75\sizeonefig,angle=-90} &
\epsfig{file=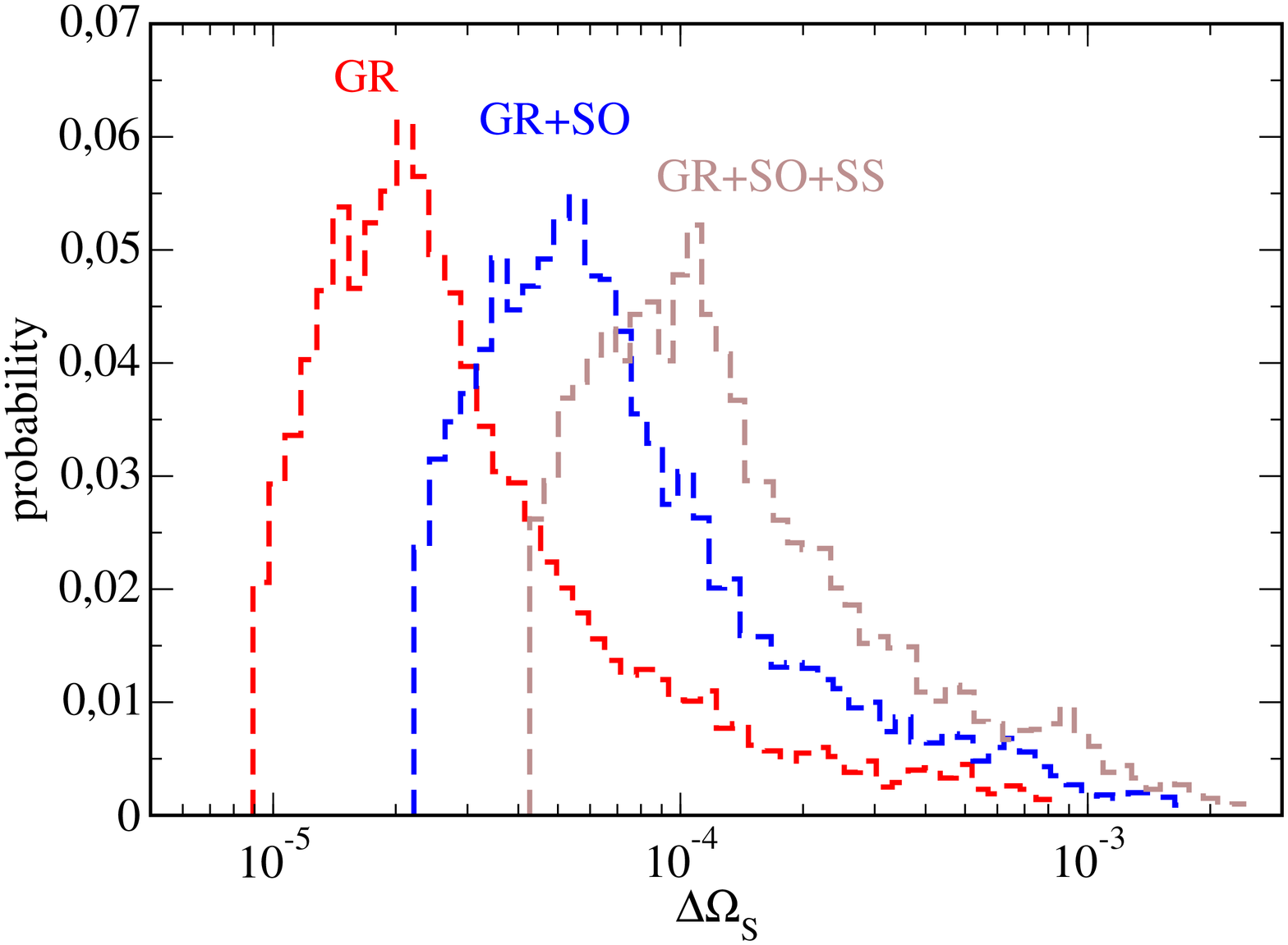,width=0.75\sizeonefig,angle=-90} \\
\end{tabular}
\epsfig{file=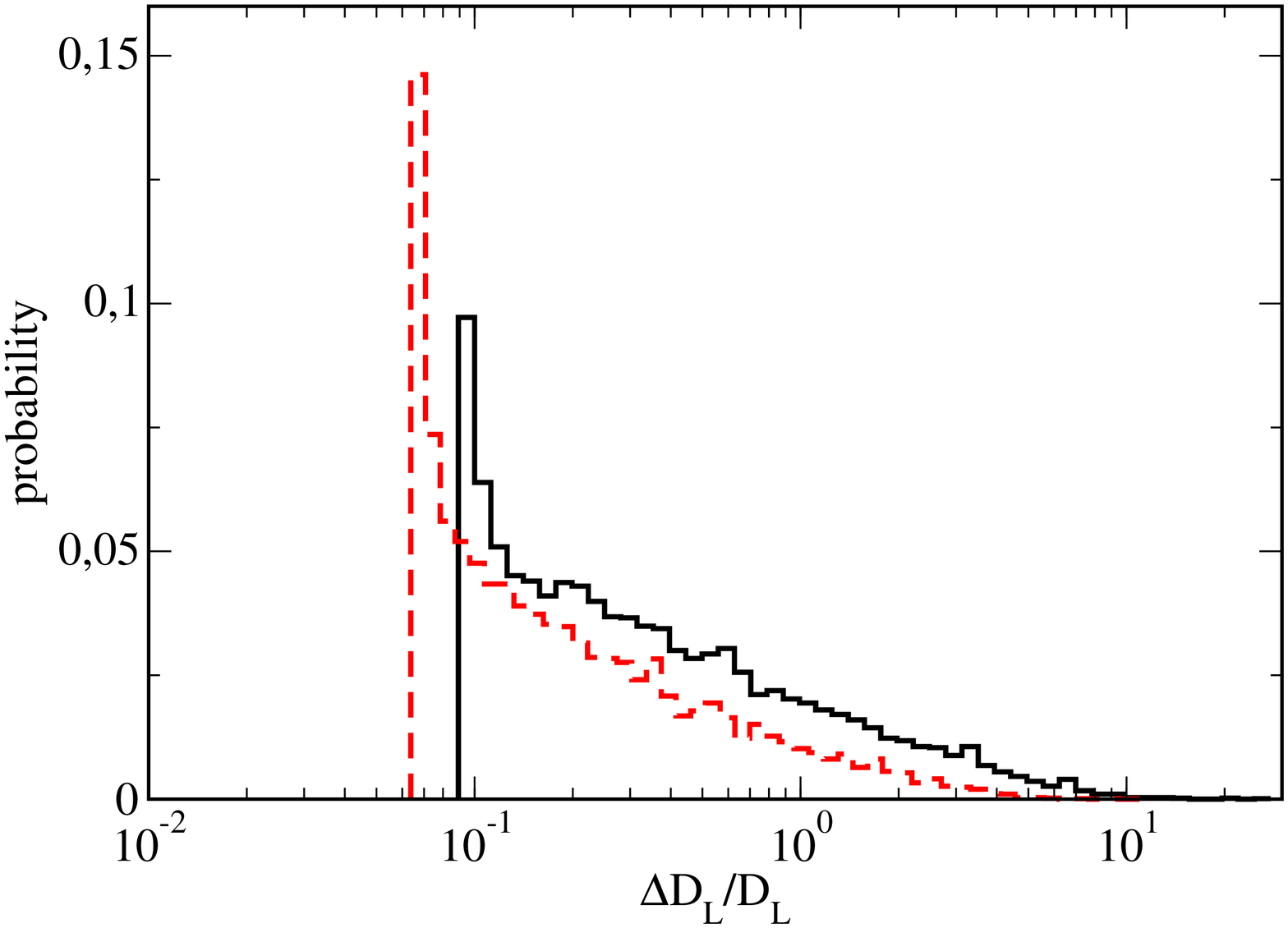,width=0.75\sizeonefig,angle=-90}
\caption{Monte Carlo simulation of $10^4$ binaries with observed total
mass $(1.4+10^3)M_\odot$ in general relativity, with single-detector
SNR=10, $\Omega_\Lambda=0.7$, $\Omega_M=0.3$.  Top: probability
distributions of the angular resolution $\Delta \Omega_S$ in steradians for one
detector (left) and two detectors (right).  In each figure, from left
to right, the histograms refer to no spins, SO included, and SO and SS
included.  Bottom: probability distributions of $\Delta D_L/D_L$ for
one detector (solid) and two detectors (dashed); $\Delta D_L/D_L$ is
essentially unaffected by the inclusion of spins, so we only show
histograms without the spin terms.
\label{histoGR3}}
\end{center}
\end{figure*}

\begin{figure*}
\begin{center}
\begin{tabular}{cc}
\epsfig{file=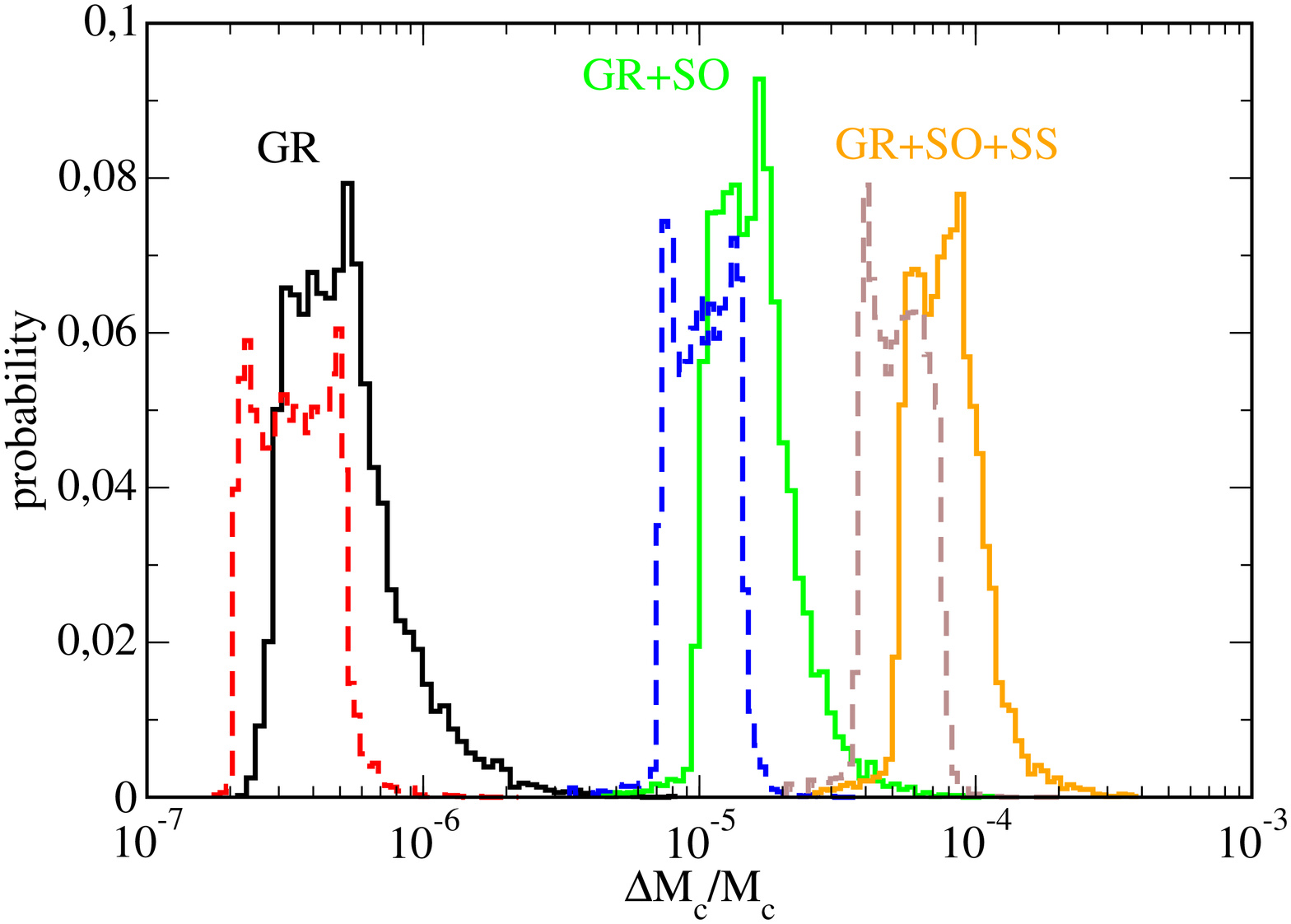,width=0.75\sizeonefig,angle=-90} &
\epsfig{file=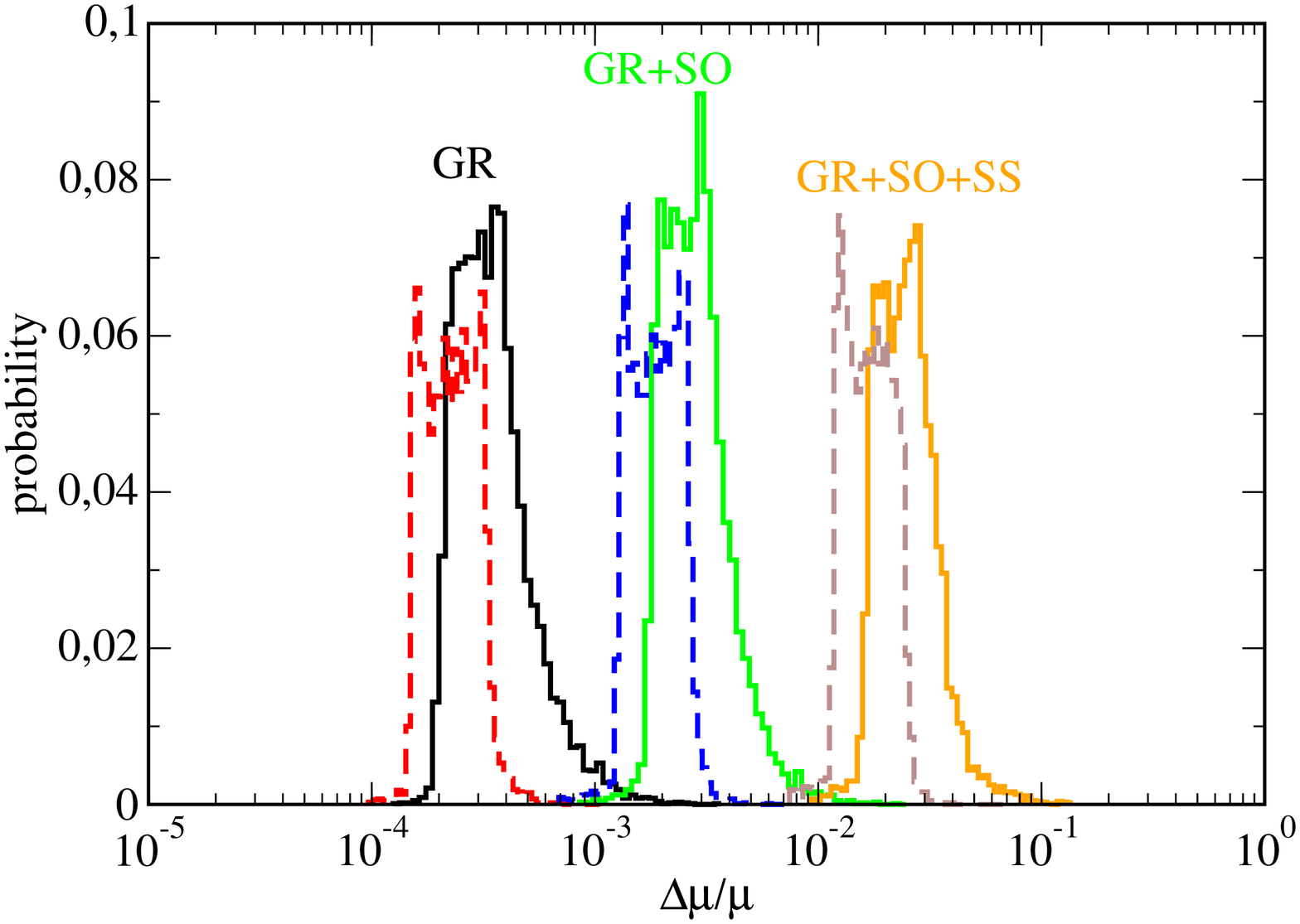,width=0.75\sizeonefig,angle=-90} \\
\epsfig{file=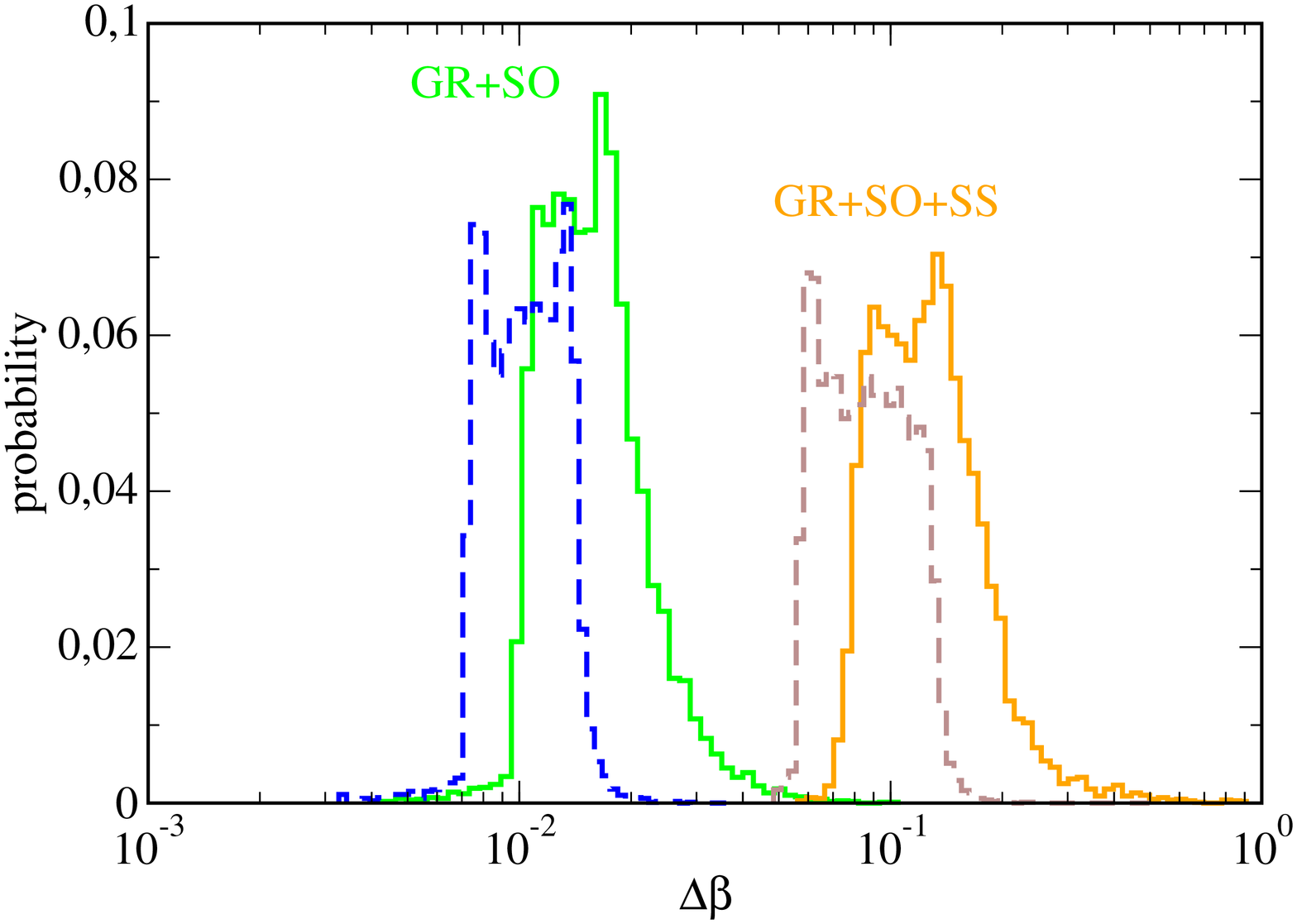,width=0.75\sizeonefig,angle=-90} &
\epsfig{file=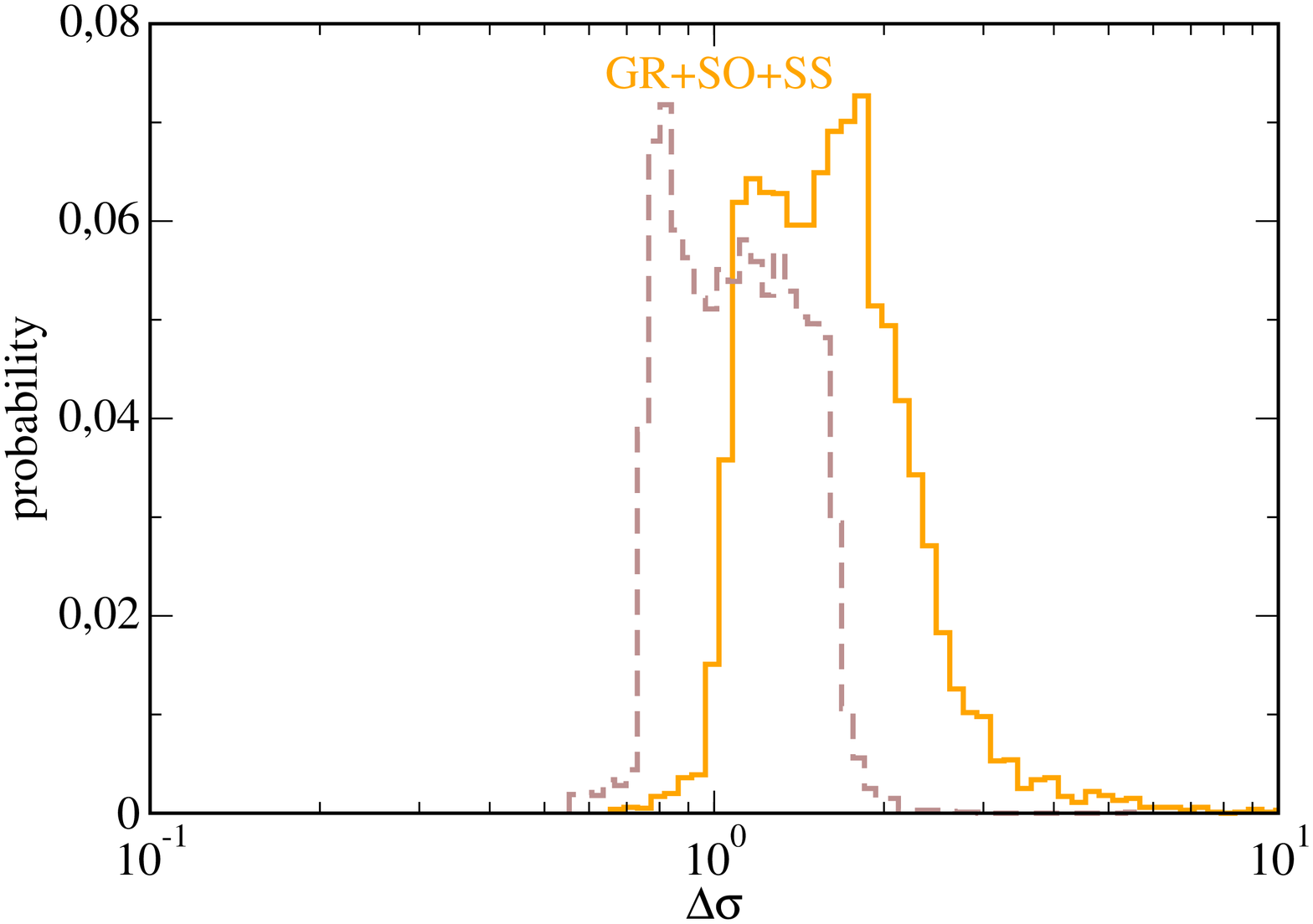,width=0.75\sizeonefig,angle=-90} \\
\end{tabular}
\epsfig{file=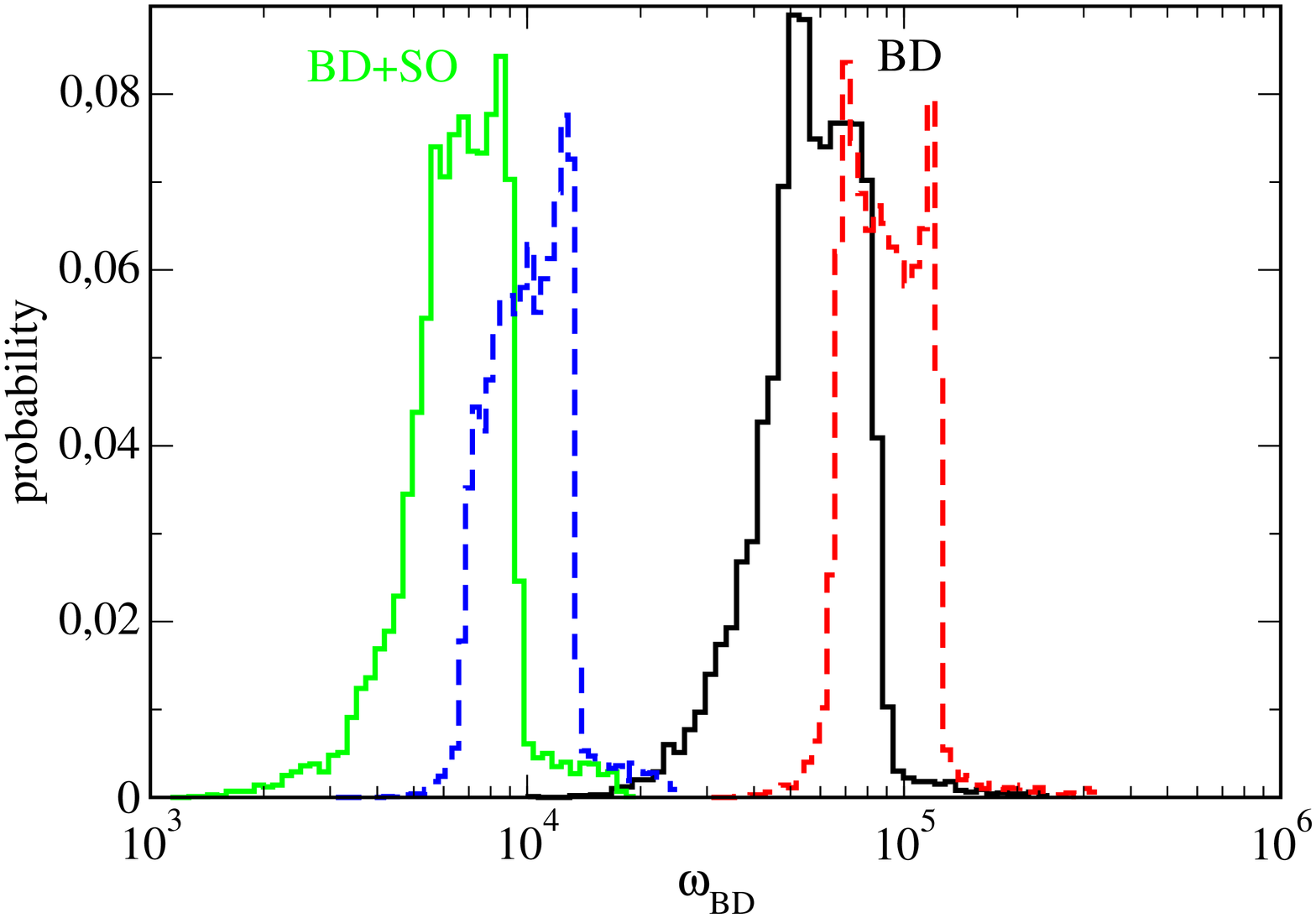,width=0.75\sizeonefig,angle=-90} 
\caption{Monte Carlo simulation of $10^4$ binaries with observed total
mass $(1.4+10^3)M_\odot$ in general relativity, with single-detector
SNR=10, $\Omega_\Lambda=0.7$, $\Omega_M=0.3$.  Top four panels:
probability distribution of the errors on the chirp mass $\Delta {\cal
M}/{\cal M}$, the reduced mass $\Delta \mu/\mu$, the SO parameter
$\Delta \beta$ and the SS parameter $\Delta \sigma$.  Bottom panel:
bound on $\omega_{\rm BD}$ when a Brans-Dicke term is included.  Solid
(dashed) lines refer to one (two) detector(s).
\label{histoGR4}}
\end{center}
\end{figure*}

\begin{figure*}
\begin{center}
\begin{tabular}{cc}
\epsfig{file=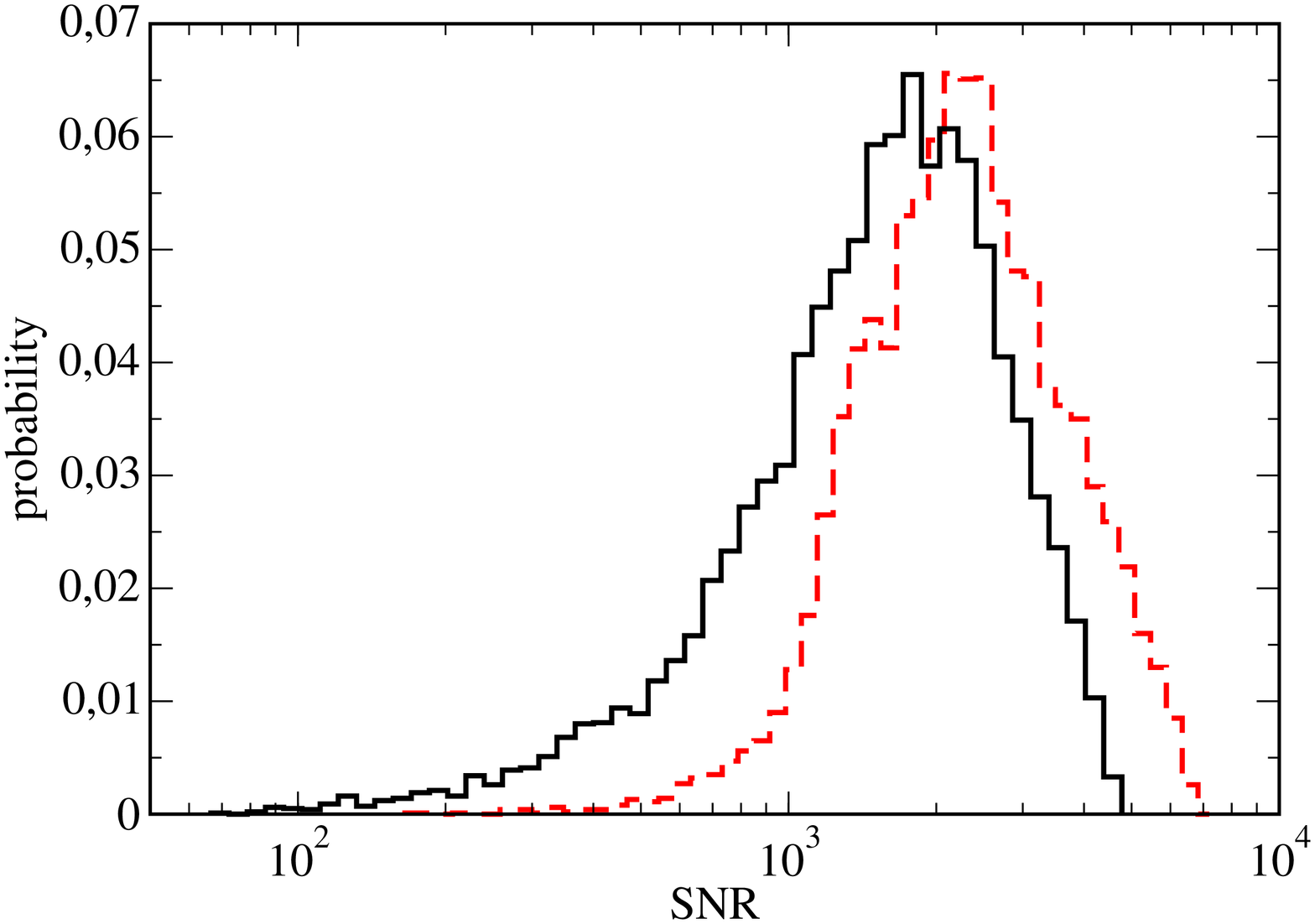,width=0.75\sizeonefig,angle=-90} &
\epsfig{file=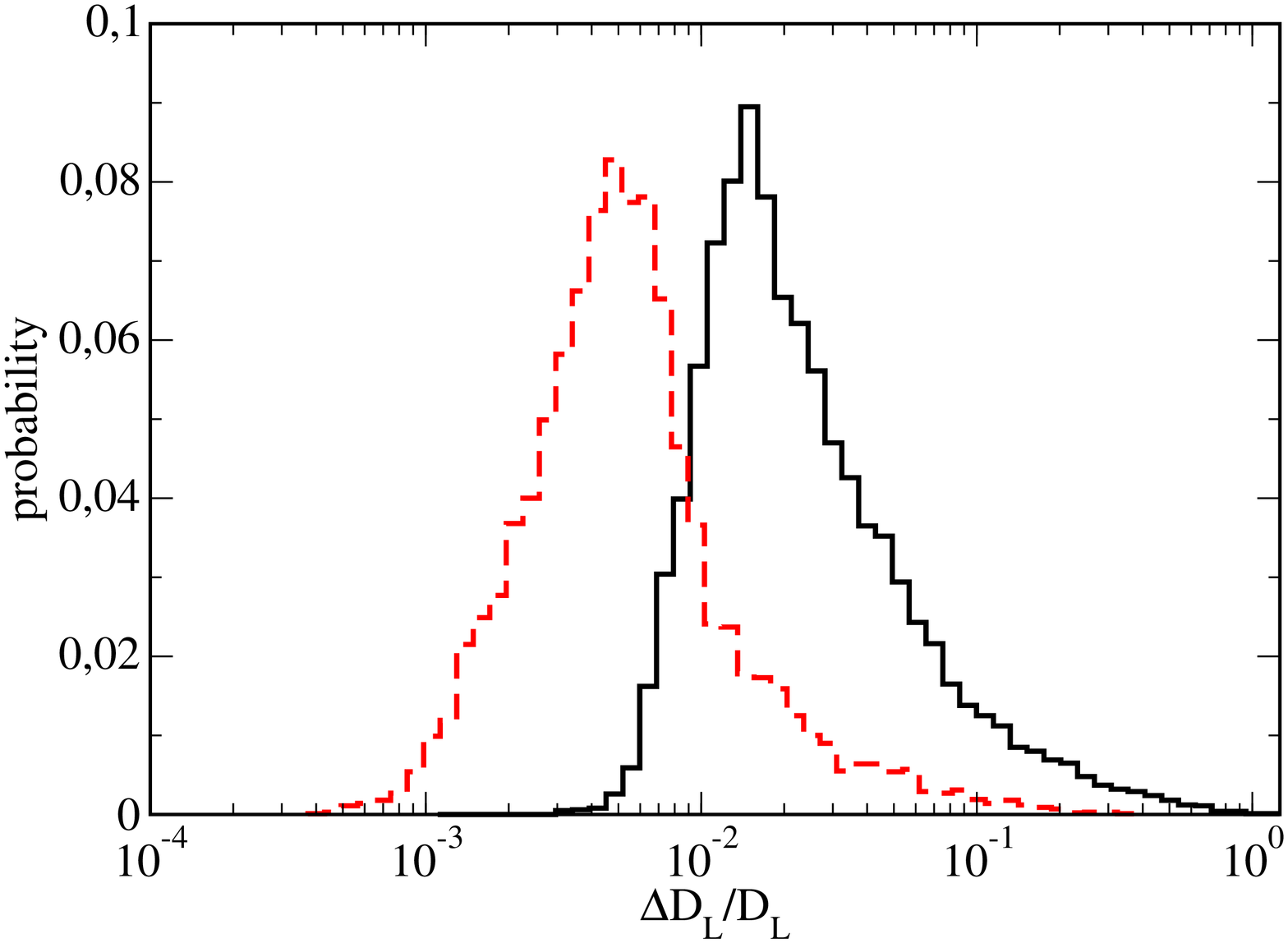,width=0.75\sizeonefig,angle=-90} \\
\epsfig{file=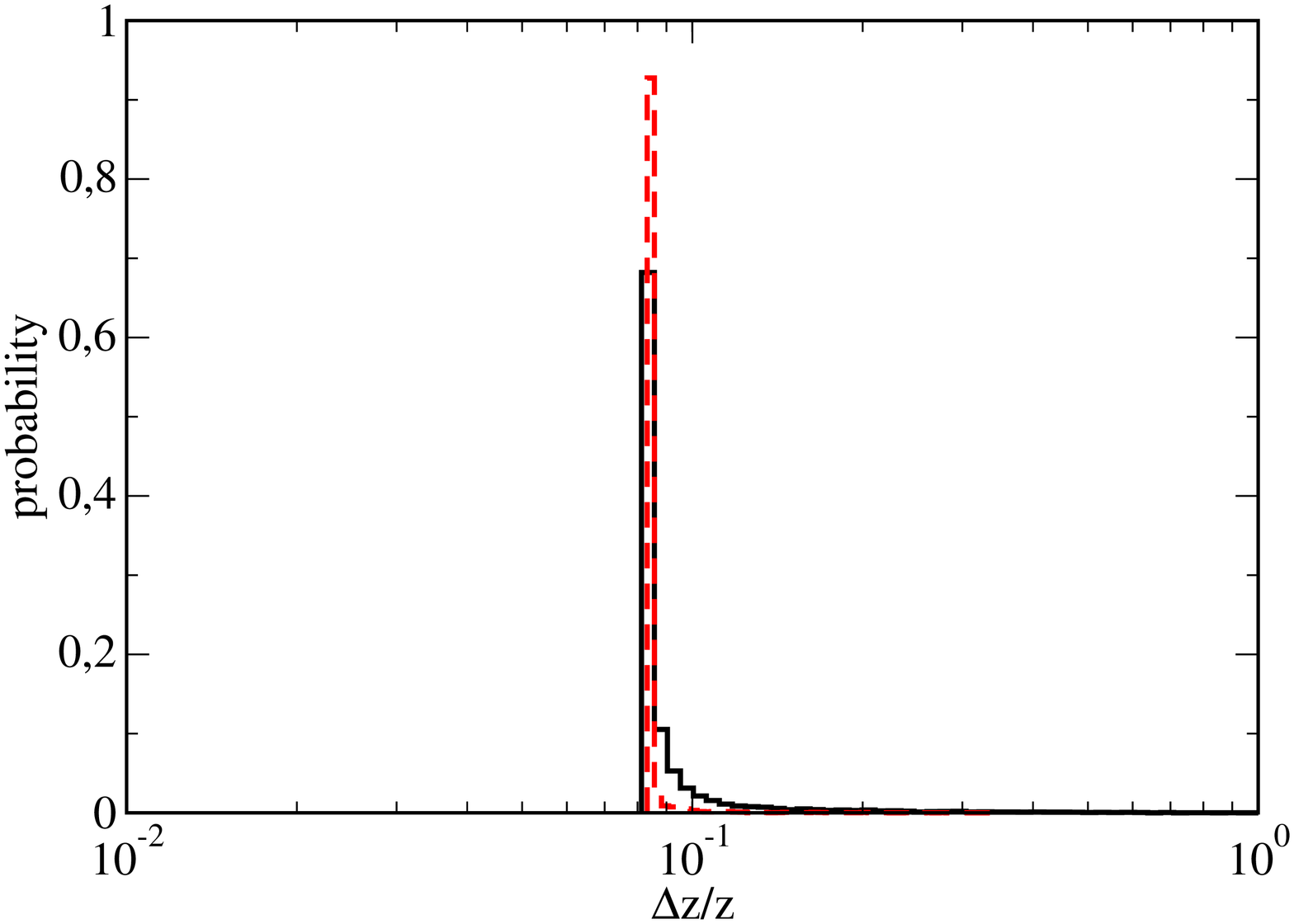,width=0.75\sizeonefig,angle=-90} &
\epsfig{file=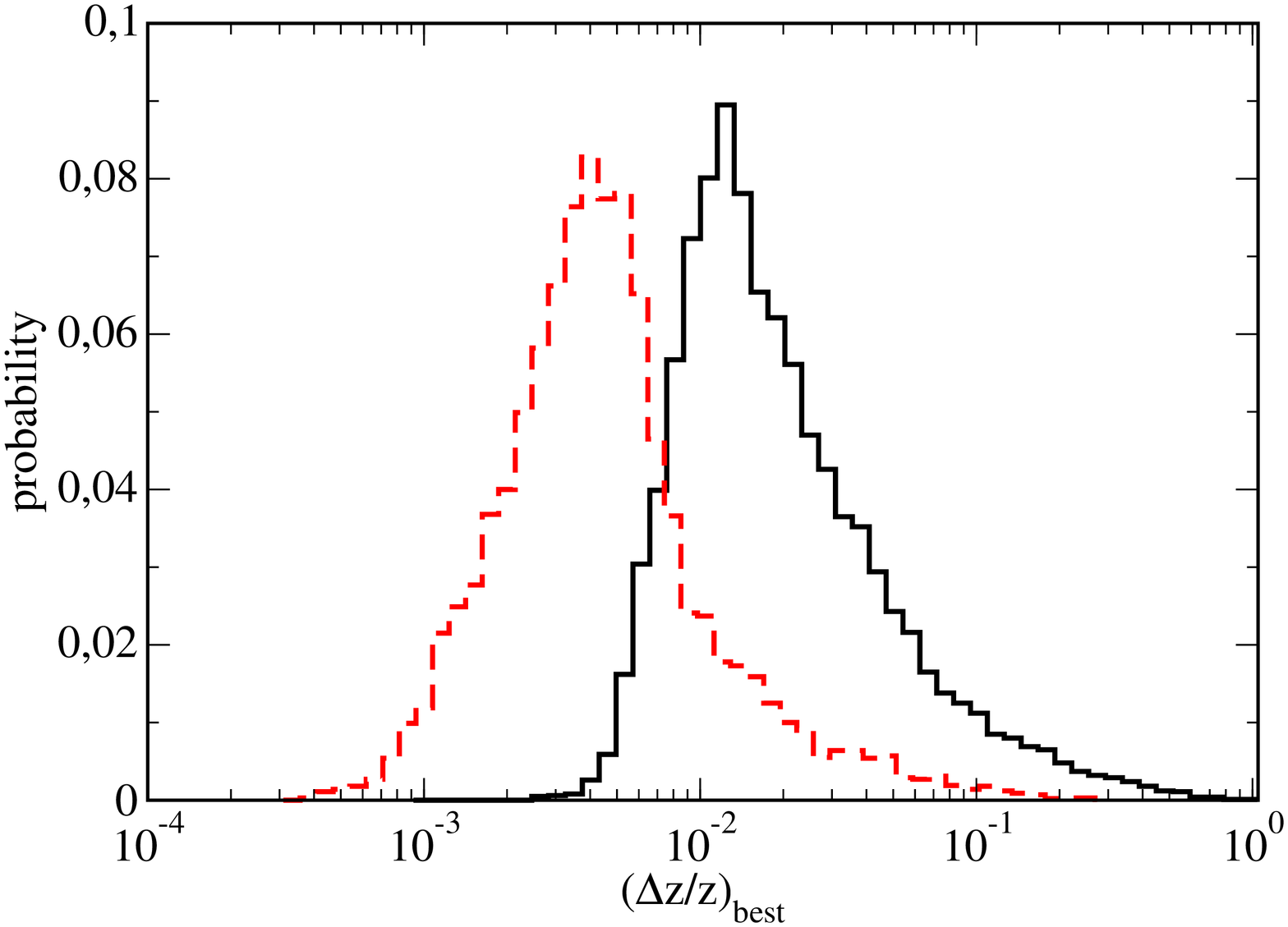,width=0.75\sizeonefig,angle=-90}\\
\end{tabular}
\epsfig{file=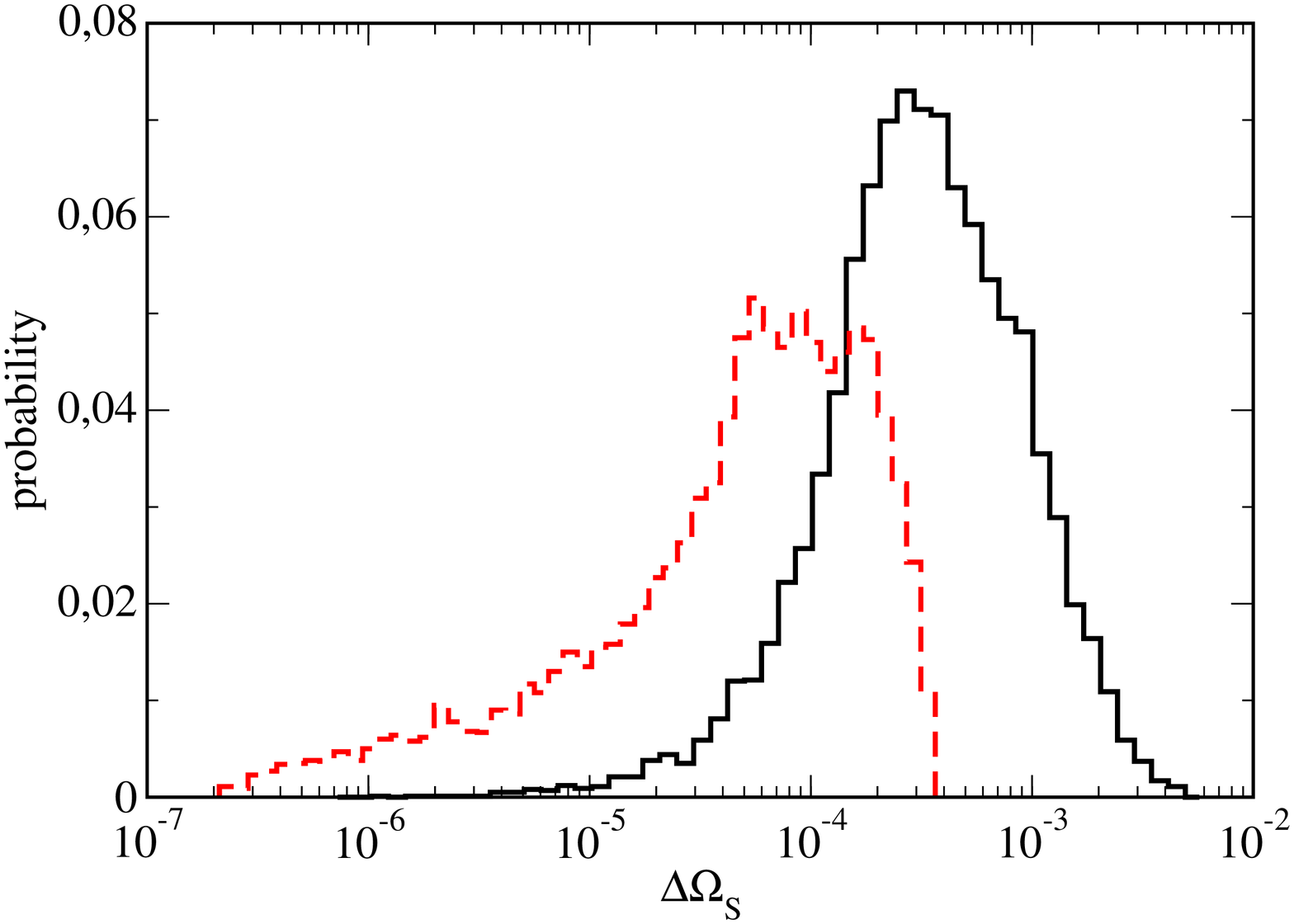,width=0.75\sizeonefig,angle=-90} 
\caption{Monte Carlo simulation of $10^4$ binaries with total mass
$(10^6+10^6)M_\odot$ in general relativity, with $D_L=3$~Gpc,
$\Omega_\Lambda=0.7$, $\Omega_M=0.3$, with no spins.  Panels show
probability distributions of the SNR, the distance determination error
$\Delta D_L/D_L$, the redshift errors $\Delta {z}/{z}$ and $(\Delta
{z}/{z})_{\rm best}$ and the angular resolution $\Delta \Omega_S$ in
steradians. Solid (dashed) lines refer to one (two) detector(s).
\label{histoGR1}}
\end{center}
\end{figure*}

\begin{figure*}
\begin{center}
\begin{tabular}{cc}
\epsfig{file=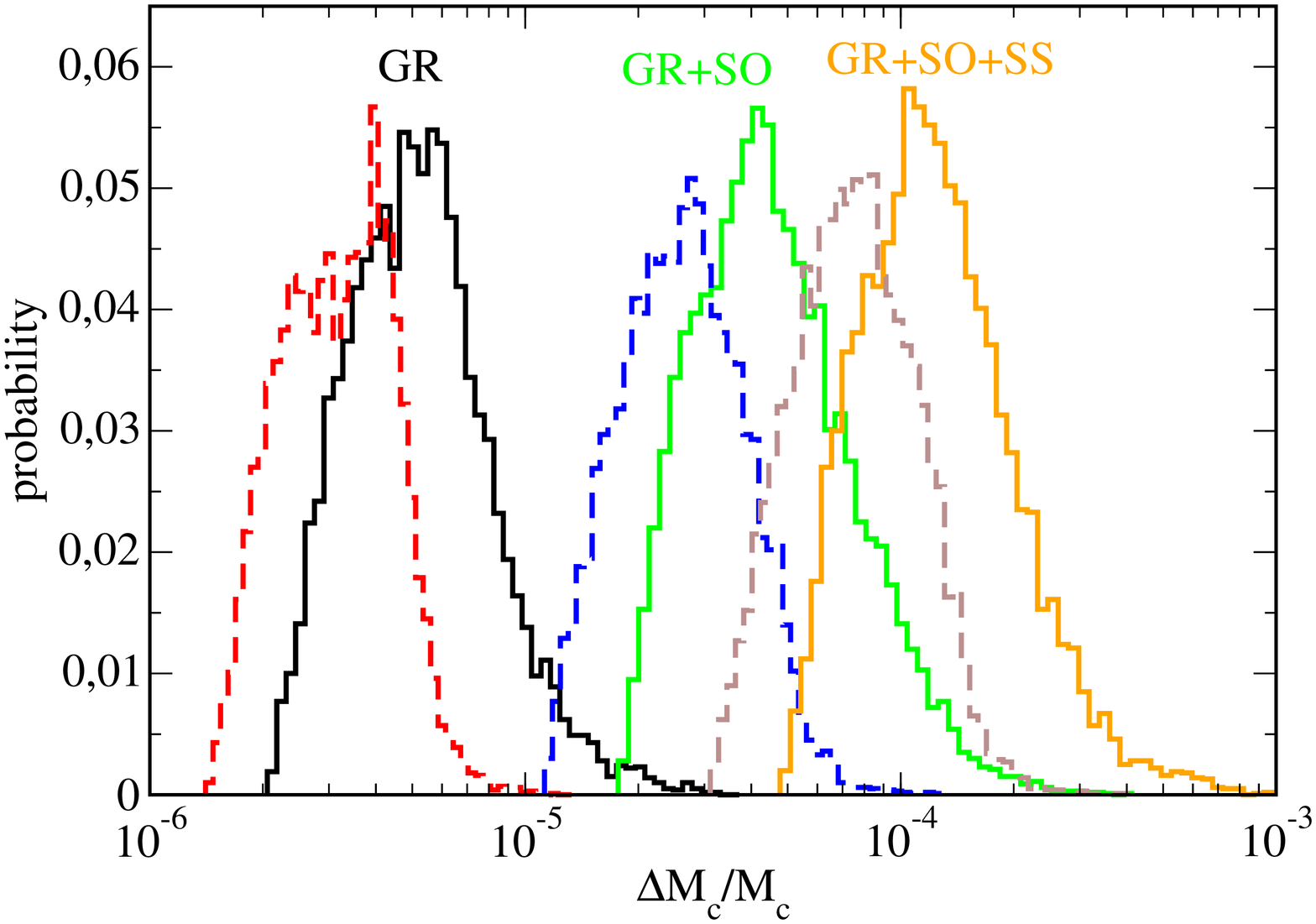,width=0.75\sizeonefig,angle=-90} &
\epsfig{file=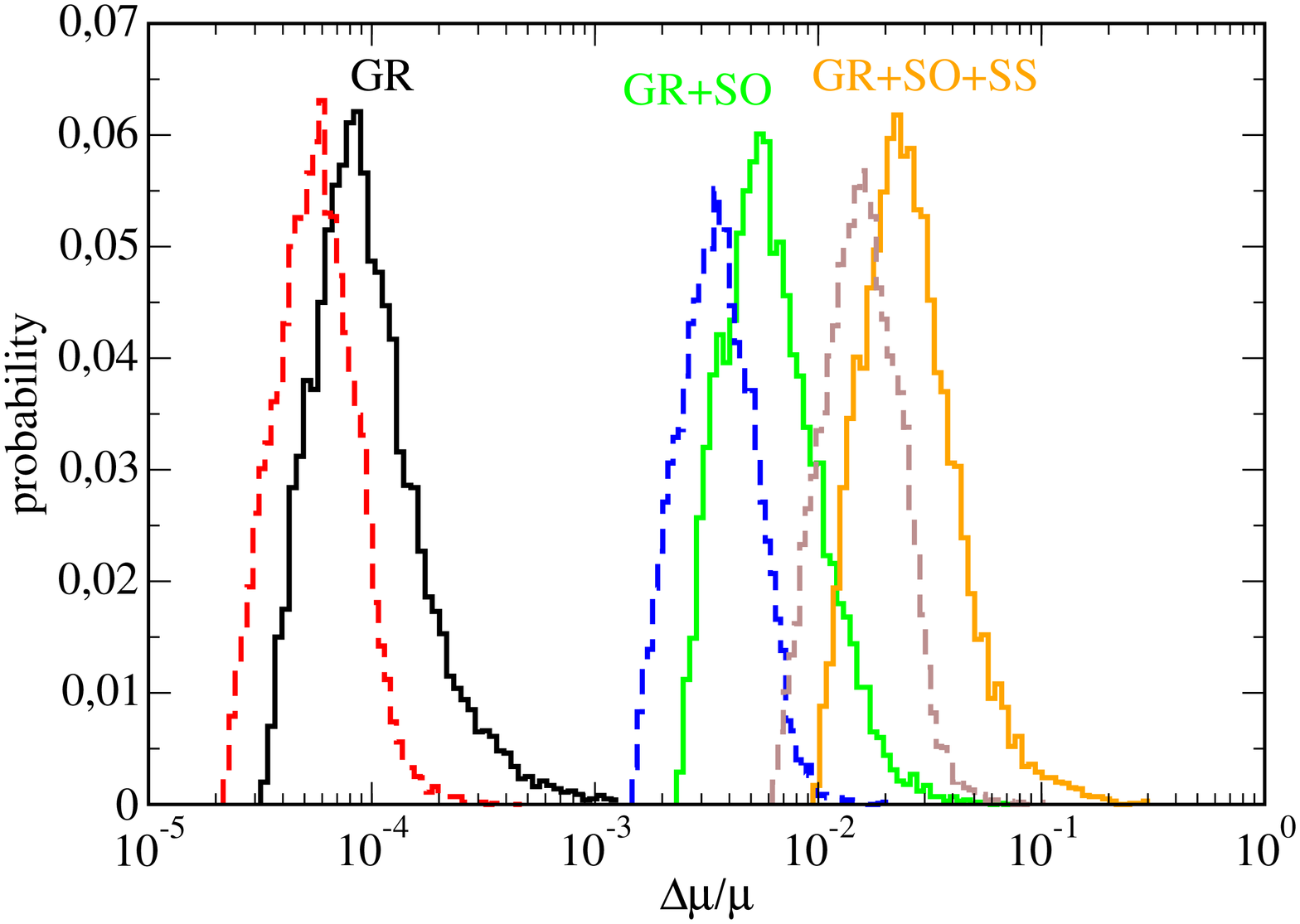,width=0.75\sizeonefig,angle=-90} \\
\epsfig{file=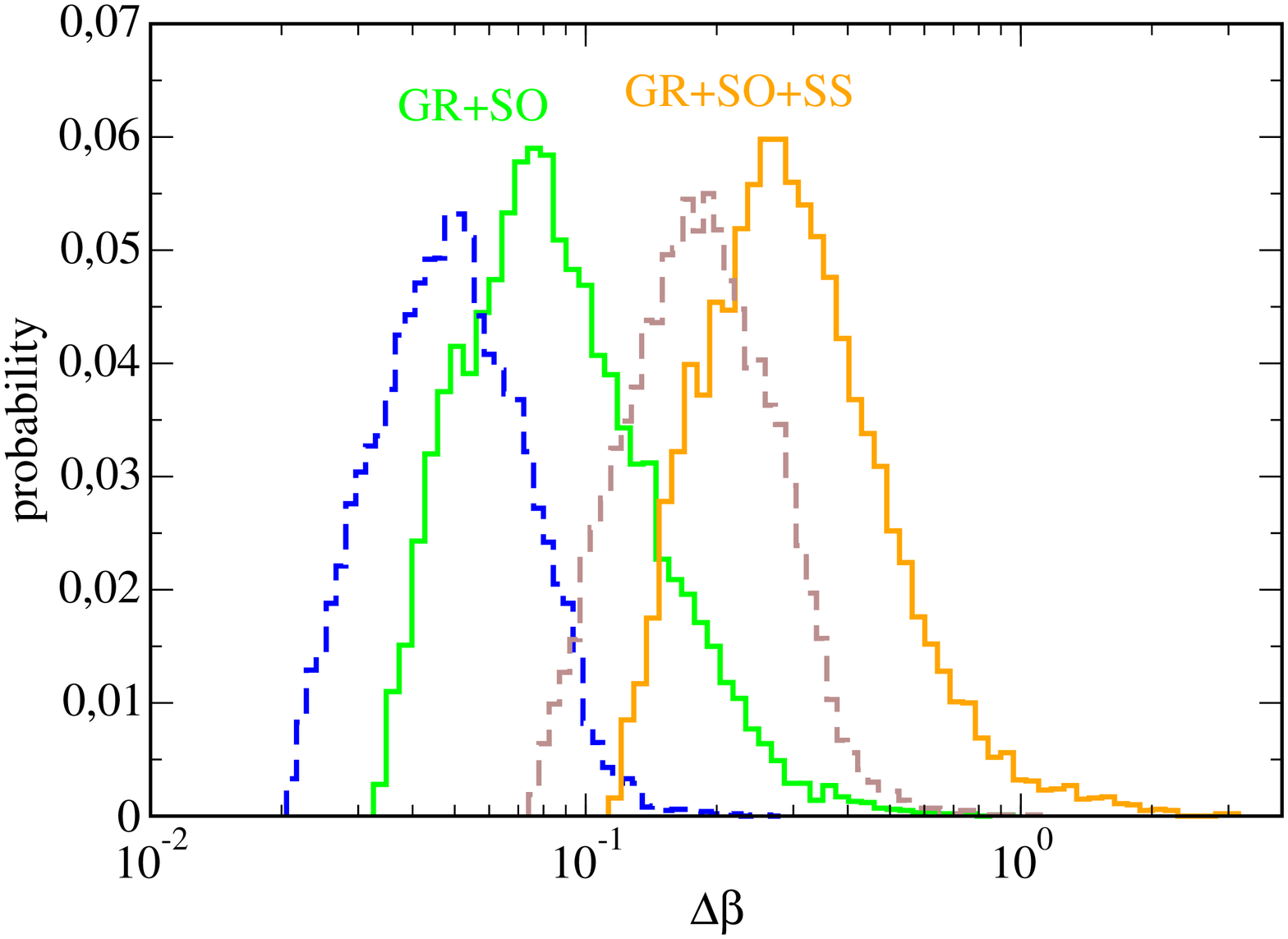,width=0.75\sizeonefig,angle=-90} &
\epsfig{file=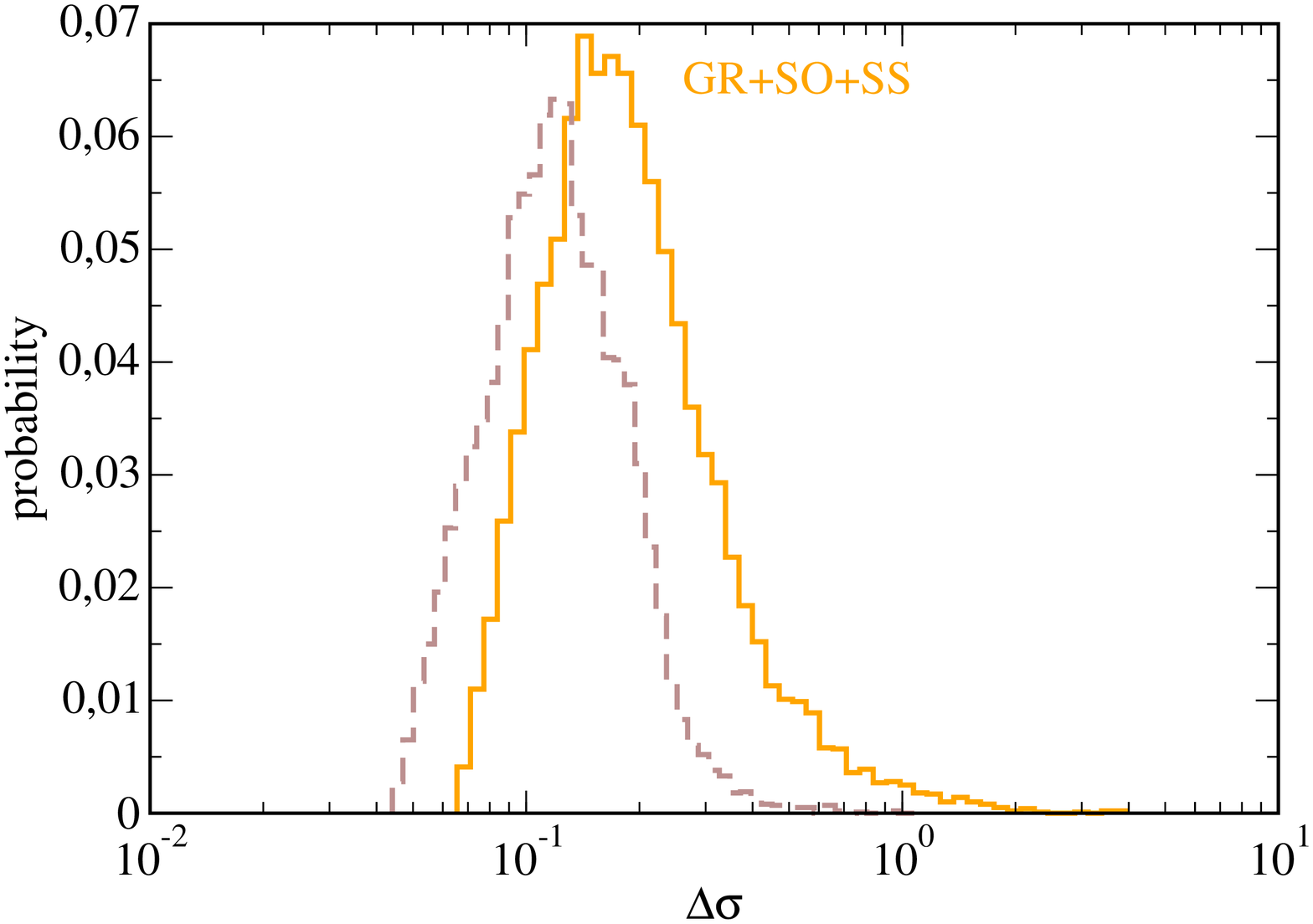,width=0.75\sizeonefig,angle=-90} \\
\end{tabular}
\epsfig{file=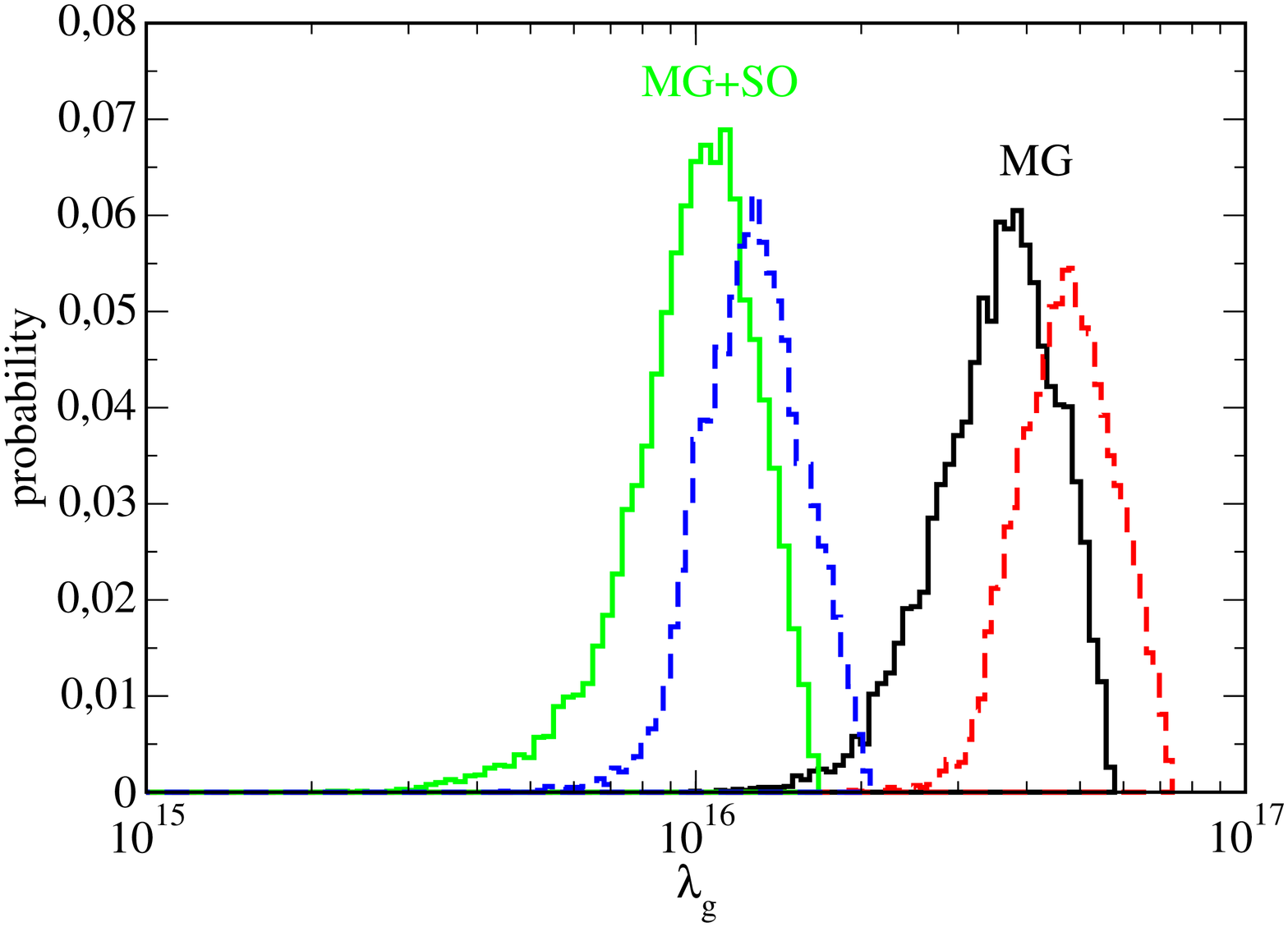,width=0.75\sizeonefig,angle=-90}
\caption{Monte Carlo simulation of $10^4$ binaries with observed total
mass $(10^6+10^6)M_\odot$ in general relativity, with $D_L=3$~Gpc,
$\Omega_\Lambda=0.7$, $\Omega_M=0.3$.  Top four panels: probability
distributions of the error on the chirp mass $\Delta {\cal M}/{\cal
M}$, the reduced mass $\Delta {\mu}/{\mu}$, the SO parameter $\Delta
\beta$, the SS parameter $\Delta \sigma$.  Bottom panel: bound on the
graviton Compton wavelength $\lambda_g$ (in km), when a massive
graviton term is included.  Solid (dashed) lines refer to one (two)
detector(s).
\label{histoGR2}}
\end{center}
\end{figure*}

\begin{figure*}
\begin{center}
\begin{tabular}{cc}
\epsfig{file=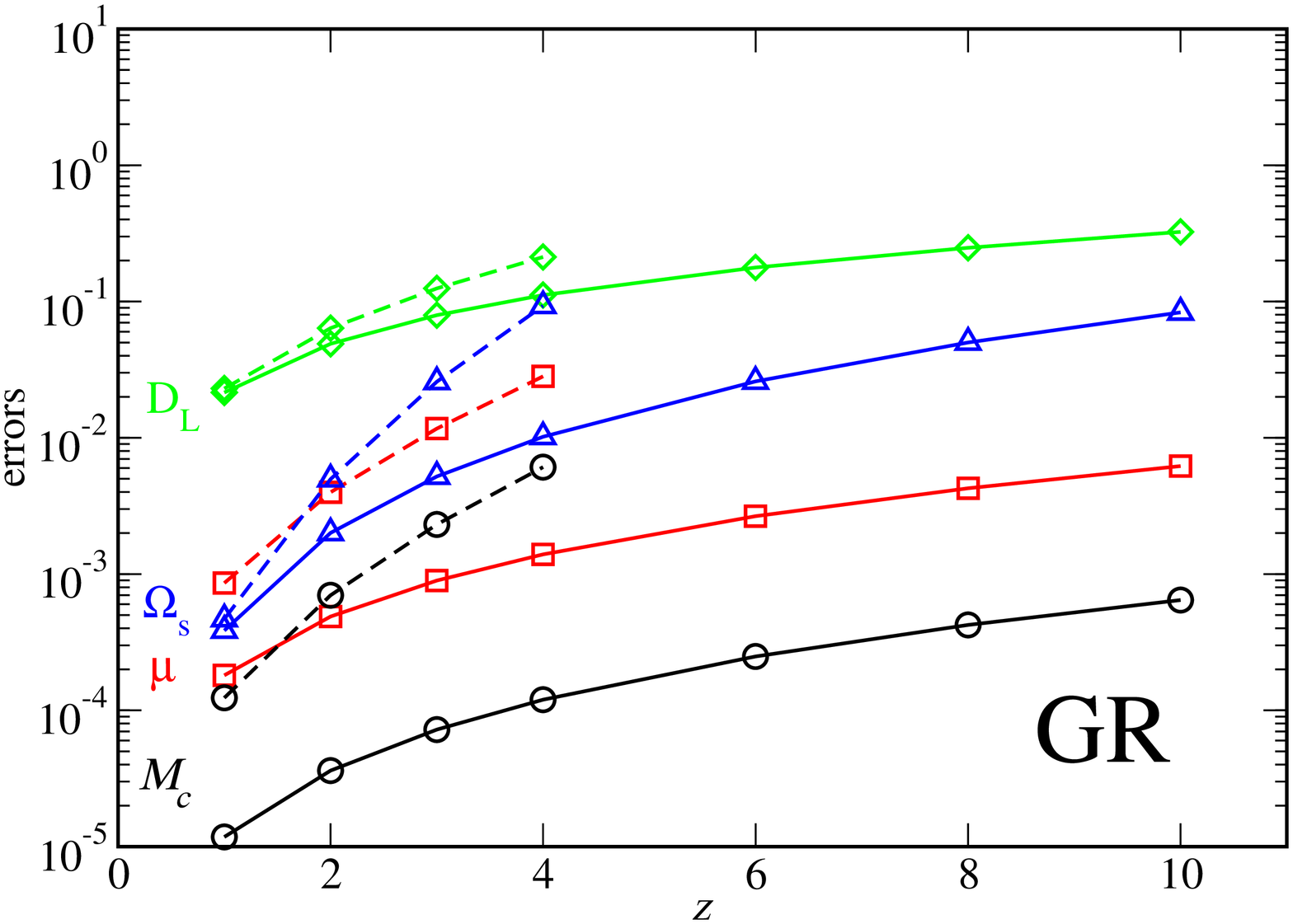,width=0.75\sizeonefig,angle=-90} &
\epsfig{file=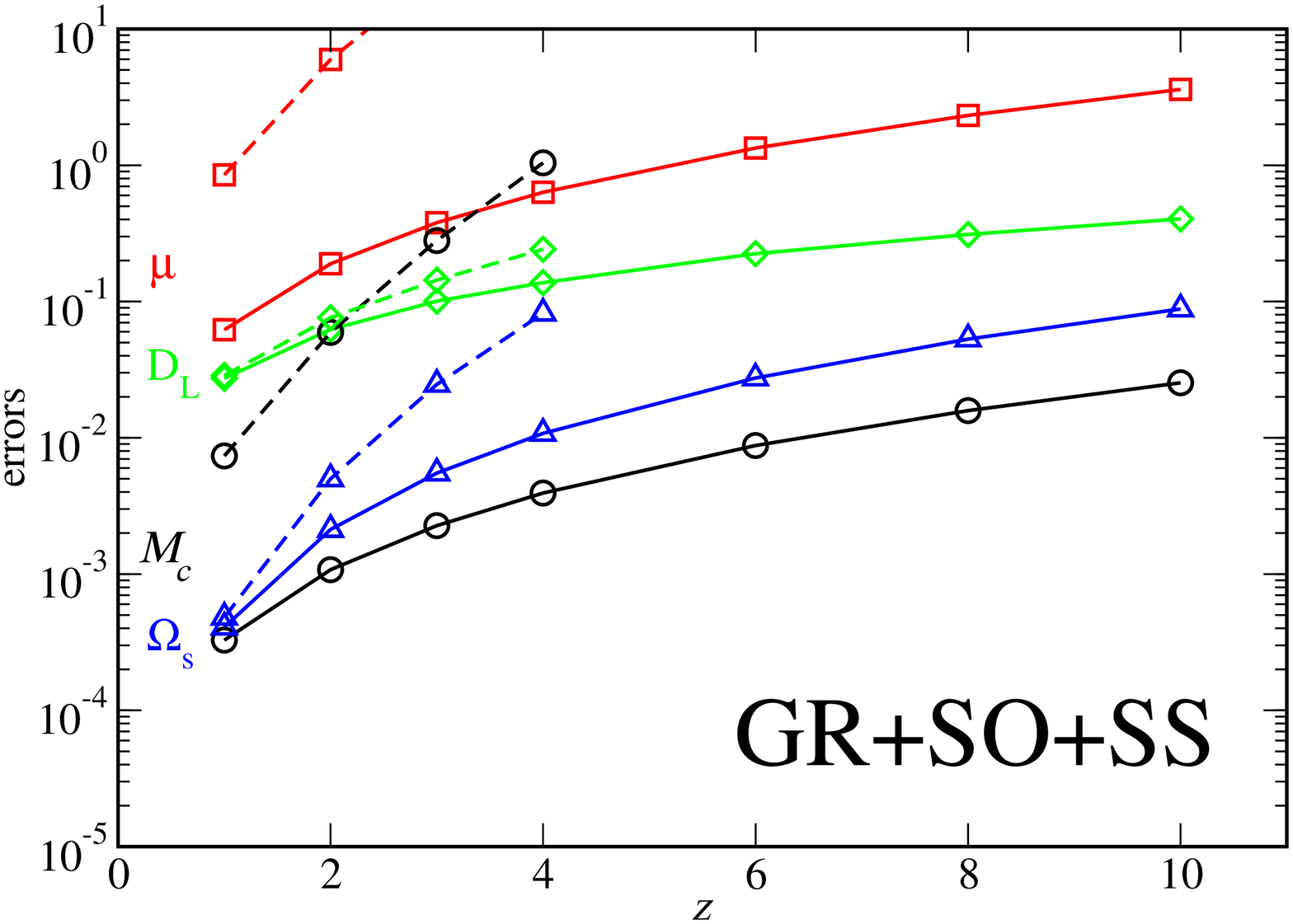,width=0.75\sizeonefig,angle=-90} \\
\end{tabular}
\caption{Errors on different parameters as a function of the redshift
$z$ for nonspinning BH-BH binaries with source mass  $(10^6+10^6)~M_\odot$
(solid lines) and $(10^7+10^7)~M_\odot$ (dashed lines).  We fix
$\Omega_\Lambda=0.7$, $\Omega_M=0.3$.  Errors are obtained by
averaging results of Monte Carlo simulations of $10^4$ binaries and
using two detectors.  Left: general relativistic non-spinning binaries
(GR); right: general relativistic binaries including SO and SS terms
(GR+SO+SS).  Circles refer to chirp mass ${\cal M}$, squares to
reduced mass $\mu$, triangles to angular resolution $\Omega_S$ in steradians,
diamonds to luminosity distance $D_L$.
\label{zeta}}
\end{center}
\end{figure*}

\begin{figure*}
\begin{center}
\begin{tabular}{cc}
\epsfig{file=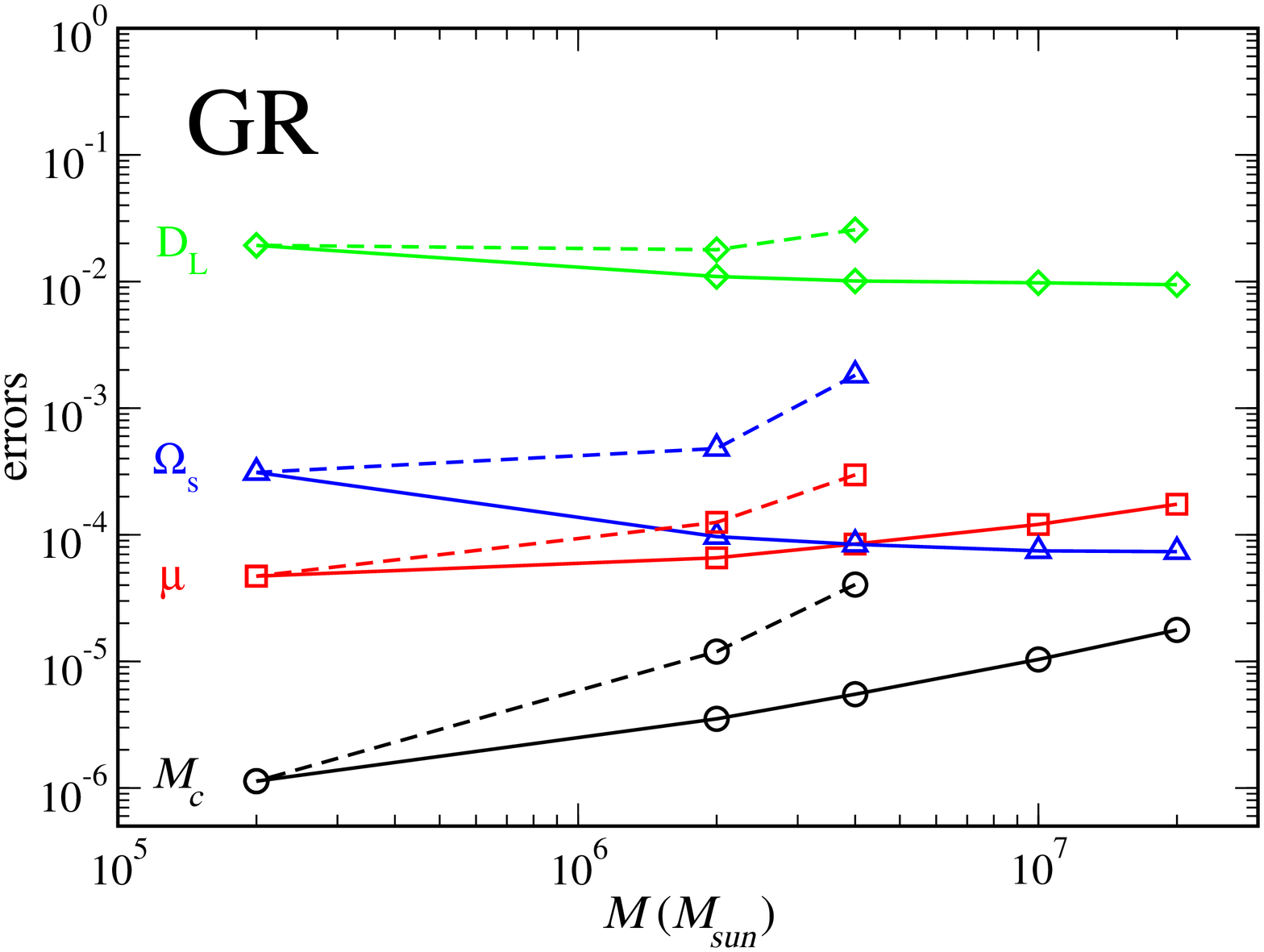,width=0.75\sizeonefig,angle=-90} &
\epsfig{file=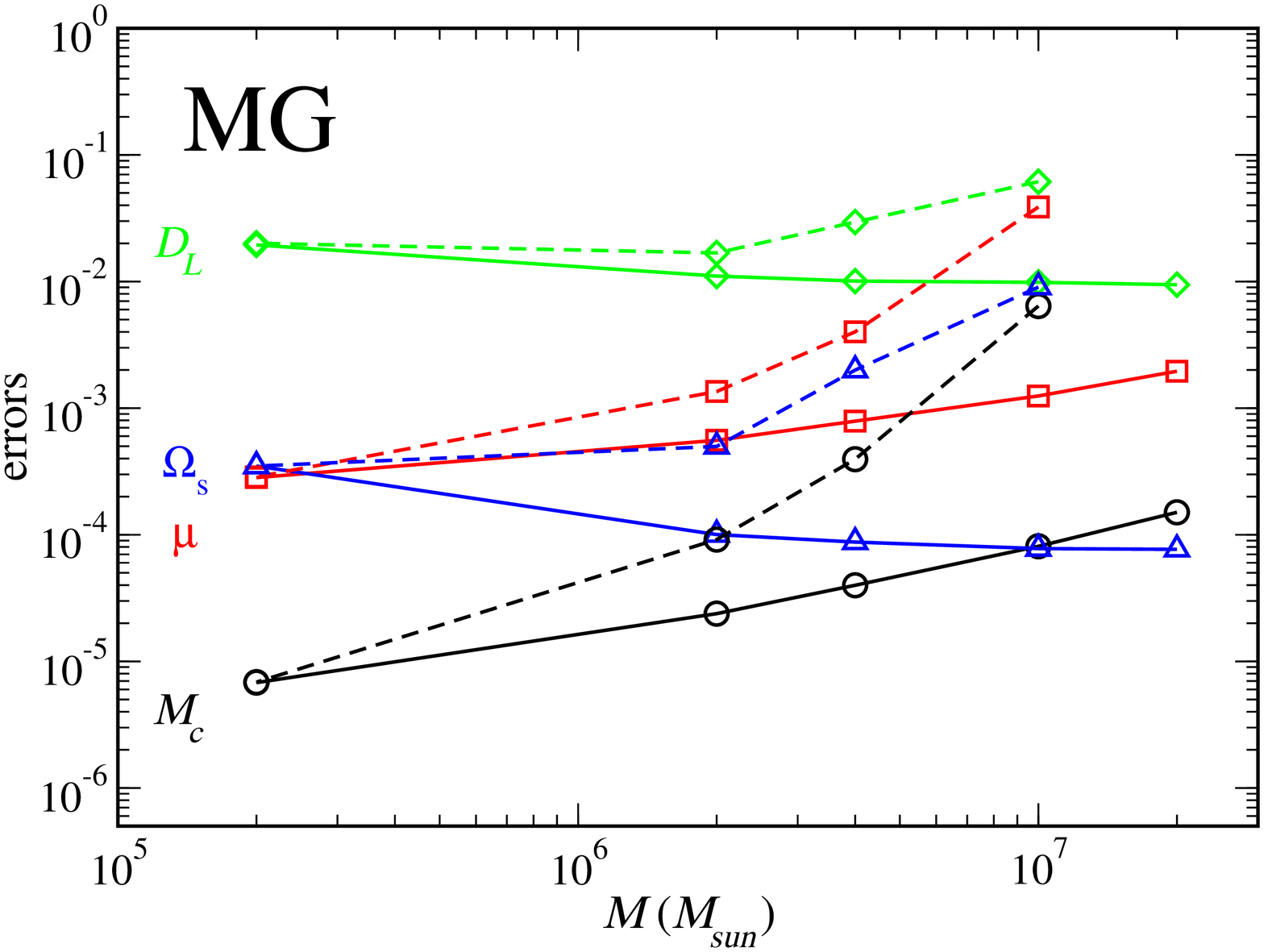,width=0.75\sizeonefig,angle=-90} \\
\end{tabular}
\caption{Errors on different parameters as a function of the total
binary mass $M$ (in solar masses) for equal-mass nonspinning BH-BH
binaries.  We fix $D_L=3$~Gpc, $\Omega_\Lambda=0.7$, $\Omega_M=0.3$.
Errors are obtained by averaging results of Monte Carlo simulations of
$10^4$ binaries and using two detectors. Left: general relativity
(GR); right: massive graviton theories (MG).  Solid lines assume that
the {\em LISA} noise can be extrapolated down to $f_{\rm
low}=10^{-5}$~Hz; dashed lines assume $f_{\rm
low}=10^{-4}$~Hz. Circles refer to chirp mass ${\cal M}$, squares to
reduced mass $\mu$, triangles to angular resolution $\Omega_S$ in
steradians, diamonds to luminosity distance $D_L$.
\label{errGRMG}}
\end{center}
\end{figure*}

\begin{figure*}
\begin{center}
\epsfig{file=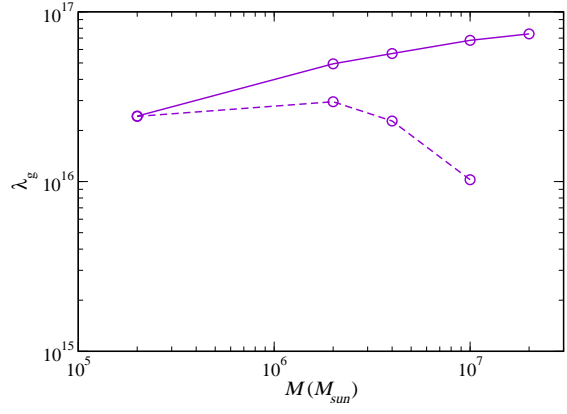,width=0.75\sizeonefig,angle=-90}
\caption{Bound on the graviton Compton wavelength as a function of the
total binary mass $M$ (in solar masses) for nonspinning equal-mass
BH-BH binaries. We fix $D_L=3$~Gpc, $\Omega_\Lambda=0.7$,
$\Omega_M=0.3$ and use both detectors.  The solid line assumes that
the {\em LISA} noise can be extrapolated down to $f_{\rm
low}=10^{-5}$~Hz; the dashed line assumes $f_{\rm low}=10^{-4}$~Hz.
\label{lambda}}
\end{center}
\end{figure*}

\end{widetext}

\end{document}